\documentclass[manuscript,screen,nonacm]{acmart}
\usepackage{graphicx}
\usepackage{enumitem}
\usepackage{multirow}
\usepackage{pifont}
\usepackage{multicol}
\usepackage{booktabs}
\usepackage{amsmath}
\usepackage[table,dvipsnames]{xcolor}
\usepackage{url}
\usepackage{xurl}
\usepackage{hyperref}
\usepackage{cleveref}
\usepackage{nth}
\usepackage{makecell}
\usepackage{subcaption}
\usepackage{svg}
\usepackage{tikz}
\usepackage{pgfplots}
\usepackage{pgfplotstable}
\usepackage{fontawesome5}
\usepackage{twemojis}
\usepackage{tcolorbox}
\usetikzlibrary{patterns, patterns.meta, calc, dateplot, matrix, decorations}
\pgfplotsset{compat=1.18}
\setlist[itemize]{leftmargin=*}
\setlist[enumerate]{leftmargin=*}

\definecolor{A5000color}{HTML}{E69F00}
\definecolor{A6000color}{HTML}{56B4E9}
\definecolor{2080color}{HTML}{CC79A7}
\definecolor{2070color}{HTML}{009E73}
\definecolor{1080color}{HTML}{0072B2}
\definecolor{P5000color}{HTML}{D55E00}
\definecolor{TitanXPcolor}{HTML}{000000}
\definecolor{TitanXcolor}{HTML}{93ab1a} %

\definecolor{grad1}{HTML}{000000}
\definecolor{grad2}{HTML}{009E73}
\definecolor{grad3}{HTML}{0072B2}
\definecolor{grad4}{HTML}{56B4E9}
\definecolor{grad5}{HTML}{F0E442}
\definecolor{grad6}{HTML}{E69F00}
\definecolor{grad7}{HTML}{D55E00}
\definecolor{grad8}{HTML}{CC79A7}

\definecolor{acces1}{HTML}{D81B60}
\definecolor{acces2}{HTML}{1E88E5}
\definecolor{acces3}{HTML}{FFC107}
\definecolor{acces4}{HTML}{004D40}

\definecolor{sweden}{HTML}{1065ab}
\definecolor{poland}{HTML}{b31529}
\definecolor{germany}{HTML}{fdb338}

\pgfplotsset{
    discard if/.style 2 args={
        x filter/.code={
            \edef\tempa{\thisrow{#1}}
            \edef\tempb{#2}
            \ifx\tempa\tempb
                
            \fi
        }
    },
    discard if not/.style 2 args={
        x filter/.code={
            \edef\tempa{\thisrow{#1}}
            \edef\tempb{#2}
            \ifx\tempa\tempb
            \else
                
            \fi
        }
    },
    discard if two/.style n args={4}{
        x filter/.code={
            \edef\tempa{\thisrow{#1}}
            \edef\tempb{#2}
            \edef\tempc{\thisrow{#3}}
            \edef\tempd{#4}
            \ifx\tempa\tempb
                \ifx\tempc\tempd
                \else
                    
                \fi
            \else
                
            \fi
        }
    },
    discard if three/.style n args={6}{
        x filter/.code={
            \edef\tempa{\thisrow{#1}}
            \edef\tempb{#2}
            \edef\tempc{\thisrow{#3}}
            \edef\tempd{#4}
            \edef\tempe{\thisrow{#5}}
            \edef\tempf{#6}
            \ifx\tempa\tempb
                \ifx\tempc\tempd
                    \ifx\tempe\tempf
                    \else
                        
                    \fi
                \else
                    
                \fi
            \else
                
            \fi
        }
    },
    common barplot/.style={
        height=2cm,
        enlarge x limits={abs=12pt},
        align=center,
        scale only axis,
        ybar,
        ymin=0,
        bar width=3mm,
        ymajorgrids=true,
        yminorgrids=true,
        ytick style={draw=none},
        xtick style={draw=none},
        legend cell align=left,
        legend style={
            /tikz/column 2/.style={
                column sep=0.3cm,
            },
            /tikz/column 4/.style={
                column sep=0.3cm,
            },
            draw=gray,
            nodes={scale=0.9},
        },
        major grid style={color=gray!20},
        yticklabel style={xshift=-10},
        xticklabel style={rotate=-30, xshift=-2pt, anchor=west, font=\small},
        draw=white,
    },
    pct labels/.style={
        nodes near coords,
        point meta=explicit,
        nodes near coords={
            \pgfmathprintnumber[
                fixed,
                precision=0,
                print sign
            ]{\pgfplotspointmeta}\%
        },
        every node near coord/.style={black, anchor=south, yshift=0, scale=0.9},
    },
}

\def\pctOf#1#2{100*(\thisrow{#1}/#2-1)}

\AtBeginDocument{%
  }

\setcopyright{acmlicensed}
\copyrightyear{2027}
\acmYear{2027}
\acmDOI{XXXXXXX.XXXXXXX}
\acmConference[FAccT '27]{ACM Conference on Fairness, Accountability, and Transparency}{June 21--24, 2027}{Porto, Portugal}
\acmISBN{978-1-4503-XXXX-X/2027/06}

\begin{document}

\title{Environmental Impact of Hardware Usage and Replacement Practices in Local AI Training}

\author{Marta López-Rauhut}
\correspondingauthor
\email{marta.lopez-rauhut@enpc.fr}
\orcid{0009-0004-7603-7069}
\affiliation{%
  \institution{LIGM, CNRS, UGE, ENPC, Institut Polytechnique de Paris}
  \city{Marne-la-Vallée}
  \country{France}
}
\affiliation{
    \institution{Université Paris-Saclay, CNRS, LISN}
    \city{Orsay}
    \country{France}
}

\author{Loic Landrieu}
\email{loic.landrieu@enpc.fr}
\orcid{0000-0002-7738-8141}
\author{Mathieu Aubry}
\email{mathieu.aubry@enpc.fr}
\orcid{0000-0002-3804-0193}
\affiliation{%
  \institution{LIGM, CNRS, UGE, ENPC, Institut Polytechnique de Paris}
  \city{Marne-la-Vallée}
  \country{France}
}

\author{Anne-Laure Ligozat}
\email{anne-laure.ligozat@lisn.fr}
\orcid{0000-0002-2188-3426}
\affiliation{%
  \institution{Université Paris-Saclay, CNRS, ENSIIE, LISN}
  \city{Orsay}
  \country{France}}

\renewcommand{\shortauthors}{López-Rauhut et al.}

\begin{abstract}
Improving the energy efficiency of AI workloads through next-generation hardware can reduce their environmental footprint, but producing this hardware also has significant impacts, and efficiency improvements can lead to rebound effects. Understanding the impacts of hardware upgrades and adopting good practices is particularly important for small-scale local AI infrastructure, such as that used by research laboratories, small companies, and individual practitioners. In this work, we investigate how hardware upgrades and usage patterns affect the environmental footprint of AI training at a local scale. Using direct power measurements, we evaluate the energy efficiency of a representative computer vision workload, training a Vision Transformer (ViT) across PyTorch versions, performance optimization techniques, and hardware setups spanning five NVIDIA GPU generations and multiple server configurations. We then assess and discuss the impact, both operational and embodied, of concrete hardware upgrade and workload evolution scenarios over a six-year period, considering three geographic locations and three environmental impact categories: climate change, water use, and abiotic resource depletion. Our study shows that the environmental benefits of hardware upgrades depend strongly on usage and workload evolution patterns, as well as on the location and environmental category considered. Our findings outline that the impact of good practices can be significant compared to hardware upgrades: software updates alone can reduce impacts or enable significant workload growth; turning off idle servers can lead to considerable impact reduction; and upgrading hardware components other than GPUs can exacerbate overall impacts for compute-bound workloads like the one we consider. Importantly, our study also points out the key importance of workload evolution patterns, and the potentially dramatic impact of increasing workloads to match hardware capacity. We hope that our analysis, based on the actual technical evolution of the last six years rather than uncertain projections, will provide insights to guide future decisions and policy design.
\end{abstract}




\maketitle

\section{Introduction}

The rapid growth of AI raises concerns about its environmental footprint, from the impacts induced during the production of computing hardware~\cite{falkFLOPsFootprintsResource2025a,MenapaceAndrea2025SAiA,wangEwasteChallengesGenerative2024a} to those of research, model training~\cite{morrisonHiddenCostThinking2026,lopez-rauhutEnvironmentalFootprintGenAI2026a,luccioniEstimatingCarbonFootprint2023}, and large-scale inference~\cite{luccioniPowerHungryProcessing2024,jeghamHowHungryAI2025,elsworth2025measuring}.

Successive generations of GPUs have brought substantial improvements in energy efficiency, reducing the energy required to execute a given workload~\cite{epoch2024mlhardwareenergyefficiency}. However, replacing older hardware with more efficient equipment is not necessarily environmentally beneficial~\cite{morandEnvironmentalImpactsLanguage2025,wrightEfficiencyNotEnough2025}. Manufacturing new GPUs consumes energy, minerals, metals, and water, contributing to greenhouse gas emissions, pollution, and resource depletion~\cite{falkMoreCarbonCradletograve2026}. Moreover, efficiency improvements may encourage greater hardware usage, offsetting their benefits through rebound effects~\cite{alcottJevonsParadox2005a,morandEnvironmentalImpactsLanguage2025}. Assessing the environmental footprint of AI therefore requires accounting for the trade-offs between hardware efficiency, manufacturing impacts, and usage and replacement practices.

While the environmental footprint of AI is often discussed in the context of frontier models trained in large data centers, much of AI research and development takes place on smaller infrastructures~\cite{glassReviewDigitalResearch2022}. Universities, research laboratories, small companies, and individual practitioners rely on local workstations and modest GPU servers to develop models, conduct experiments, and train the next generation of AI practitioners. Unlike large AI providers, these actors rarely have dedicated teams to optimize hardware utilization and energy efficiency. They must nevertheless make consequential decisions about their computing infrastructure: whether to purchase newer GPUs, upgrade other components, or improve existing systems through software optimizations.

A central question is therefore whether, and under what conditions, the operational savings offered by newer hardware compensate for the environmental impacts of its manufacturing and the shortened lifespan of the equipment it replaces. Crucially, the answer depends on which environmental impacts are considered: an upgrade may reduce carbon emissions while increasing water use or mineral resource depletion. This trade-off must also account for alternatives to replacement: software optimizations, framework updates, and GPU power capping may offer substantial efficiency gains without requiring new equipment.

In this work, we investigate how software, hardware, and usage choices have shaped the environmental footprint of local AI training over the past six years. We first measure the energy efficiency and throughput of a representative local computer vision workload, namely training a Vision Transformer for ImageNet classification, and then systematically evaluate the impact of PyTorch optimizations and versions, and of hardware components including CPUs, memory, storage, and five generations of NVIDIA GPUs with power capping.

We then use these measurements to retrospectively assess the environmental consequences of different hardware replacement and usage choices, and workload evolution patterns over a six-year period. We consider different workload and usage scenarios, three geographic locations, and three environmental impact categories: climate change, water use, and mineral and metal resource depletion. This allows us to examine how the balance between operational savings and manufacturing impacts varies with hardware generations, usage patterns, and local electricity mixes.

Our results highlight: (i) the importance of good practices, such as regular software updates and performance optimizations, limiting hardware upgrades to key components, and turning off idle servers; (ii) the magnitude of embodied impacts; and (iii) the moderate environmental savings brought by hardware efficiency improvements compared to potential rebound effects. Although specific to the hardware, software, and workloads considered, these findings offer insights into the consequences of past practices that can help inform future infrastructure decisions and policies.

\section{Related Work}

Previous studies have already highlighted the opportunities to reduce operational and embodied impacts through conscious hardware and software choices~\cite{wuSustainableAIEnvironmental2022a, wuEfficiencyScalingAI2024,makinSustainableAITraining2025}, focusing on the greenhouse gas emissions of industry-scale AI development and deployment. We bring a new perspective by translating this viewpoint to local AI infrastructures, and enrich the discussion by incorporating two other impact categories: water use and abiotic resource depletion. In this section, we summarize existing work first on the \emph{energy efficiency} of AI, then on its \emph{environmental impacts}, emphasizing the few works that specifically address the environmental impacts of \emph{hardware replacement} cycles.

\subsection{Energy Efficiency of AI}

The energy efficiency of AI workloads depends on the interplay between hardware, software, model architecture, and algorithmic choices. A substantial body of work benchmarks the influence of these factors for training and inference tasks, particularly in the fields of language modeling~\cite{nelapatlaSurveyEnergyEfficiencyMechanisms2026,samsiWordsWattsBenchmarking2023a} and computer vision~\cite{mavromatisComputingLimitsEmpirical2024,yangHiddenJoulesEvaluating2025}. Beyond the efficiency of particular workloads, some studies have assessed energy consumption trade-offs across the life cycle of AI models, such as using more energy-intensive training strategies to reduce inference consumption. 

Network architecture is often one of the main objects of study in \emph{training} efficiency benchmarks, across the domains of language modeling~\cite{enskatEvaluatingMFUProxy2026,liuGreenAIExploring2024} and computer vision~\cite{mavromatisComputingLimitsEmpirical2024,aquino-britezEnergyConsumptionIndex2025,cortesEnergyVariationTraining2026}. Instead of comparing model architectures and training strategies, we fix our task to training a Vision Transformer (ViT)~\cite{dosovitskiyImageWorth16x162021} for image classification. The ViT has been a reference architecture in computer vision for the last six years, and we believe its training to be representative of workloads that can be run on a local machine. Besides model architecture comparisons, research on computer vision training efficiency investigates factors that we also assess in this work: software framework choices~\cite{liEvaluatingEnergyEfficiency2016a, wangBenchmarkingPerformanceEnergy2020}, lower precision floating point formats~\cite{wangBenchmarkingPerformanceEnergy2020,tchakouteEnergyAwareDeepLearning2025,enskatEvaluatingMFUProxy2026}, and hardware accelerator choices~\cite{wangBenchmarkingPerformanceEnergy2020,mayrAIApplicationBenchmarking2026}.

\emph{Inference} efficiency has also attracted increasing attention in the field of language modeling, following the large-scale deployment of LLMs~\cite{samsiWordsWattsBenchmarking2023a,aliAssessingSustainabilityLLM2026,fernandezEnergyConsiderationsLarge2025}. In the field of computer vision, inference efficiency benchmarks assess similar factors as training benchmarks: model architecture~\cite{yaoEvaluatingAnalyzingEnergy2021}, deep learning framework~\cite{malenzaExploringEnergyConsumption2025,alizadehGreenAIPreliminary2024}, floating point precision, and hardware~\cite{liuMetricsEvaluationsComputational2025,yangHiddenJoulesEvaluating2025,yaoEvaluatingAnalyzingEnergy2021}.

Another line of research looks beyond a single training or inference task, and instead examines energy trade-offs across different stages of the \emph{model life cycle}~\cite{mehlinEnergyefficientDeepLearning2023,khanGreenAITechniques2026}, considering techniques such as pruning, distillation, and neural architecture search~\cite{liFullCycleEnergyConsumption2021}, pre-training strategies~\cite{gowdaWattWhatRethinking2024}, and multi-task learning~\cite{geisslerPowerTrainingHow2024}.

\subsection{Environmental Impacts of AI}

Several of the works above link energy efficiency to a reduction in greenhouse gas emissions, which is a valid but incomplete viewpoint on AI sustainability~\cite{schwartzGreenAI2020}. AI sustainability requires a \emph{life cycle assessment} perspective taking into account a range of environmental impact categories beyond climate change, as well as the embodied impacts of AI hardware infrastructure~\cite{plociennikLifeCycleAssessment2025,dokicSustainabilityAISystematic2024}, arising from the raw material extraction, manufacturing, transportation, and end of life of hardware equipment.

Published environmental assessments of AI systems have addressed the impacts induced during the \emph{R\&D and training} phases of specific models, with older studies limited to greenhouse gas emissions impacts~\cite{strubellEnergyPolicyConsiderations2019,luccioniEstimatingCarbonFootprint2023,pattersonCarbonEmissionsLarge2021}, and more recent work incorporating additional impact categories, such as water use~\cite{morrisonHolisticallyEvaluatingEnvironmental2025a,morrisonHiddenCostThinking2026,liMakingAILess2025a} and abiotic resource depletion~\cite{lopez-rauhutEnvironmentalFootprintGenAI2026a}. 

A parallel body of work targets the environmental impacts of model \emph{inference}, given that they can quickly match training impacts after model deployment~\cite{jeghamHowHungryAI2025}. Several studies assess the operational greenhouse gas emissions~\cite{luccioniPowerHungryProcessing2024} and water footprint~\cite{jeghamHowHungryAI2025,minozaMLEcoLyzerQuantifyingEnvironmental2025} of AI inference, with others extending their assessment to incorporate embodied emissions due to hardware manufacturing~\cite{elsworth2025measuring,liEcoServeDesigningCarbonAware2025,nguyenSustainableLargeLanguage2024}, and mineral and metal depletion impacts~\cite{berthelotEstimatingEnvironmentalImpact2024}.

Finally, some studies focus specifically on the impacts of \emph{AI hardware} manufacturing, from just greenhouse gas emissions~\cite{schneiderLifeCycleEmissionsAI2025,guptaChasingCarbonElusive2022} to assessments spanning sixteen environmental impact categories~\cite{falkMoreCarbonCradletograve2026}, passing through detailed mineral and metal depletion impact breakdowns~\cite{falkFLOPsFootprintsResource2025a}. 

On the topic of \emph{hardware replacement}, one study investigates the trade-off between acquiring more energy-efficient GPU models to reduce operational emissions, and the additional impacts induced through hardware production~\cite{baeuerle5green}, identifying carbon break-even points for GPU fleet replacement at the data center scale across various AI workload types, locations, and GPU generations. While this study provides valuable insights, it cannot be easily translated at the scale of individual workstations and small research infrastructures, which are the focus of our study. We evaluate concrete hardware replacement and workload evolution scenarios for a single tower server, accounting for operational and embodied impacts in different geographic locations, analyzing three environmental impact categories: climate change, water use, and abiotic resource depletion.

\section{Methodology}

\subsection{Workload}
To provide a consistent benchmark for AI training efficiency, we consider a standard computer vision workload: training a Vision Transformer (ViT)~\cite{dosovitskiyImageWorth16x162021} for image classification on ImageNet-1k. The ViT has been a standard architecture in the field for the last six years and is representative of modern transformer-based architectures used for many tasks. While datasets depend heavily on the target application, ImageNet-1k remains a reference and can be seen as representative of medium-sized datasets, sufficiently large to train ViT architectures but small enough to easily handle on local workstations. More precisely, we train a ViT-B/16 with standard hyperparameters and data augmentation, detailed in \cref{sec:A_baseline_details}. We use PyTorch for training, given that it is the dominant framework for deep learning research and development.

\subsection{Default configuration}
\label{sec:baseline}

We vary hardware in our experiment, but unless stated otherwise we use an isolated Dell Precision 7920 server equipped with a single NVIDIA RTX A6000 GPU (48 GB), two Intel Xeon Silver 4208 CPUs, four 16 GB RAM modules (2400 MT/s), one 4 TB Dell 6KR2M SATA hard disk drive, and two 2 TB SK Hynix PC801 NVMe solid-state drives. By default, the dataset is stored on an NVMe drive. The server runs Ubuntu 22.04.5 LTS (kernel 6.8.0-134-generic) with NVIDIA driver 580.173.02. Our default software environment uses Python 3.12.12, PyTorch 2.9.1 (CUDA 12.6), and Torchvision 0.24.1.

\subsection{Measurements and metrics}

We monitor power consumption at one-second intervals using a physical wattmeter that directly measures the electricity consumption of the entire server. We simultaneously log separate CPU, GPU, and RAM power consumption and additional system metrics using CodeCarbon~\cite{codecarbon}.

For each experiment, we monitor the first training epoch, typically lasting between thirty minutes and three hours. We exclude the first ten minutes to avoid initialization effects and compute average power consumption and throughput over a stable interval of at least fifteen minutes. We experimentally verify that these measurements are representative of subsequent epochs.

We focus our analysis on training \emph{throughput} and energy \emph{efficiency}. Given $B$ the batch size, $i$ the number of iterations in the measured interval, and $t$ the interval duration in seconds, we define throughput and energy efficiency as follows:
\hfill
\begin{minipage}[b]{0.39\textwidth}
\[\text{Throughput}=\frac{B\times i}{t}\]
\end{minipage}
\hfill
\begin{minipage}[b]{0.59\textwidth}
\[\text{Efficiency}=\frac{B\times i}{E}; \quad E=\frac{P_{t_1}+P_{t_n}}{2} +\sum_{i=t_2}^{t_{n-1}}P_{t_i},\]
\end{minipage}
with $E$ the energy in joules consumed during the measured interval, estimated by trapezoidal integration of the power consumption measurements $P_{t_i}$ in the temporal interval $t_1..t_n$. Energy efficiency, measured in images per joule (img/J), is the main metric we consider, but we occasionally consider its trade-off with throughput, in images per second (img/s). It should be noted that we assess raw sample processing efficiency, and not full training efficiency: some of the settings that we consider, such as batch size variations and mixed precision, may impact time to convergence.

\section{Hardware and Software Impact on Energy Efficiency}

In this section, we investigate how software and hardware affect energy efficiency. We first examine efficiency gains achievable through software in \cref{sec:pytorch_optimizations}, then those related to hardware and its coupling to software in \cref{sec:gpu}.

\begin{figure}[t]
    \centering
    \input{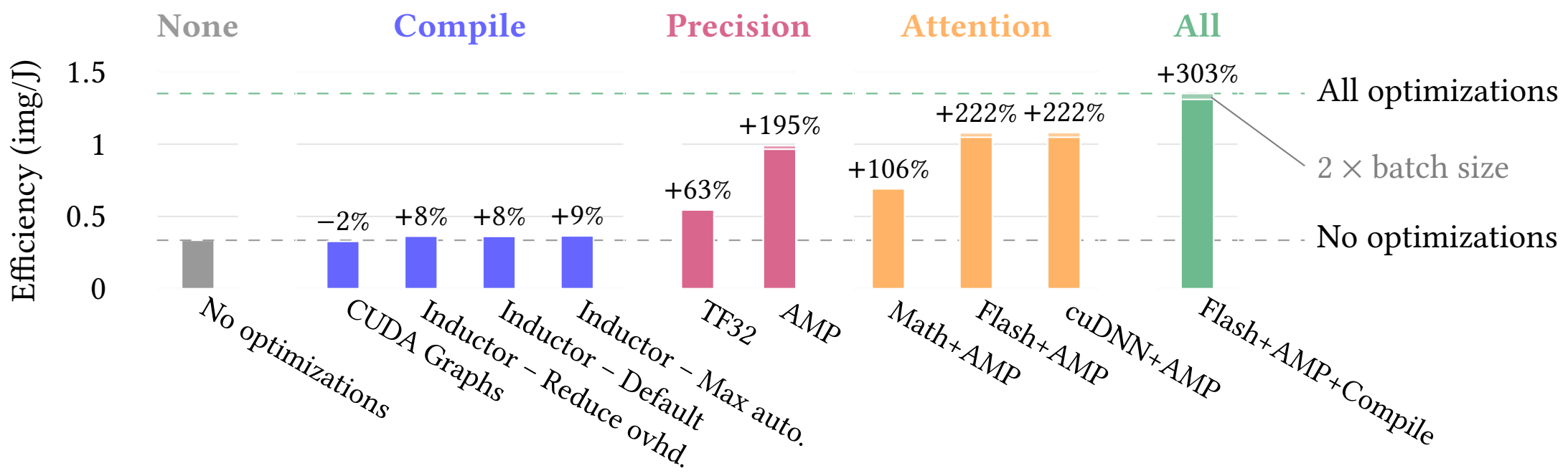}

\pgfplotsset{
    optimization panel/.append style={ymax=\optiYmaxEff, table/x=exp_name, table/y=img_per_j},
    pct labels/.append style={table/meta expr={\pctOf{img_per_j}{\noOptiEff}}},
}

\begin{tikzpicture}
    \begin{axis}[
        no opti panel,
        ylabel={Efficiency (img/J)},
    ]
        \addplot[fill=gray!80] table {\baselinedata};
    \end{axis}

    \draw[dashed, gray!80] (rel axis cs:0.5,\noOptiEff/\optiYmaxEff) -- (rel axis cs:14.5,\noOptiEff/\optiYmaxEff) node[right, black] {No optimizations};
    \draw[dashed, ForestGreen!60] (rel axis cs:0.5,\allOptiEff/\optiYmaxEff) -- (rel axis cs:14.5,\allOptiEff/\optiYmaxEff) node[right, black] {All optimizations};

    \begin{axis}[
        compile panel,
    ]
        \addplot[fill=blue!60, pct labels] table {\compiledata};
    \end{axis}

    \begin{axis}[
        precision panel,
    ]
        \addplot[fill=purple!40, pct labels]
            table[y=img_per_j_large_bs, meta expr={\pctOf{img_per_j_large_bs}{\noOptiEff}}] {\precisiondata};
        \addplot[fill=purple!60] table {\precisiondata};
    \end{axis}

    \begin{axis}[
        attention panel,
    ]
        \addplot[fill=orange!40, pct labels]
            table[y=img_per_j_large_bs, meta expr={\pctOf{img_per_j_large_bs}{\noOptiEff}}] {\attentiondata};
        \addplot[fill=orange!60] table {\attentiondata};
    \end{axis}

    \begin{axis}[
        all opti panel,
    ]
        \addplot[fill=ForestGreen!40, pct labels]
            table[y=img_per_j_large_bs, meta expr={\pctOf{img_per_j_large_bs}{\noOptiEff}}] {\optidata};
        \addplot[fill=ForestGreen!60] table {\optidata};
    \end{axis}

    \draw[gray] (rel axis cs:13.36,1.325/\optiYmaxEff) -- (rel axis cs:14.5,0.85/\optiYmaxEff) node[right, gray] {2 $\times$ batch size};
\end{tikzpicture}
    \caption{\textbf{Impact of PyTorch optimizations on training efficiency.} We measure efficiency at batch size 256, and additionally at batch size 512 when made possible by the lower memory usage of Automatic Mixed Precision (AMP) (bar extensions). Throughput follows a similar trend (\cref{fig:optimization_throughput} in \cref{sec:C_throughput}).}
    \label{fig:optimization_efficiency}
    \Description{Fully described in the text.}
\end{figure}

\subsection{Impact of Software-Side Optimizations} \label{sec:pytorch_optimizations}

We first evaluate the efficiency gains related to the optimizations outlined in the PyTorch performance tuning guide~\cite{PyTorchPerformanceTuningGuide} that are most relevant to our workload, then analyze the efficiency that can be attained by successive PyTorch versions.

\paragraph{PyTorch optimizations}

In \cref{fig:optimization_efficiency}, we report the impact of different optimizations, taking as reference a configuration with no optimizations, i.e., no model compilation, Float32 precision, and the default PyTorch attention based on Meta's xFormers library~\cite{rabeSelfattentionDoesNot2022}.

PyTorch's \texttt{torch.compile} translates model operations into optimized GPU kernels and reduces launch overheads. We compare four compilation configurations~\cite{TorchcompilePyTorch214}: \emph{CUDA Graphs}, \emph{Inductor -- Default}, \emph{Inductor -- Reduce overhead}, and \emph{Inductor -- Max autotune}. The results, in \cref{fig:optimization_efficiency} (\textcolor{blue!60}{\textbf{Compile}} bars), show that the different TorchInductor compile modes improve efficiency for our workload by approximately $8\%$. CUDA Graphs, which only reduce kernel launch overheads, have no positive impact on efficiency for our reference ViT training experiment.

Using lower-precision floating-point formats can accelerate computation and reduce memory usage. PyTorch currently supports two main approaches for reduced-precision computation: \emph{TensorFloat32 (TF32)} and \emph{Automatic Mixed Precision (AMP)}. As shown in \cref{fig:optimization_efficiency} (\textcolor{purple!60}{\textbf{Precision}} bars), they improve efficiency by $63\%$ and $188\%$, respectively. AMP also reduces memory bandwidth requirements by storing intermediate results in lower precision. Its smaller memory footprint allows us to double the batch size, increasing the gain up to $195\%$ (bar extensions in \cref{fig:optimization_efficiency}).

PyTorch provides several backends for computing scaled dot-product attention, with different levels of optimization~\cite{Torchnnfunctionalscaled_dot_product_attentionPyTorch214}: \emph{Math}, a C++ implementation, and two versions of FlashAttention-2~\cite{flashAttention}, \emph{Flash} and \emph{cuDNN}. Since both \emph{Flash} and \emph{cuDNN} require AMP, we compare all attention backends with AMP enabled. As can be seen in \cref{fig:optimization_efficiency} (\textcolor{orange!60}{\textbf{Attention}} bars), \emph{Flash} and \emph{cuDNN} provide an additional and similar boost compared to AMP.

Taken together, TorchInductor default compilation, AMP, and Flash attention boost efficiency by more than $300\%$, and thus we assume in the following sections that by default these optimizations are used whenever possible.
\begin{figure}[t]
    \centering
    \input{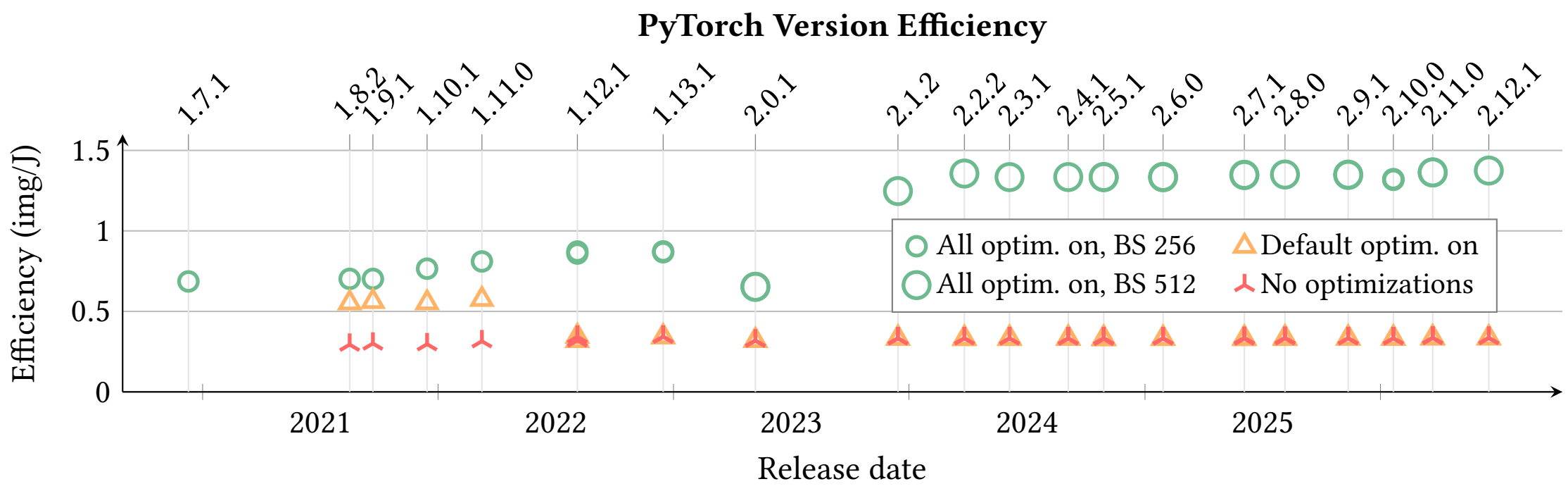}

\begin{tikzpicture}
    \begin{axis}[
        release year axis,
        ylabel={Efficiency (img/J)},
    ]
    \end{axis}
    \begin{axis}[
        release plot axis,
        title={\textbf{PyTorch Version Efficiency}},
        ymax=1.6,
    ]
        \pytorchOptiSmallBatch{img_per_j}\label{plot:pytorch_all_optis}
        \pytorchDefault{img_per_j}\label{plot:pytorch_default_optis}
        \pytorchOptiLargeBatch{img_per_j}
        \pytorchNoOpti{img_per_j}\label{plot:pytorch_no_optis}
        \legend{{All optim. on, BS 256}, Default optim. on, {All optim. on, BS 512}, No optimizations}
    \end{axis}
\end{tikzpicture}
    \caption{\textbf{Impact of PyTorch versions on training efficiency.} We compare default settings to all applicable optimizations enabled or disabled. Larger green markers indicate a batch size of 512, which we use whenever possible instead of 256. Overlapping markers correspond to different CUDA builds of the same version. Throughput follows a similar trend (\cref{fig:pytorch_throughput} in \cref{sec:C_throughput}).}
    \label{fig:pytorch_efficiency}
    \Description{Fully described in the text.}
\end{figure}

\paragraph{Temporal evolution with PyTorch releases} \label{sec:pytorch_versions}

To ground these optimizations in a temporal perspective, we evaluate in \cref{fig:pytorch_efficiency} how the training efficiency of our Vision Transformer has evolved across PyTorch releases, from version 1.7 (late 2020) to 2.12 (mid-2026). For each release, we consider three configurations: \emph{all optimizations on}, enabling all applicable optimizations mentioned above, i.e., AMP in all versions, plus compilation and Flash attention starting in version 2.0; \emph{default optimizations on}, retaining PyTorch's default settings; and \emph{no optimizations}, explicitly disabling all optimizations. We always select the latest patch version for each release (e.g., 2.2.2 for 2.2). We use the largest batch size possible between 256 and 512, which is 256 for all settings without AMP and all versions before 2.0, as well as 2.10. For version 1.7, even batches of size 256 are too large to fit in memory without AMP. Also note that we tested all available CUDA builds compatible with our hardware and driver, up to CUDA 13.0, but found no significant performance differences.

With all optimizations disabled (\ref{plot:pytorch_no_optis}), efficiency has remained remarkably stable across PyTorch releases. In contrast, default settings (\ref{plot:pytorch_default_optis}) have changed substantially with, surprisingly, a performance drop in PyTorch 1.12. Indeed, in previous versions, TF32 was enabled by default, making them approximately $85\%$ more efficient. PyTorch later disabled TF32 by default to favor numerical accuracy~\cite{CUDASemanticsPyTorch,pytorchAddHighLevelRoot}. 

From PyTorch 1.7 to 1.13, using AMP (\ref{plot:pytorch_all_optis}), we can see efficiency progressively improve, achieving a $27\%$ gain over two years. PyTorch 2.0 introduced Flash attention as well as model compilation as a beta feature, but initially reduced efficiency by $25\%$. The more mature implementation in PyTorch 2.1 instead improved efficiency by $43\%$ relative to 1.13. FlashAttention-2, introduced in PyTorch 2.2, could be responsible for a further $9\%$ improvement, after which efficiency remained largely stable.

\begin{tcolorbox}[title=\textbf{Key Takeaways}, colback=gray!15, colframe=gray]
\begin{itemize}
    \item \textbf{Optimization}: the combination of Automatic Mixed Precision (AMP), compilation, and Flash attention boosts efficiency by a factor of 4. Optimizing code is thus critical.  
    \item \textbf{PyTorch evolution}: default PyTorch behavior changed with version 1.12, disabling all optimizations. While efficiency without optimizations did not evolve from 2021 to 2026, performance with optimizations regularly increased from 2021 to 2024 and regular updates were thus important even with fixed optimized code. 
\end{itemize}
\end{tcolorbox}

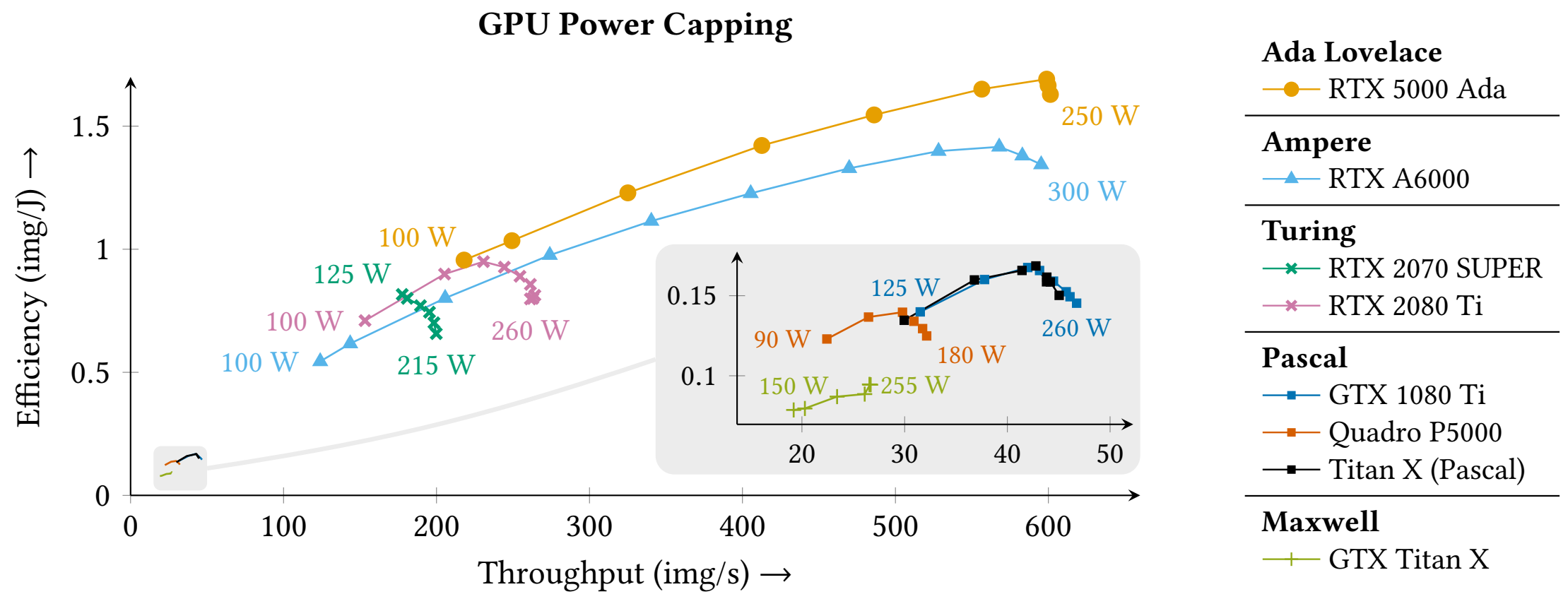
\begin{figure}[t]
    \centering
    \begin{tikzpicture}
\begin{axis}[
    name=mainplot,
    axis lines=left,
    title={\textbf{GPU Power Capping}},
    width={0.73\textwidth},
    height=5.5cm,
    ymax=1.7,
    xmax=660,
    xmin=0,
    ymin=0,
    xlabel={Throughput (img/s) $\rightarrow$},
    ylabel={Efficiency (img/J) $\rightarrow$},
    point meta=explicit,
]
\coordinate (bottomSW) at (axis cs:15,0.02);
\coordinate (bottomN) at (axis cs:15,0.2);
\coordinate (bottomE) at (axis cs:50, 0.11);
\coordinate (bottomNE) at (axis cs:50,0.2);
\fill[gray!15, rounded corners=2] (bottomSW) rectangle (bottomNE);
\addplot[mark=*, mark size=2pt, line width=0.5pt, color=A5000color] table [
    x=img_per_s,
    y=img_per_j,
    meta=power_limit_w,
    col sep=comma,
    discard if two={batch_size}{256}{compile}{True},
]{data/A5000.csv};
\label{plot:A5000}
\node[below right, scale=0.9, A5000color] at (axis cs:601.14, 1.6293) {250 W};
\node[above left, scale=0.9, A5000color] at (axis cs:220.23, 0.96531) {100 W};

\addplot[mark=triangle*, mark size=2pt, line width=0.5pt, color=A6000color] table [
    x=img_per_s,
    y=img_per_j,
    meta=power_limit_w,
    col sep=comma,
    discard if two={batch_size}{512}{compile}{True},
]{data/A6000.csv};
\label{plot:A6000}
\node[below, scale=0.9, A6000color, yshift=-3] at (axis cs:625.29, 1.3643) {300 W};
\node[left, scale=0.9, A6000color, xshift=-3] at (axis cs:123.98, 0.54415) {100 W};

\addplot[mark=x, mark size=2pt, line width=0.5pt, color=2070color, mark options={line width=1pt}] table [
    x=img_per_s,
    y=img_per_j,
    meta=power_limit_w,
    col sep=comma,
    discard if two={batch_size}{64}{compile}{True},
]{data/2070_super.csv};
\label{plot:2070SUPER}
\node[below, scale=0.9, 2070color, yshift=-3] at (axis cs:199.84, 0.65725) {215 W};
\node[above left, scale=0.9, 2070color] at (axis cs:177.59, 0.81625) {125 W};

\addplot[mark=x, mark size=2pt, line width=0.5pt, color=2080color, mark options={line width=1pt}] table [
    x=img_per_s,
    y=img_per_j,
    meta=power_limit_w,
    col sep=comma,
    discard if two={compile}{True}{batch_size}{64},
]{data/2080Ti.csv};
\label{plot:2080Ti}
\node[below, scale=0.9, 2080color, yshift=-3] at (axis cs:261.35, 0.79806) {260 W};
\node[left, scale=0.9, 2080color, xshift=-3] at (axis cs:153.24, 0.71001) {100 W};

\addplot[mark=\empty, line width=0.5pt, 1080color] table [
    x=img_per_s,
    y=img_per_j,
    meta=power_limit_w,
    col sep=comma,
    discard if not={batch_size}{32},
]{data/1080Ti.csv};

\addplot[mark=\empty, line width=0.5pt, P5000color] table [
    x=img_per_s,
    y=img_per_j,
    meta=power_limit_w,
    col sep=comma,
    discard if not={batch_size}{16},
]{data/P5000.csv};

\addplot[mark=\empty, line width=0.5pt, TitanXPcolor] table [
    x=img_per_s,
    y=img_per_j,
    meta=power_limit_w,
    col sep=comma,
    discard if not={batch_size}{32},
]{data/TitanXPascal.csv};

\addplot[mark=\empty, line width=0.5pt, TitanXcolor] table [
    x=img_per_s,
    y=img_per_j,
    meta=power_limit_w,
    col sep=comma,
    discard if not={batch_size}{16},
]{data/TitanX.csv};
\coordinate (insetPosition) at (rel axis cs:0.52,0.05);

\coordinate (topSW) at (rel axis cs:0.52,0.05);
\coordinate (topW) at (rel axis cs:0.52,0.325);
\coordinate (topNE) at (rel axis cs:1.0,0.6);
\fill[gray!15, rounded corners=5] (topSW) rectangle (topNE);
\draw[very thick, gray!15] (bottomE) to [bend right=5] (topW);

\end{axis}
\node [shape=rectangle, align=center, scale=0.9, anchor=north west](legend_table1) at ([xshift=25,yshift=14]mainplot.north east) {
    \begin{tabular}{r@{ }l}
        \multicolumn{2}{l}{\textbf{Ada Lovelace}} \\
        \ref{plot:A5000} & RTX 5000 Ada \\ \midrule
        \multicolumn{2}{l}{\textbf{Ampere}} \\
        \ref{plot:A6000} & RTX A6000 \\\midrule
        \multicolumn{2}{l}{\textbf{Turing}} \\
        \ref{plot:2070SUPER} & RTX 2070 SUPER \\
        \ref{plot:2080Ti} & RTX 2080 Ti \\\midrule
        \multicolumn{2}{l}{\textbf{Pascal}} \\
        \ref{plot:1080ti} & GTX 1080 Ti \\
        \ref{plot:P5000} & Quadro P5000 \\
        \ref{plot:TitanXP} & Titan X (Pascal) \\\midrule
        \multicolumn{2}{l}{\textbf{Maxwell}} \\
        \ref{plot:TitanX} & GTX Titan X \\
    \end{tabular}
};

\begin{axis}[
    at={(insetPosition)},anchor={outer south west},
    width={0.35\textwidth},
    height=3.2cm,
    axis lines=left,
    enlarge y limits=0.1,
    enlarge x limits=0.2,
    label style={font=\small},
    tick label style={font=\small},
    point meta=explicit,
]

\addplot[mark=square*, mark size=1pt, line width=0.5pt, color=1080color] table [
    x=img_per_s,
    y=img_per_j,
    meta=power_limit_w,
    col sep=comma,
    discard if not={batch_size}{32},
]{data/1080Ti.csv};
\label{plot:1080ti}
\node[below, scale=0.8, 1080color, yshift=-2pt] at (axis cs:46.742, 0.14534) {260 W};
\node[above, scale=0.8, xshift=-5pt, yshift=2pt, 1080color] at (axis cs:31.518, 0.1399) {125 W};

\addplot[mark=square*, mark size=1pt, line width=0.5pt, color=P5000color] table [
    x=img_per_s,
    y=img_per_j,
    meta=power_limit_w,
    col sep=comma,
    discard if not={batch_size}{16},
]{data/P5000.csv};
\label{plot:P5000}
\node[below right, scale=0.8, P5000color] at (axis cs:32.144, 0.12490) {180 W};
\node[left, scale=0.8, P5000color, xshift=-2pt] at (axis cs:22.428, 0.12298) {90 W};

\addplot[mark=square*, mark size=1pt, line width=0.5pt, color=TitanXPcolor] table [
    x=img_per_s,
    y=img_per_j,
    meta=power_limit_w,
    col sep=comma,
    discard if not={batch_size}{32},
]{data/TitanXPascal.csv};
\label{plot:TitanXP}

\addplot[mark=+, mark size=2pt, line width=0.5pt, color=TitanXcolor] table [
    x=img_per_s,
    y=img_per_j,
    meta=power_limit_w,
    col sep=comma,
    discard if not={batch_size}{16},
]{data/TitanX.csv};
\label{plot:TitanX}
\node[right, scale=0.8, TitanXcolor] at (axis cs:26.692, 0.094418) {255 W};
\node[above, scale=0.8, TitanXcolor, yshift=2pt] at (axis cs:19.196, 0.078658) {150 W};

\end{axis}
\end{tikzpicture}%
    \caption{\textbf{GPU power capping.} Throughput and efficiency when power capping five generations of NVIDIA GPUs with all supported PyTorch \emph{optimizations enabled} at their \emph{best batch size}, i.e., the one that maximizes efficiency and throughput. We power cap each GPU in intervals of 20 W ranging from the maximum to the minimum allowed power caps, always explicitly including the latter. We do not show power caps more than 10 W beyond the thermal design power of each card. Additional results can be found in \cref{sec:D_gpu_power_capping_all}.}
    \Description{Line graph showing throughput in images per second on the X axis against efficiency in images per joule on the Y axis for Maxwell (Titan X), Pascal (GTX 1080 Ti, Quadro P5000, Titan X Pascal), Turing (RTX 2070 SUPER, RTX 2080 Ti), Ampere (RTX A6000), and Ada Lovelace (RTX 5000 Ada) GPUs. Labels along each line indicate power limits in watts, with higher power limit values on the top right and lower values in the bottom left of the graph. Throughput ranges from 125 to 600 for Turing, Ampere, and Ada Lovelace cards, and from 19 to 47 for the rest. Efficiency ranges from 0.5 to 1.7 for Turing, Ampere, and Ada Lovelace cards, and from 0.08 to 0.17 for the rest. Analyzed in the text.}
    \label{fig:gpu_power_capping}
\end{figure}

\begin{figure}[t]
    \centering
    \input{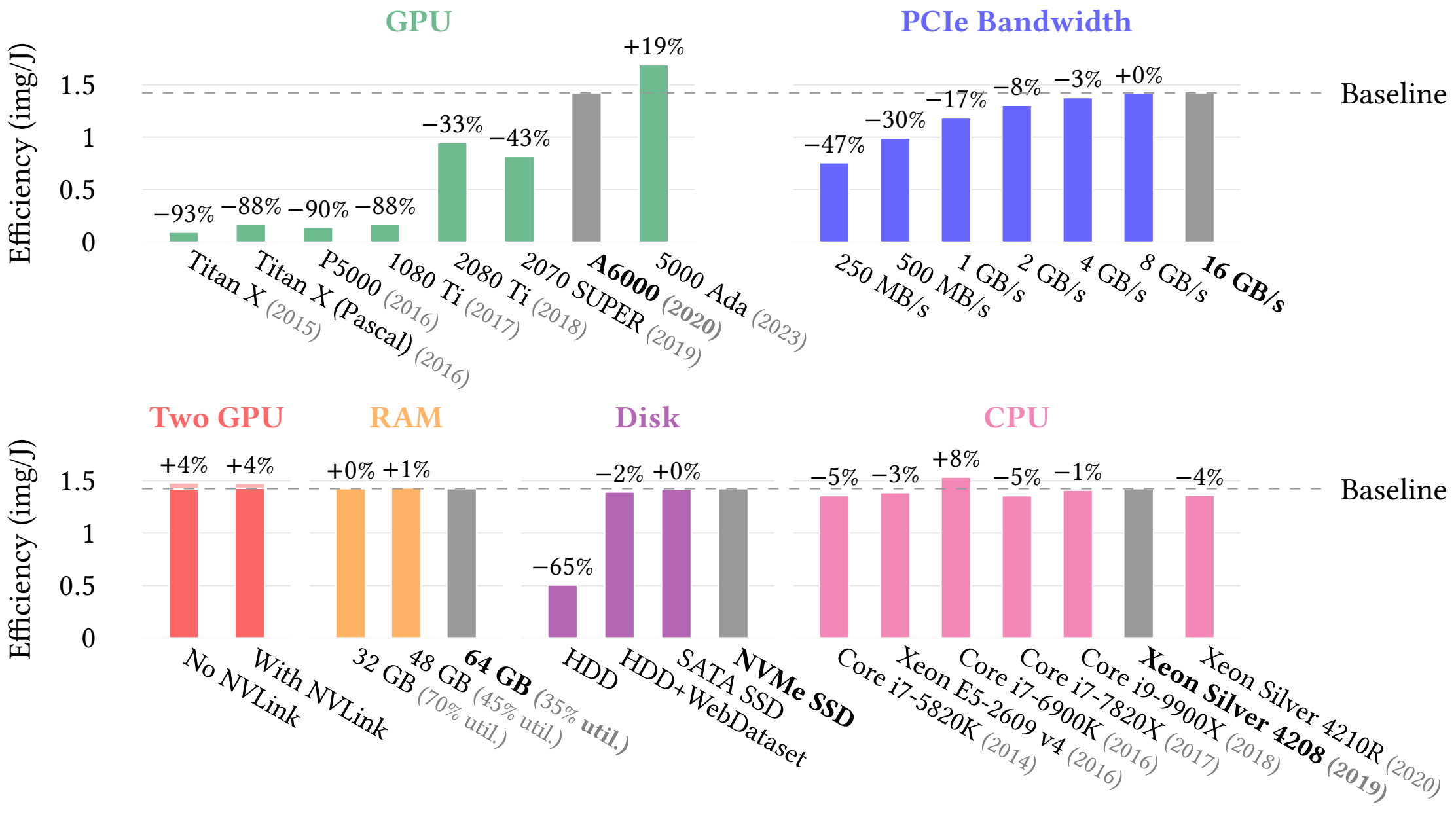}

\pgfplotsset{
    hardware panel/.append style={table/x=exp_name, table/y=img_per_j},
    pct labels/.append style={table/meta expr={\pctOf{img_per_j}{\baselineEff}}},
}

\begin{tikzpicture}
    \begin{axis}[
        gpu panel,
        ymax=\effYmax,
        ylabel={Efficiency (img/J)},
    ]
        \addplot[fill=ForestGreen!60, pct labels] table {\gpudata};
        \addplot[reference bar] coordinates {(a6000_512,\baselineEff)};
        \coordinate (baseline_start) at (rel axis cs: 0., \baselineEff/\effYmax);
    \end{axis}

    \begin{axis}[
        two gpu panel,
        ymax=\effYmax,
        ylabel={Efficiency (img/J)},
    ]
        \addplot[fill=red!30, pct labels]
            table[y=img_per_j_large_bs, meta expr={\pctOf{img_per_j_large_bs}{\baselineEff}}] {\twogpudata};
        \addplot[fill=red!60] table {\twogpudata};
        \coordinate (baseline_start2) at (rel axis cs: 0., \baselineEff/\effYmax);
    \end{axis}

    \begin{axis}[
        ram panel,
        ymax=\effYmax,
    ]
        \addplot[fill=orange!60, pct labels] table {\ramdata};
        \addplot[reference bar] coordinates {(64gb,\baselineEff)};
    \end{axis}

    \begin{axis}[
        disk panel,
        ymax=\effYmax,
    ]
        \addplot[fill=violet!60, pct labels] table {\diskdata};
        \addplot[reference bar] coordinates {(nvme_ssd,\baselineEff)};
    \end{axis}

    \begin{axis}[
        cpu panel,
        ymax=\effYmax,
    ]
        \addplot[fill=magenta!60, pct labels] table {\cpudata};
        \addplot[reference bar] coordinates {(moravia,\baselineEff)};
    \end{axis}

    \begin{axis}[
        pcie panel,
        ymax=\effYmax,
    ]
        \addplot[fill=blue!60, pct labels] table {\pciedata};
        \addplot[reference bar] coordinates {(x16gen3,\baselineEff)};
    \end{axis}

    \draw[dashed, gray!80] (baseline_start) -- ++(12, 0) node[right, black] {Baseline};
    \draw[dashed, gray!80] (baseline_start2) -- ++(12, 0) node[right, black] {Baseline};
\end{tikzpicture}
    \caption{\textbf{Impact of hardware on training efficiency.} Our baseline is the \emph{power-capped RTX A6000} with all PyTorch \emph{optimizations enabled} at \emph{batch size 512}. In the \textcolor{ForestGreen!60}{\textbf{GPU}} plot, we show power-capped performance at best batch size with all supported optimizations enabled (peak efficiency in \cref{fig:gpu_power_capping}). \textcolor{blue!60}{\textbf{PCIe Bandwidth}} bar labels indicate the nominal unidirectional bandwidth as dictated by PCIe generation and number of lanes. Bar extensions in the \textcolor{red!60}{\textbf{Two GPU}} plot show efficiency when doubling the batch size. Throughput results can be found in \cref{fig:hardware_throughput} of \cref{sec:C_throughput}.}
    \label{fig:hardware_efficiency}
    \Description{Fully described in the text.}
\end{figure}

\subsection{Impact of Hardware} \label{sec:gpu}

In this section, we evaluate the impact on efficiency of hardware configurations, including GPUs, PCIe bandwidth, multi-GPU with and without an NVLink connector, RAM, disks, and CPUs. The results are summarized in \cref{fig:hardware_efficiency}.

\paragraph{GPU}
We compare the efficiency of eight different graphics cards, which span five generations of NVIDIA GPUs released between 2015 and 2023: Maxwell, Pascal, Turing, Ampere, and Ada Lovelace (see \cref{tab:gpu_inventory} in \cref{sec:B_hardware_specifications}). Three main factors impact the efficiency of these GPUs: the software optimizations they enable, the batch size used, and power capping.

When supported, we use the PyTorch optimizations introduced in \cref{sec:pytorch_optimizations}. However, only the newer Ampere and Ada Lovelace generations support all of them. Turing cards support compilation and AMP but not Flash attention. The remaining GPUs do not support compilation, and enabling Automatic Mixed Precision induces a roughly $2\times$ slowdown due to Float16 being considerably less performant than Float32 (see \cref{fig:gpu_flops} in \cref{sec:B_hardware_specifications}). We thus use them without any optimization. 

GPU power capping is a well-known technique to reduce consumption at the data center scale~\cite{freyBenchmarkingResourceUsage2022, youZeusUnderstandingOptimizing2023a, mayrAIApplicationBenchmarking2026}. Power capping consists in limiting the maximum operating power of the GPU, potentially achieving efficiency gains at the cost of throughput. The optimal power cap that achieves maximum efficiency is task- and hardware-dependent. To approximate the optimal power cap for our workload, we start from the maximum power cap allowed by the NVIDIA System Management Interface, and perform a sweep in intervals of 20 W until the minimum power limit.

\Cref{fig:gpu_power_capping} summarizes the effect of power capping when using the \emph{best} batch size for each GPU, i.e., the one that achieves highest efficiency and throughput. Note that for recent GPUs in our inventory, the best batch size is typically the largest one that fits in memory, but this is not true for our older GPUs (see \cref{fig:gpu_efficiency_bsize} in \cref{sec:D_gpu_power_capping_all}). Power capping achieves higher efficiency gains for our Turing and Pascal GPUs compared to our newer cards, increasing the efficiency for the RTX A6000 and 5000 Ada by only $4\%$ to $5\%$.

\Cref{fig:gpu_power_capping} also outlines that our GPUs of the Ampere and Ada Lovelace generations have considerably higher efficiency and throughput than those of the preceding Turing generation ($\approx2\times$), which itself is significantly more efficient than the previous Pascal generation ($\approx5\times$). These increases result from the combination of broader support for PyTorch optimizations and significant hardware performance improvements (see \cref{tab:gpu_inventory,fig:gpu_flops} in \cref{sec:B_hardware_specifications}).

In summary, the choice of GPU model has a critical impact on energy efficiency, as also visualized in \cref{fig:hardware_efficiency} (\textcolor{ForestGreen!60}{\textbf{GPU}} bars). However, this impact is much more significant when a GPU upgrade enables a critical software optimization, such as compilation and AMP for Turing, or Flash attention for Ampere. Although power capping or limiting the batch size do not significantly boost the efficiency of our newer GPUs, related work shows improvements of up to $45\%$ for the contemporary Hopper H100 SXM when setting a power limit during ViT training~\cite{mayrAIApplicationBenchmarking2026}. Despite measurement protocol differences, this result suggests that the effect of power capping is GPU-specific, rather than determined by GPU release date alone.

\paragraph{PCIe bandwidth}

Consumer-grade GPUs use the PCIe interface standard to transfer data to and from system memory. Each new generation of the standard doubles the nominal bandwidth, starting at approximately 250 MB/s unidirectional bandwidth for PCIe Gen 1, for a single lane. We simulate a range of data-transfer bandwidths by placing the graphics card in a PCIe slot on the motherboard with either one, eight, or sixteen lanes and by artificially limiting the PCIe interface to Gen 1, Gen 2, or Gen 3, the latter being the latest standard supported by our baseline server.

Efficiency starts decreasing at one-fourth of the original nominal bandwidth of 16 GB/s ($-3\%$ efficiency). We empirically observe a bandwidth of 3 GB/s at this turning point, coupled with a decrease in GPU utilization indicating that this data transfer rate is not enough to sustain peak GPU utilization for our experiment. Efficiency drops by a maximum of $47\%$ when dividing the original bandwidth by 64, reaching 250 MB/s (\textcolor{blue!60}{\textbf{PCIe Bandwidth}} bars in \cref{fig:hardware_efficiency}).

\paragraph{Multi-GPU}

Training on multiple GPUs is a common strategy used to speed up training and increase effective GPU memory, enabling larger batch sizes and parameter counts. At the same time, overheads such as inter-GPU communication and synchronization operations introduce energy bloat~\cite{chungReducingEnergyBloat2024a}. We only explore this scenario in a local two-GPU setting, by adding a second RTX A6000 to our server. For the same effective batch size of 512 (256 per GPU), efficiency stays the same. Doubling the effective batch size brings a minor $4\%$ efficiency increase (\textcolor{red!60}{\textbf{Two GPU}} bar extensions in \cref{fig:hardware_efficiency}). In a multi-GPU setup, GPU-to-GPU communication can be sped up with specialized hardware, such as NVLink connectors. We do not observe any improvement when connecting both GPUs with an NVLink at batch sizes 512 or 1024 (\cref{fig:hardware_efficiency}). However, as batch size decreases, the multi-GPU scenario becomes less efficient than the single-GPU baseline, with the NVLink mitigating the gap.

\paragraph{RAM capacity}

RAM capacity can be a limiting factor when training larger models or using larger batch sizes, but typically does not have a strong impact on efficiency as long as capacity is sufficient: the power draw of a single RAM module is low~\cite{karyakinAnalysisMemoryPower2017,karyakinDimmStoreMemoryPower2019}, meaning that for standard workstations with four to eight RAM modules (\cref{tab:server_inventory}), system RAM is not the main contributor to energy consumption. We evaluate efficiency after removing one, two, and three 16 GB RAM modules from the server (\textcolor{orange!60}{\textbf{RAM}} bars in \cref{fig:hardware_efficiency}) and observe that efficiency remains constant even when memory utilization increases up to $70\%$ with 32 GB of RAM, and that 16 GB is not enough to run our experiment. Having an excessive number of RAM modules, e.g., to fill all twenty-four slots in our server, could have a detrimental impact on efficiency by adding a significant constant power draw during training, but we do not evaluate this scenario due to a lack of available resources.

\paragraph{Disks}\label{sec:disks}

By default, we store the ImageNet-1k dataset in a PCIe NVMe solid-state drive (SSD). We compare this setting to storing the dataset in a SATA SSD, and in a SATA mechanical hard disk drive (HDD) (\textcolor{violet!60}{\textbf{Disk}} bars in \cref{fig:hardware_efficiency}). Efficiency is the same for both SSDs regardless of the protocol used (SATA or NVMe), whereas the HDD is $65\%$ less efficient. To mitigate the high latency of the HDD, we test using the WebDataset format. WebDataset reads data sequentially and then shuffles samples in system memory, which can provide significant speedups for mechanical disks. This software-side solution almost completely bridges the efficiency gap. This is particularly interesting because the embodied carbon footprint of an SSD is more than an order of magnitude higher than that of an HDD~\cite{tannu2023dirty}, and can actually account for a large part of the carbon footprint of a server. Note, however, that HDDs have higher abiotic depletion and toxicity impacts~\cite{kim2019environmental}.

\paragraph{CPU}

After graphics cards, CPUs are the second most energy-hungry hardware components in AI training workloads~\cite{geisslerPowerTrainingHow2024}. We compare the impact on efficiency of seven CPUs (detailed in \cref{tab:cpu_inventory} in \cref{sec:B_hardware_specifications}) in \cref{fig:hardware_efficiency} (\textcolor{magenta!60}{\textbf{CPU}} bars). Since swapping CPUs is a delicate operation and most of the CPUs we test are not compatible with the motherboard socket of our default server, we instead run our experiment on the servers described in \cref{tab:server_inventory} in \cref{sec:B_hardware_specifications}. The previous sections indicate that the RAM capacity and PCIe bandwidth of these servers have no significant impact on the efficiency of our baseline. We use eight data-loading workers, each prefetching two batches of data of size 512 and applying standard data augmentation. For this workload, we observe that CPU performance has only a minor impact on energy efficiency. Interestingly, efficiency is not correlated with CPU age: we observe similar performance for servers equipped with CPUs dating from 2014 to 2020. Only the i7-6900K setup provides an $8\%$ efficiency boost with respect to our baseline (\cref{fig:hardware_efficiency}), managing to feed the GPU while consuming only 65 W on average compared to the 83 W of our default configuration (\cref{tab:cpu_inventory} in \cref{sec:B_hardware_specifications}).

\begin{tcolorbox}[title=\textbf{Key Takeaways}, colback=gray!15, colframe=gray]
\begin{itemize}
    \item \textbf{GPUs and optimization}: GPUs significantly impact efficiency for our workload, especially when they enable new software optimizations (AMP and compilation for Turing, and Flash attention for Ampere cards).
    \item \textbf{Other hardware:} other hardware has limited impact on the efficiency of our compute-bound workload, hinting that any upgrade should be limited to the GPU in such cases.
    \item \textbf{HDD/SSD}: using a software-based solution, WebDataset, makes our workload as efficient on HDD as on SSD. Using an HDD could reduce embodied carbon emissions, but increase abiotic depletion impacts.
\end{itemize}
\end{tcolorbox}

\section{Environmental Impact of Upgrade and Workload Evolution Scenarios} \label{sec:hardware_replacement}

The efficiency results described in the previous section alone do not suffice to assess the environmental footprint of AI workloads. To relate them to concrete usage, we estimate the environmental impact of different workload evolution and upgrade scenarios for training on a local server during a period of six years, from January 2021 to December 2026. We choose 2021 as a starting date since it aligns with the appearance of the Vision Transformer~\cite{dosovitskiyImageWorth16x162021}. We explore the influence of hardware and software upgrade choices during this period, observing the trade-off between the embodied impacts of producing new hardware and the operational savings enabled by efficiency improvements. We consider three workload evolution scenarios across three locations with distinct water and carbon intensity profiles (south-central Sweden, Germany, and Poland), and six hardware and software upgrade scenarios.

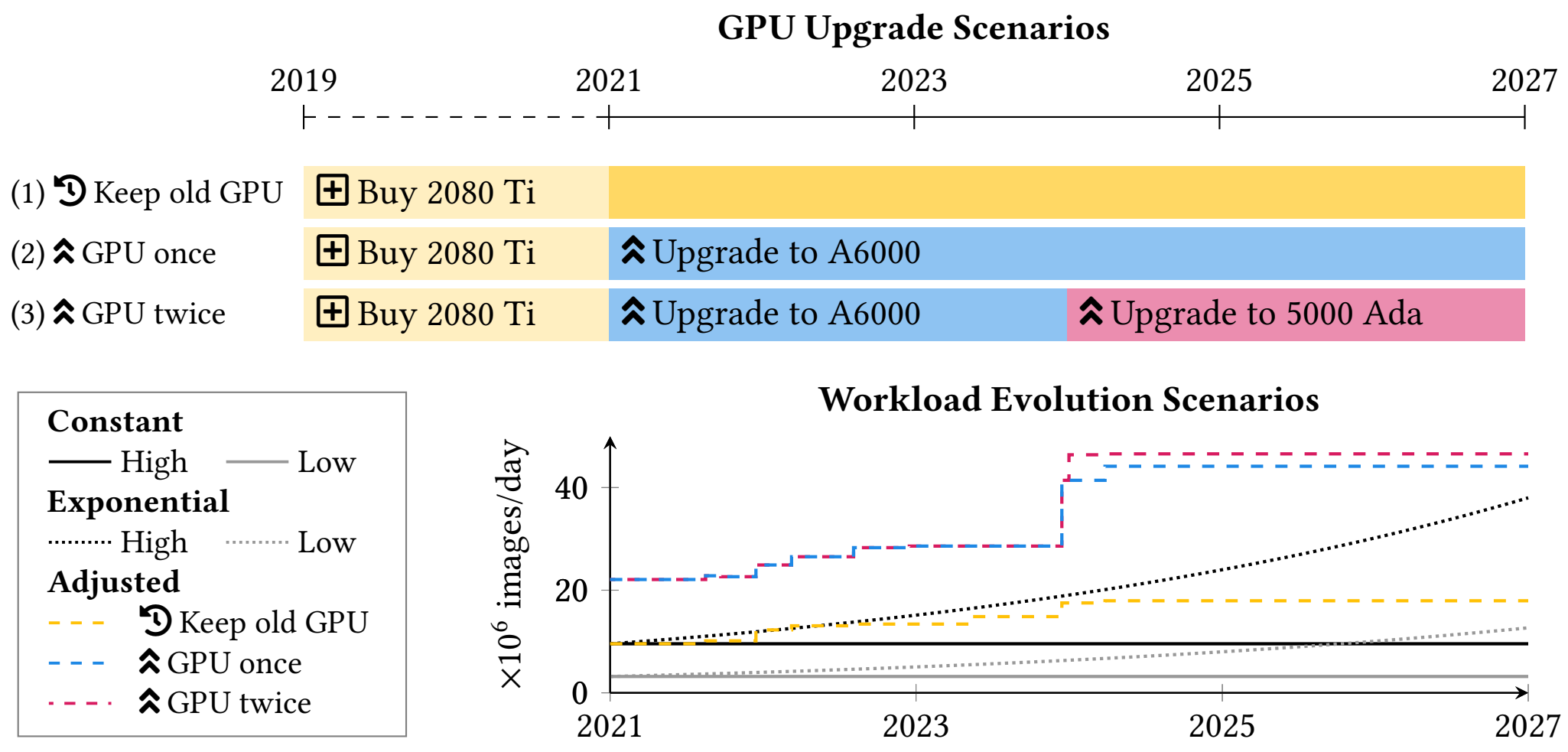
\begin{figure}
    \centering
    \begin{tikzpicture}

\def\mplier{1.1}
\def\timelineyshift{-15pt}
\def\topyshift{25pt}
\def\rownameshift{-75pt}
\def\rectheight{0.45}
\def\rectmiddle{0.225}

\coordinate (start) at (0,0);
\coordinate (timeline_start) at (0, \topyshift);

\draw[dashed] (timeline_start) -- ++(2.4 * \mplier,0);
\draw ($(timeline_start) + (2.4 * \mplier, 0)$) -- ++(7.2 * \mplier,0);

\foreach \x in {0, 2.4, 4.8, 7.2, 9.6}{
   \draw (\x * \mplier,\topyshift+3pt) -- (\x * \mplier,\topyshift-3pt);
}
\draw (timeline_start) node[above=3pt] {2019};
\draw (2.4 * \mplier,\topyshift) node[above=3pt, align=center] {2021};
\draw (4.8 * \mplier,\topyshift) node[above=3pt, align=center] {2023};
\draw (7.2 * \mplier,\topyshift) node[above=3pt, align=center] {2025};
\draw (9.6 * \mplier,\topyshift) node[above=3pt, align=center] {2027};

\node at ($(timeline_start) + (4.8 * \mplier, 0.75)$) {\textbf{GPU Upgrade Scenarios}};

\def\counter{0}
\draw ([xshift=\rownameshift, yshift=\counter*\timelineyshift]0, \rectmiddle) node[anchor=west, scale=0.9] {(1) \faHistory\hspace{2pt}Keep old GPU};
\fill[acces3, opacity=0.25] ([yshift=\counter*\timelineyshift]0 * \mplier, 0) rectangle ([yshift=\counter*\timelineyshift]9.6 * \mplier,\rectheight);
\fill[acces3, opacity=0.5] ([yshift=\counter*\timelineyshift]2.4 * \mplier, 0) rectangle ([yshift=\counter*\timelineyshift]9.6 * \mplier,\rectheight);
\draw ([yshift=\counter*\timelineyshift]0 * \mplier, \rectmiddle) node[anchor=west] {\faPlusSquare[regular] Buy 2080 Ti};

\def\counter{1}
\draw ([xshift=\rownameshift, yshift=\counter*\timelineyshift]0, \rectmiddle) node[anchor=west, scale=0.9] {(2) \faAngleDoubleUp\hspace{2pt}GPU once};
\fill[acces3, opacity=0.25] ([yshift=\counter*\timelineyshift]start) rectangle ([yshift=\counter*\timelineyshift]2.4 * \mplier,\rectheight);
\fill[acces2!50] ([yshift=\counter*\timelineyshift]2.4 * \mplier, 0) rectangle ([yshift=\counter*\timelineyshift]9.6 * \mplier,\rectheight);
\draw ([yshift=\counter*\timelineyshift]2.4 * \mplier,\rectmiddle) node[anchor=west] {\faAngleDoubleUp\hspace{2pt}Upgrade to A6000};
\draw ([yshift=\counter*\timelineyshift]0, \rectmiddle) node[anchor=west] {\faPlusSquare[regular] Buy 2080 Ti};

\def\counter{2}
\draw ([xshift=\rownameshift, yshift=\counter*\timelineyshift]0, \rectmiddle) node[anchor=west, scale=0.9] {(3) \faAngleDoubleUp\hspace{2pt}GPU twice};
\fill[acces3, opacity=0.25] ([yshift=\counter*\timelineyshift]start) rectangle ([yshift=\counter*\timelineyshift]2.4 * \mplier,\rectheight);
\fill[acces2!50] ([yshift=\counter*\timelineyshift]2.4 * \mplier, 0) rectangle ([yshift=\counter*\timelineyshift]6.0 * \mplier,\rectheight);
\fill[acces1!50] ([yshift=\counter*\timelineyshift]6.0 * \mplier, 0) rectangle ([yshift=\counter*\timelineyshift]9.6 * \mplier,\rectheight);
\draw ([yshift=\counter*\timelineyshift]2.4 * \mplier,\rectmiddle) node[anchor=west] {\faAngleDoubleUp\hspace{2pt}Upgrade to A6000};
\draw ([yshift=\counter*\timelineyshift]6.0 * \mplier,\rectmiddle) node[anchor=west] {\faAngleDoubleUp\hspace{2pt}Upgrade to 5000 Ada};
\draw ([yshift=\counter*\timelineyshift]0, \rectmiddle) node[anchor=west] {\faPlusSquare[regular] Buy 2080 Ti};

\pgfplotstableread{
year throughput
2021-01-01 122.92
2021-08-17 130.22
2021-12-15 157.75
2022-03-10 167.98
2022-08-05 172.26
2023-05-08 191.0
2023-12-15 225.27
2024-03-27 230.71
2027-01-01 230.71
}\throughputextend

\pgfplotstableread{
year throughput
2021-01-01 283.96
2021-08-17 293.38
2021-09-22 291.02
2021-12-15 320.4
2022-03-10 341.04
2022-08-05 363.81
2022-12-16 367.72
2023-12-15 532.6
2024-03-27 567.83
2027-01-01 567.83
}\throughupgradeonce

\pgfplotstableread{
year throughput
2021-01-01 283.96
2021-08-17 293.38
2021-09-22 291.02
2021-12-15 320.4
2022-03-10 341.04
2022-08-05 363.81
2022-12-16 367.72
2023-12-15 532.6
2024-01-01 596.57
2024-03-27 598.86
2027-01-01 598.86
}\throughputupgradetwice

\def\startThr{122.92}
\begin{axis}[
    name=mainplot,
    at={(2.65cm, -4.1cm)},
    anchor=south west,
    title={\textbf{Workload Evolution Scenarios}},
    title style={yshift=-3},
    width={0.63\textwidth},
    height=3.8cm,
    axis lines=left,
    ymin=0,
    ymax=50,
    xmin=0,
    xmax=2191,
    xtick={0,730,1461,2191},
    xticklabels={2021, 2023, 2025, 2027},
    ylabel={$\times10^6$ images/day},
    legend cell align=left,
]
\addplot[black, thick, domain=0:2191, samples=2] {0.9 * \startThr * 3600 * 24 / 1e6};\label{plot:constant}
\addplot[black!40, thick, domain=0:2191, samples=2] {0.3 * \startThr * 3600 * 24 / 1e6};\label{plot:constant2}
\addplot[black, densely dotted, thick, domain=0:2191, samples=200] {(0.9 * \startThr * 3600 * 24 / 1e6) * (2^(1/(3*365))^x)};\label{plot:doubling}
\addplot[black!40, densely dotted, thick, domain=0:2191, samples=200] {(0.3 * \startThr * 3600 * 24 / 1e6) * (2^(1/(3*365))^x)};\label{plot:doubling2}
\end{axis}

\begin{axis}[
    at={(mainplot.north west)},
    anchor=north west,
    width={0.63\textwidth},
    height=3.8cm,
    axis lines=left,
    ymin=0,
    ymax=50,
    date ZERO=2021-01-01,
    xmin=2021-01-01,
    xmax=2027-01-01,
    date coordinates in=x,
    xticklabels={},
    xtick={2021-01-01, 2023-01-01, 2025-01-01, 2027-01-01},
    xticklabel style={},
]
\addplot[mark=\empty, dashed, dash phase=1.75pt, thick, mark size=2pt, acces1, const plot] table [
    x=year,
    y expr={0.9 * \thisrow{throughput} * 3600 * 24 / 1e6},
]{\throughputupgradetwice};\label{plot:upgrade_twice}
\addplot[mark=\empty, dashed, thick, mark size=2pt, acces2, const plot] table [
    x=year,
    y expr={0.9 * \thisrow{throughput} * 3600 * 24 / 1e6},
]{\throughupgradeonce};\label{plot:upgrade_once}
\addplot[mark=\empty, dashed, thick, mark size=2pt, acces3, const plot] table [
    x=year,
    y expr={0.9 * \thisrow{throughput} * 3600 * 24 / 1e6},
]{\throughputextend};\label{plot:extend}
\end{axis}
\node [draw=gray, shape=rectangle, align=center, scale=0.9, anchor=north east](legend_table) at ([xshift=-50pt, yshift=11pt]mainplot.north west) {
    \begin{tabular}{r@{ }lr@{ }l}
        \multicolumn{4}{l}{\textbf{Constant}} \\
        \ref{plot:constant} & High & \ref{plot:constant2} & Low\\
        \multicolumn{4}{l}{\textbf{Exponential}} \\
        \ref{plot:doubling} & High & \ref{plot:doubling2} & Low \\
        \multicolumn{4}{l}{\textbf{Adjusted}} \\
        \ref{plot:extend} & \multicolumn{3}{l}{\faHistory\hspace{2pt}Keep old GPU}\\
        \ref{plot:upgrade_once} & \multicolumn{3}{l}{\faAngleDoubleUp\hspace{2pt}GPU once} \\
        \ref{plot:upgrade_twice} & \multicolumn{3}{l}{\faAngleDoubleUp\hspace{2pt}GPU twice} \\
    \end{tabular}
};

\end{tikzpicture}
    \caption{\textbf{GPU upgrade and workload evolution scenarios.} Increases in the \emph{adjusted} workload evolution scenarios at points other than the GPU upgrade in 2024 are due to PyTorch updates. We estimate environmental impacts starting in 2021.
    }
    \label{fig:hardware_replacement_timeline}
    \Description{Diagram showing a timeline from 2019 to 2027 with three horizontal bars on the top, and a line plot spanning the same time period on the bottom. The three top bars are labeled: 1. Keep old GPU; 2. Upgrade GPU once; 3. Upgrade GPU twice. All three bars start in 2019 with "Buy 2080 Ti". Bar 1 shows no changes. Bar 2 indicates "Upgrade to A6000" in 2021. Bar 3 indicates "Upgrade to A6000" in 2021 and "Upgrade to 5000 Ada" in 2024. The period between 2019 and 2021 is shown differently to indicate that environmental impacts for these two years are not considered. The bottom line plot is titled "Workload Evolution Scenarios" and shows a 2021 to 2027 timeline on the X axis against the number of training images processed per day on the Y axis, ranging from 0 to 50 million. The plot contains seven lines: two constant lines at 9.6 and 3.2 million images per day respectively, two exponential lines starting at 9.6 and 3.2 million images per day respectively and doubling every three years, and three dashed lines in the colors of the three top bars. Line 1 starts at 9.6 million images per day and increases in irregular steps up to 17.9 million images per day. Lines 2 and 3 start at 22.1 million images per day and slightly increase in irregular steps until the end of 2023, where they both increase significantly up to 44.2 and 46.6 million images per day, stepwise once again.}
\end{figure}

\subsection{Scenario Presentation}
\paragraph{Hardware and software upgrades}

All scenarios start with our server (described in \cref{sec:baseline} and \cref{tab:server_inventory} in \cref{sec:B_hardware_specifications}) equipped with a single RTX 2080 Ti, both bought in 2019. Our analysis then focuses on the choices made in the 2021--2026 period, during which the server runs a ViT training workload daily following our proposed workload evolution scenarios. 

By default, PyTorch is updated every time a new version brings efficiency improvements, from version 1.7.1 until version 2.2.2 (\cref{fig:pytorch_efficiency}). Besides software updates, we consider the following three GPU upgrade scenarios (\cref{fig:hardware_replacement_timeline}, top), always assuming all supported PyTorch optimizations enabled, power-capped GPUs, and maximum batch size:
 
\begin{enumerate}
    \item \emph{\faHistory\hspace{2pt}Keep old GPU}: Keep using the original 2080 Ti until the end of 2026. 
    \item \emph{\faAngleDoubleUp\hspace{2pt}Upgrade GPU once}: Replace the 2080 Ti by an A6000 in 2021.
    \item \emph{\faAngleDoubleUp\hspace{2pt}Upgrade GPU twice}: Replace the 2080 Ti by an A6000 in 2021, then replace the A6000 by a 5000 Ada in 2024.
\end{enumerate}

In addition, we outline the impact of good practices with three more scenarios:
\begin{enumerate}
    \setcounter{enumi}{3}
    \item \emph{\faHistory\hspace{2pt}Keep old PyTorch}: Omit all PyTorch updates in (1).
    \item \emph{\faPowerOff\hspace{2pt}Keep server on}: Do not turn the server off when not running the workload in (1).
    \item \emph{\faAngleDoubleUp\hspace{2pt}Upgrade server and GPU twice}: Purchase a full server every time there is a GPU upgrade in (3).
\end{enumerate}

\paragraph{Workload evolution}
            
We define a daily workload as the number of training images processed per day.
We compare three workload evolution scenarios (\cref{fig:hardware_replacement_timeline}, bottom):
\begin{enumerate}
    \item \emph{Constant:} the number of images processed per day remains constant throughout the six-year period. This corresponds to a rather frugal scenario in which efficiency improvements serve to complete work faster, and not to increase the amount of completed work.
    \item \emph{Exponential:} the workload increases exponentially over time, so that it doubles every three years. This scenario reflects the increasing popularity of AI, with the growing workload potentially requiring the acquisition of more powerful graphics cards.
    \item \emph{Adjusted:} after each hardware or software upgrade, the workload is adjusted to represent a daily utilization of $90\%$ at the throughput of the new configuration (i.e., the server is active $90\%$ of the time). This corresponds to a case where hardware is used as much as possible and efficiency improvements serve to increase the workload.
\end{enumerate}
For both the \emph{constant} and \emph{exponential} evolution patterns, we consider two initial utilization levels of the RTX 2080 Ti: \emph{high} ($90\%$) and \emph{low} ($30\%$).

\begin{table}[t]
    \centering
    \caption{\textbf{Accumulated environmental impacts in Germany \raisebox{-2pt}{\twemoji[height=0.35cm]{flag: Germany}}.} Accumulated global warming potential, water use, and abiotic depletion potential impacts for our computer vision training workload sustained over the 2021--2026 period on a local server. We estimate operational impacts considering the daily average carbon intensity and water scarcity footprint over time in Germany.}
    \label{tab:germany_impacts}
    \resizebox{\columnwidth}{!}{
    \begin{tabular}{llr@{ }l!{\hspace{6pt}}r@{ }l!{\hspace{6pt}}r@{ }l!{\hspace{6pt}}r@{ }l!{\hspace{6pt}}r@{ }l!{\hspace{6pt}}r@{ }l!{\hspace{4pt}}}
\multicolumn{14}{c}{\textbf{Global Warming Potential} (kg CO$_2$eq)} \\ \toprule
\multirow{3}[2]{*}{\makecell[c]{Workload\\Evolution}} & 
\multirow{3}[2]{*}{\makecell[c]{Initial\\Util.}} & 
\multicolumn{12}{c}{Upgrade Scenario} \\ \cmidrule(lr){3-14}
 &  & \multicolumn{2}{c}{\makebox[0pt][c]{\hspace*{-3pt}\makecell[c]{\faHistory\hspace{2pt}Keep\\old GPU}\hspace*{3pt}}} & \multicolumn{2}{c}{\makebox[0pt][c]{\hspace*{-3pt}\makecell[c]{\faHistory\hspace{2pt}Keep old\\PyTorch}\hspace*{3pt}}} & \multicolumn{2}{c}{\makebox[0pt][c]{\hspace*{-3pt}\makecell[c]{\faPowerOff\hspace{2pt}Keep\\server on}\hspace*{3pt}}} & \multicolumn{2}{c}{\makebox[0pt][c]{\hspace*{-3pt}\makecell[c]{\faAngleDoubleUp\hspace{2pt}Upgrade\\GPU once}\hspace*{3pt}}} & \multicolumn{2}{c}{\makebox[0pt][c]{\hspace*{-3pt}\makecell[c]{\faAngleDoubleUp\hspace{2pt}Upgrade\\GPU twice}\hspace*{3pt}}} & \multicolumn{2}{c}{\makebox[0pt][c]{\hspace*{-2pt}\makecell[c]{\faAngleDoubleUp\hspace{2pt}Upgrade\\Server+GPU}\hspace*{2pt}}} \\
\midrule
\multirow{2}{*}{Constant} & Low & 1,956 & \textcolor{gray}{\small{(ref)}} & 2,409 & \textcolor{Red}{\small{$+23\%$}} & 3,213 & \textcolor{Red}{\small{$+64\%$}} & 1,958 & \textcolor{gray}{\small{$+0\%$}} & 2,146 & \textcolor{Red}{\small{$+10\%$}} & 4,506 & \textcolor{Red}{\small{$+130\%$}} \\
 & High & 3,872 & \textcolor{gray}{\small{(ref)}} & 5,232 & \textcolor{Red}{\small{$+35\%$}} & 4,490 & \textcolor{Red}{\small{$+16\%$}} & 3,373 & \textcolor{ForestGreen}{\small{$-13\%$}} & 3,482 & \textcolor{ForestGreen}{\small{$-10\%$}} & 5,842 & \textcolor{Red}{\small{$+51\%$}} \\
\cmidrule(lr){1-14}
\multirow{2}{*}{Exponential} & Low & 2,844 & \textcolor{gray}{\small{(ref)}} & \multicolumn{2}{c}{-} & 3,808 & \textcolor{Red}{\small{$+34\%$}} & 2,589 & \textcolor{ForestGreen}{\small{$-9\%$}} & 2,704 & \textcolor{ForestGreen}{\small{$-5\%$}} & 5,064 & \textcolor{Red}{\small{$+78\%$}} \\
 & High & \multicolumn{2}{c}{-} & \multicolumn{2}{c}{-} & \multicolumn{2}{c}{-} & 5,264 & \textcolor{gray}{\small{(ref)}} & 5,156 & \textcolor{ForestGreen}{\small{$-2\%$}} & 7,516 & \textcolor{Red}{\small{$+43\%$}} \\
\midrule
\multicolumn{2}{l}{Of which embodied:} & 994 & \textcolor{gray}{\small{(ref)}} & 994 & \textcolor{gray}{\small{$+0\%$}} & 994 & \textcolor{gray}{\small{$+0\%$}} & 1,247 & \textcolor{Red}{\small{$+26\%$}} & 1,474 & \textcolor{Red}{\small{$+48\%$}} & 3,834 & \textcolor{Red}{\small{$+286\%$}}\\
\bottomrule
\multicolumn{14}{c}{} \\
\multicolumn{14}{c}{\textbf{Water Use} (dm$^3$ world-eq)} \\ \toprule
\multirow{2}{*}{Constant} & Low & 64 & \textcolor{gray}{\small{(ref)}} & 95 & \textcolor{Red}{\small{$+49\%$}} & 149 & \textcolor{Red}{\small{$+133\%$}} & 47 & \textcolor{ForestGreen}{\small{$-26\%$}} & 45 & \textcolor{ForestGreen}{\small{$-30\%$}} & 45 & \textcolor{ForestGreen}{\small{$-30\%$}} \\
 & High & 191 & \textcolor{gray}{\small{(ref)}} & 285 & \textcolor{Red}{\small{$+49\%$}} & 233 & \textcolor{Red}{\small{$+22\%$}} & 141 & \textcolor{ForestGreen}{\small{$-26\%$}} & 133 & \textcolor{ForestGreen}{\small{$-30\%$}} & 133 & \textcolor{ForestGreen}{\small{$-30\%$}} \\
\cmidrule(lr){1-14}
\multirow{2}{*}{Exponential} & Low & 126 & \textcolor{gray}{\small{(ref)}} & \multicolumn{2}{c}{-} & 190 & \textcolor{Red}{\small{$+51\%$}} & 91 & \textcolor{ForestGreen}{\small{$-27\%$}} & 83 & \textcolor{ForestGreen}{\small{$-34\%$}} & 83 & \textcolor{ForestGreen}{\small{$-34\%$}} \\
 & High & \multicolumn{2}{c}{-} & \multicolumn{2}{c}{-} & \multicolumn{2}{c}{-} & 273 & \textcolor{gray}{\small{(ref)}} & 250 & \textcolor{ForestGreen}{\small{$-8\%$}} & 250 & \textcolor{ForestGreen}{\small{$-8\%$}} \\
\midrule
\multicolumn{2}{l}{+ Embodied:} & 3.3e5 & \textcolor{gray}{\small{(ref)}} & 3.3e5 & \textcolor{gray}{\small{$+0\%$}} & 3.3e5 & \textcolor{gray}{\small{$+0\%$}} & 4.0e5 & \textcolor{Red}{\small{$+21\%$}} & 4.5e5 & \textcolor{Red}{\small{$+38\%$}} & 1.2e6 & \textcolor{Red}{\small{$+279\%$}}\\
\bottomrule
\multicolumn{14}{c}{} \\
\multicolumn{14}{c}{\textbf{Abiotic Depletion Potential} (g Sbeq)} \\ \toprule
All & All & 84 & \textcolor{gray}{\small{(ref)}} & 84 & \textcolor{gray}{\small{$+0\%$}} & 84 & \textcolor{gray}{\small{$+0\%$}} & 92 & \textcolor{Red}{\small{$+10\%$}} & 100 & \textcolor{Red}{\small{$+19\%$}} & 308 & \textcolor{Red}{\small{$+266\%$}} \\
\bottomrule
\end{tabular}

    }
\end{table}

\subsection{Environmental Assessment Methodology}

We estimate operational and embodied global warming potential, water use, and abiotic depletion potential.

\paragraph{Operational impacts}

We derive daily \emph{active} electricity consumption by dividing the daily workload, in number of images, by the efficiency of the current hardware setup, in images per joule. We consider that the server is either turned \emph{off} or \emph{idle} the rest of the time, depending on the scenario. The server consumes 0.2 W when turned off and 80 W when idle. To obtain global warming potential impacts, we multiply the daily electricity consumption by the average daily carbon intensity, including power plant life-cycle emissions~\cite{ElectricityMaps}. We analogously obtain water use impacts by multiplying by the daily water scarcity footprint per kilowatt-hour~\cite{melguizoWattnetMatchingElectricity2026}, illustrating contextualized water stress instead of raw water consumption. We replace daily carbon intensity and water scarcity for dates after the time of writing (November 2026) with values from 2025. We omit operational abiotic depletion potential as it can be considered negligible~\cite{morandMLCAToolMachine2024,morandEnvironmentalImpactsLanguage2025}, and we do not account for any additional impacts related to the infrastructure housing the server, such as cooling, ventilation, or security systems.

\paragraph{Embodied impacts}
We estimate embodied impacts leveraging the Resilio database~\cite{ResilioDatabaseIT} and the ADEME embodied GPU impact calculator~\cite{lees-perasso2026lcagpu, HubbloGraphicsCard} with data from TechPowerUp~\cite{TechPowerUp2026} (details in \cref{sec:E1_embodied_impacts}). These figures are intended to give an order of magnitude, not to be a rigorous life cycle assessment. 
We assume that the lifespan of the hardware starts the moment it is purchased, and ends for all hardware in January 2027. Since we study the 2021 to 2026 period, this implies that we allocate $75\%$ of the embodied impacts of the server and 2080 Ti purchased in 2019, and $100\%$ of the impacts of the remaining hardware. While the most common practice is to decide for a fixed lifespan for hardware, we aim to reflect that the hardware bought has already caused environmental harm and could be used, and it is a choice to stop using it and replace it in some scenarios.

\subsection{Analysis}

\begin{figure}[t]
    \centering
    \input{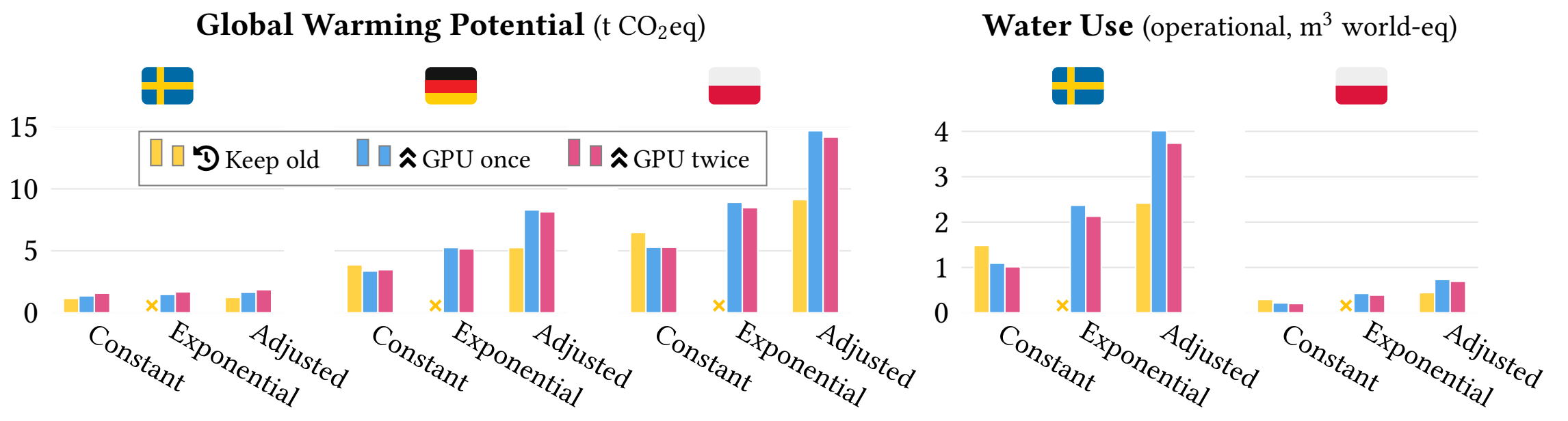}
\pgfplotstableread{
scenario poland germany sweden
constant              6.48031   3.87185    1.1605
exponential                 0         0         0
adjusted              9.12635   5.26124   1.24331
}\keepoldcarbon

\pgfplotstableread{
scenario poland germany sweden
constant             0.294062  0.191024   1.48535
exponential                 0         0         0
adjusted             0.444885  0.287019   2.42053
}\keepoldwater

\pgfplotstableread{
scenario poland germany sweden
constant                5.286   3.37306   1.37006
exponential           8.91087   5.26437   1.48407
adjusted              14.6823   8.30106   1.65996
}\upgradeoncecarbon

\pgfplotstableread{
scenario poland germany sweden
constant             0.217136  0.141117   1.09893
exponential           0.42897  0.272617   2.37216
adjusted             0.735474  0.474245   4.01371
}\upgradeoncewater

\pgfplotstableread{
scenario poland germany sweden
constant              5.28396   3.48208   1.58945
exponential           8.48489   5.15581   1.68993
adjusted               14.172   8.14913   1.86336
}\upgradetwicecarbon

\pgfplotstableread{
scenario poland germany sweden
constant             0.204344  0.133092   1.01372
exponential          0.391655   0.24969   2.12947
adjusted             0.693833  0.447983   3.73999
}\upgradetwicewater

\begin{tikzpicture}
    \begin{axis}[gwp sweden panel]
        \addplot[fill=acces3!75] table[
            y expr={\thisrow{sweden}},
            x=scenario,
        ] {\keepoldcarbon};
        \addplot[fill=acces2!75] table[
            y expr={\thisrow{sweden}},
            x=scenario,
        ] {\upgradeoncecarbon};
        \addplot[fill=acces1!75] table[
            y expr={\thisrow{sweden}},
            x=scenario,
        ] {\upgradetwicecarbon};
        \addplot[infeasible scenario cross] coordinates {(exponential,0)};
    \end{axis}

    \begin{axis}[gwp germany panel]
        \addplot[fill=acces3!75] table[
            y expr={\thisrow{germany}},
            x=scenario,
        ] {\keepoldcarbon};
        \addplot[fill=acces2!75] table[
            y expr={\thisrow{germany}},
            x=scenario,
        ] {\upgradeoncecarbon};
        \addplot[fill=acces1!75] table[
            y expr={\thisrow{germany}},
            x=scenario,
        ] {\upgradetwicecarbon};
        \addplot[infeasible scenario cross] coordinates {(exponential,0)};
    \end{axis}

    \begin{axis}[gwp poland panel]
        \addplot[fill=acces3!75] table[
            y expr={\thisrow{poland}},
            x=scenario,
        ] {\keepoldcarbon};
        \addplot[fill=acces2!75] table[
            y expr={\thisrow{poland}},
            x=scenario,
        ] {\upgradeoncecarbon};
        \addplot[fill=acces1!75] table[
            y expr={\thisrow{poland}},
            x=scenario,
        ] {\upgradetwicecarbon};
        \addplot[infeasible scenario cross] coordinates {(exponential,0)};
        \carbonLegend
    \end{axis}
    \gwpHeading

    \begin{axis}[water sweden panel]
        \addplot[fill=acces3!75] table[
            y expr={\thisrow{sweden}},
            x=scenario,
        ] {\keepoldwater};
        \addplot[fill=acces2!75] table[
            y expr={\thisrow{sweden}},
            x=scenario,
        ] {\upgradeoncewater};
        \addplot[fill=acces1!75] table[
            y expr={\thisrow{sweden}},
            x=scenario,
        ] {\upgradetwicewater};
        \addplot[infeasible scenario cross] coordinates {(exponential,0)};
    \end{axis}

    \begin{axis}[water poland panel]
        \addplot[fill=acces3!75] table[
            y expr={\thisrow{poland}},
            x=scenario,
        ] {\keepoldwater};
        \addplot[fill=acces2!75] table[
            y expr={\thisrow{poland}},
            x=scenario,
        ] {\upgradeoncewater};
        \addplot[fill=acces1!75] table[
            y expr={\thisrow{poland}},
            x=scenario,
        ] {\upgradetwicewater};
        \addplot[infeasible scenario cross] coordinates {(exponential,0)};
    \end{axis}
    \waterHeading
    \trimCarbonPlotBoundingBox
\end{tikzpicture}
    \caption{\textbf{Global warming potential and operational water use across GPU upgrade scenarios.} Accumulated global warming potential (GWP) and operational water use (WU) impacts for our computer vision training workload sustained over the 2021--2026 period on our local server for three workload evolution scenarios: \emph{high constant}, \emph{high exponential}, and \emph{adjusted}, i.e., always keeping the current graphics card at $90\%$ utilization (\cref{fig:hardware_replacement_timeline}). Operational water use in Germany is similar to that in Poland.}
    \label{fig:carbon_all_scenarios}
    \Description{Three bar plots on the left side under the title "Global Warming Potential (tonnes of carbon dioxide equivalent)", and two bar plots on the right side under the title "Water Use (operational, cubic metres world equivalent)". The three plots on the left show accumulated greenhouse gas emissions in Sweden, Germany and Poland for three workload evolution scenarios (constant, exponential, and adjusted) and three GPU upgrade scenarios (keep old, upgrade once, and upgrade twice). The Y axis ranges from 0 to 15. Emissions in Sweden stay under 2 in all scenarios. Emissions in Germany range between 3.4 and 8.3 tonnes of carbon dioxide equivalent. Emissions in Poland range between 5.3 and 14.7 tonnes of carbon dioxide equivalent. The two plots on the right show accumulated water use in Sweden and Poland for the same scenarios. The Y axis ranges from 0 to 4.1. Water use in Sweden ranges between 1 and 4. Water use in Poland stays under 0.8 in all scenarios. Figure analyzed in the text.}
\end{figure}

\Cref{tab:germany_impacts} summarizes accumulated embodied and operational impacts in Germany for all upgrade scenarios (1 through 6) and for the hardware-independent \emph{constant} and \emph{exponential} workload evolution cases, and \cref{fig:carbon_all_scenarios} compares impacts between countries for GPU-only upgrades (1 through 3) in the \emph{high} utilization and \emph{adjusted} workload scenarios. We report complete results in \cref{sec:E2_all_env_assessment_results}.

\paragraph{Importance of good practices}

\Cref{tab:germany_impacts} quantifies in concrete cases the impacts of three bad practices:

\begin{enumerate}
    \setcounter{enumi}{3}
    \item \emph{\faHistory\hspace{2pt}Keep old PyTorch}: even without any GPU upgrades (which would enable key optimizations), omitting PyTorch updates increases total global warming potential by more than $20\%$ and operational water use by $49\%$ for the \emph{constant} workloads. Optimizations also make it possible to support the \emph{low exponentially growing} workload without upgrading hardware.
    \item \emph{\faPowerOff\hspace{2pt}Keep server on}: leaving the servers running continuously instead of turning them off when unused increases global warming potential (GWP) and water use (WU) at the compute location significantly for low workloads ($+64\%$ GWP and $+133\%$ WU for a \emph{constant low} utilization) and still noticeably for higher workloads ($+16\%$ GWP and $+22\%$ WU at a \emph{constant high} utilization).
    \item \emph{\faAngleDoubleUp\hspace{2pt}Upgrade server and GPU twice}: upgrading full servers twice instead of just graphics cards increases embodied impacts by almost $4\times$ across the three environmental impact indicators, showing the difference in magnitude between full-server and single-component production impacts~\cite{ResilioDatabaseIT,Datavizta}.
\end{enumerate}

\paragraph{Impact trade-offs}

The three environmental impact indicators that we consider have different profiles when it comes to assessing operational/embodied trade-offs: embodied global warming potential can be offset through operational savings, abiotic depletion potential is dominated by embodied impacts, and water use is context-dependent: operational water savings do not compensate for water stress at manufacturing locations.

\Cref{tab:germany_impacts} indicates that for hardware-independent workload evolution scenarios (\emph{constant} and \emph{exponential}), upgrading hardware always reduces water stress at the computation location because of efficiency gains, while it is always detrimental at locations involved in hardware production and disposal. Most importantly, embodied water use is at least three orders of magnitude larger than operational water use. Similarly, the impact of hardware upgrades on abiotic depletion potential is always adverse, as increased efficiency cannot compensate for the minerals and metals extracted for hardware production. Even though we do not assess other impact categories, it should be noted that many of these metals are carcinogenic or highly toxic, posing ecological and health hazards during their extraction and processing~\cite{falkFLOPsFootprintsResource2025a}, which is why we advocate for hardware frugality.

Global warming potential shows more nuanced results. In regions with moderate carbon intensity like Germany, upgrading the GPU once is moderately beneficial in our workload scenarios ($-13\%$, and $-9\%$ in the \emph{high constant} and \emph{low exponential} scenarios, respectively), while updating a second time can be harmful ($+10\%$ in the \emph{low constant} scenario). Upgrading at least once is necessary to handle a \emph{high exponentially growing} workload, but even in this case upgrading a second time only provides a marginal reduction ($-2\%$).

\paragraph{Location and workload}
\Cref{fig:carbon_all_scenarios} shows total global warming potential (GWP) and operational water use (WU) for different countries and our three workload evolution scenarios at \emph{high} utilization. Two strong effects dominate: the differences between countries, and the much higher impacts in the \emph{adjusted} scenario. In countries with high carbon intensity (e.g., Poland), GWP can be $3\times$ to $9\times$ greater than in countries with low carbon intensities (e.g., Sweden), but WU can conversely be lower. In high-carbon-intensity countries, GPU upgrades can lower total emissions, for example, by 1.2 t CO$_2$eq in Poland for a \emph{constant high} workload (\cref{fig:carbon_all_scenarios}, third plot). However, more than hardware upgrade choices themselves, user behavior after upgrading can affect impacts very significantly in carbon-intensive countries. For example, in a workload evolution scenario where the daily duration of use of the hardware remains the same (\emph{adjusted}), instead of the daily number of processed images being fixed (\emph{constant}), emissions increase by 8.9 t CO$_2$eq in Poland and 4.7 t CO$_2$eq in Germany, effectively trumping any carbon savings from GPU upgrades (\cref{fig:carbon_all_scenarios}, left). In Sweden, this rebound effect increases emissions minimally, but exacerbates water stress (\cref{fig:carbon_all_scenarios}, right). These observations quantify and highlight that beyond efficiency gains, keeping usage moderate is key to reducing environmental impacts.

\begin{tcolorbox}[title=\textbf{Key Takeaways}, colback=gray!15, colframe=gray]
\begin{itemize}
    \item \textbf{Importance of embodied impacts}: embodied emissions dominate in countries with low carbon intensity, embodied water use can be three orders of magnitude higher than operational water use, and upgrading hardware has workload-independent impacts, including abiotic resource depletion.
    \item \textbf{Rebound effects may overshadow efficiency gains}: while in some scenarios the efficiency gains brought by better hardware can lower global warming potential and water stress, these gains are very moderate compared to the effects of usage evolution, for example, if workloads increase to match hardware capacity.
\end{itemize}
\end{tcolorbox}

\section{Conclusion}

We investigated the energy efficiency of local AI training infrastructure through direct power measurements of a standard computer vision workload, varying software optimizations, framework versions, and hardware configurations. Our results show that substantial efficiency gains can be achieved without replacing hardware. PyTorch updates and optimizations considerably improved energy efficiency, while software solutions alleviated data-loading bottlenecks that might otherwise motivate hardware upgrades. Although newer GPUs delivered major efficiency gains, replacing other workstation components generally offered limited benefits relative to their manufacturing impacts.

Building on these measurements, we retrospectively assessed the environmental consequences of hardware replacement and usage scenarios over the 2021--2026 period, considering different geographic locations and environmental categories. Our analysis highlights why energy efficiency alone is insufficient to assess the environmental benefits of hardware upgrades. The balance between operational savings and embodied impacts depends strongly on workload evolution, usage patterns, location, and the environmental impact category considered. Simple practices, such as keeping software up to date or switching off idle equipment, can be as consequential as hardware upgrades. In contrast, increasing workloads to exploit additional computing capacity can offset the savings from more efficient GPUs. Although specific to our experimental setting and the period studied, these findings illustrate the importance of considering software, hardware, and usage practices together, offering insights to inform future infrastructure management and environmental policies.

\begin{acks}
This work was supported by the ANR project SHARP ANR-23-PEIA-0008 in the context of the PEPR IA.

The authors would like to thank all the colleagues at the IMAGINE lab who lent their hardware for the purposes of the efficiency assessment in this work, and who provided valuable insights and recommendations: Ségolène Albouy, Sonat Baltacı, Paul Borne--Pons, Gatien Chenu, Victor Hoffmann, Syrine Kalleli, Paul Kervegan, Fei Meng, Yohann Perron, and Nicolás Violante.
\end{acks}

\section*{Generative AI Usage Statement}

We used Copilot with GPT-6 Luna and Codex with GPT-5.6-Sol medium and GPT-6.1-Sol medium for coding assistance and bug fixing, and with MAI-Code-1.1-Flash for spelling and grammar checking. Code refactoring, cleaning, and verification were supported by DeepSeek V4.1 Flash 1017 served on CoreWeave Forge. We leveraged ChatGPT for assistance with related work search, hardware benchmark tool and design suggestions, troubleshooting while setting up old servers, and graph plotting with TikZ. Finally, we made use of the AI overview feature integrated in Google web search, powered by the Gemini-3 family. The conception of the experiments and hardware replacement scenarios that we present in this work was done exclusively by the authors with no AI assistance, inspired and shaped by discussions with peers.

\section*{Ethical Considerations Statement}

Our assessment of the trade-off between operational and embodied global warming potential is based on a particular reference AI training workload running on a specific infrastructure, and thus should not be generalized carelessly. Additionally, our work could potentially influence the behavioral patterns of small institutions and research labs in terms of hardware upgrades. Such changes could lead to unexpected rebound effects that we do not account for in this study, such as more training workloads migrating to data centers.

\section*{Environmental Footprint}

In total, our wattmeter readings spanning the period between the initial stages of the project (December \nth{10}, 2025) and the last experiment (November \nth{23}, 2026) add up to 1{,}169 kWh, with an average power consumption of 140 W. Considering local carbon intensity and water scarcity with a granularity of fifteen minutes, this energy consumption resulted in a global warming potential of 40 kg CO$_2$eq and a water use of 50 dm$^3$ world-eq, not including embodied impacts.

\bibliographystyle{ACM-Reference-Format}
\bibliography{references}

\appendix
\crefalias{section}{appendix}
\crefalias{Section}{Appendix}
\section{Baseline Details}\label{sec:A_baseline_details}

We adopt a standard data augmentation strategy for ViT training~\cite{steinerHowTrainYour2022a, beyerBetterPlainViT2022}: random resized cropping to size 224 with bilinear interpolation, random horizontal flipping with probability 0.5, two rounds of RandAugment with magnitude 10, ImageNet-1k normalization, and MixUp with $\alpha = 0.2$. We use v1 of the Torchvision \texttt{transforms} library and apply transforms on PIL images, not tensors. For data loading, we employ eight persistent workers with pinned memory, a prefetch factor of 2, and batch size either 256 (without mixed precision) or 512 (with mixed precision). We do not cache the dataset to observe differences in disk read speed. We isolate a stratified subset of $1\%$ of the ImageNet-1k training split as a validation subset for sanity checks, and use the remainder for training.

We use the AdamW optimizer with weight decay 1e-4 and an initial learning rate of 1e-3. We apply a cosine annealing learning rate schedule, preceded by a linear warmup over 1e4 training steps starting from 1e-5 and ending at 1e-3. We clip gradient norms to 1 during training.
    
For our codebase, we leverage the \href{https://github.com/pytorch/vision/tree/main/references/classification}{PyTorch vision repository} and adapt it to make the training loop closer to the original implementation~\cite{dosovitskiyImageWorth16x162021}, to enable additional optimization options, and to ensure backward compatibility with previous versions of PyTorch. We disable periodic validation and checkpointing, but keep the logging of iteration timestamps and a set of basic metrics (iteration time, data-loading time, and allocated memory) after confirming that there is no noticeable overhead.

\begin{table}[ht]
    \centering
    \caption{\textbf{GPU inventory.} Our baseline is the RTX A6000. Thermal design power (TDP) is the maximum amount of power that the GPU is allowed to draw under a heavy load~\cite{NvidiaPowerPrimer}. The optimizations that we consider are: Automatic Mixed Precision (AMP), TensorFloat32 (TF32), model compilation (compile), Flash attention (Flash). $^*$AMP introduces a roughly $2\times$ slowdown, so we do not enable mixed precision for these cards.}
    \resizebox{\columnwidth}{!}{
        \begin{tabular}{lrlcccccc}
    \toprule
    \multirow{2}[1]{*}{GPU Model} & \multirow{2}[1]{*}{Release Date} & \multirow{2}[1]{*}{Architecture} & \multirow{2}[1]{*}{\makecell[c]{VRAM\\(GB)}} & \multirow{2}[1]{*}{\makecell[c]{TDP\\(W)}} & \multicolumn{4}{c}{Supported Optimizations} \\ \cmidrule(lr){6-9}
      & & & & & AMP & Compile & TF32 & Flash \\ \midrule
    RTX 5000 Ada Generation & Aug 2023 & Ada Lovelace & 32 & 250 & \ding{51} & \ding{51} & \ding{51} & \ding{51} \\
    \textbf{RTX A6000} & Dec 2020 & Ampere & 48 & 300 & \ding{51} & \ding{51} & \ding{51} & \ding{51} \\
    GeForce RTX 2070 SUPER & Jul 2019 & Turing & 8 & 215 & \ding{51} & \ding{51} & & \\
    GeForce RTX 2080 Ti & Sep 2018 & Turing & 11 & 250 & \ding{51} & \ding{51} & & \\
    GeForce GTX 1080 Ti & Mar 2017 & Pascal & 11 & 250 & \ding{51}$^*$ & & & \\
    Quadro P5000 & Oct 2016 & Pascal & 16 & 180 & \ding{51}$^*$ & & & \\
    Titan X (Pascal) & Aug 2016 & Pascal & 12 & 250 & \ding{51}$^*$ & & &\\
    GeForce GTX Titan X & Mar 2015 & Maxwell 2.0 & 12 & 250 & \ding{51}$^*$ & & & \\
    \bottomrule
\end{tabular}

    }
    \label{tab:gpu_inventory}
\end{table}

\begin{figure}[ht]
    \centering
    \pgfplotstableread{
gpu float16 float32 matrixfp16 matrixbf16 matrixtf32
A5000 65.28 65.28 261.1 261.1 130.6
A6000 38.71 38.71 154.8 154.8 77.4
2070 18.12 9.062 36.3 0 0
2080Ti 26.9 13.45 53.8 0 0
}\teraflops

\pgfplotstableread{
gpu float16 float32
A5000 65.28 65.28
A6000 38.71 38.71
2070 18.12 9.062
2080Ti 26.9 13.45
1080Ti 0.1772 11.3
P5000 0.1386 8.9
TitanXPascal 0.1715 11
TitanX 0 6.6048
}\teraflopsold

\begin{tikzpicture}
    \pgfplotsset{
        shared efficiency chart/.style={
            height=2.8cm,
            scale only axis,
            align=center,
            ybar,
            bar width=2mm,
            xtick=data,
            ymajorgrids=true,
            yminorgrids=true,
            legend cell align=left,
            draw=white,
            legend style={
                /tikz/column 2/.style={
                    column sep=0.2cm,
                },
                nodes={scale=0.9}
            },
            major grid style={color=gray!20},
            ytick style={draw=none},
            xtick style={draw=none},
        }
    }
    \begin{axis}[
        name=mainplot,
        ylabel={TFLOP/s},
        shared efficiency chart,
        width=0.28\textwidth,
        title=\textbf{Tensor Core Performance},
        symbolic x coords={A5000, A6000, 2070, 2080Ti},
        xticklabels={5000\\Ada, A6000, 2070\\SUPER, 2080 Ti},
        xticklabel style={font=\small, yshift=10, align=center},
    ]
        \addplot[fill=acces1] table[
            y=matrixfp16,
            x=gpu,
        ] {\teraflops};
        \addplot[fill=acces2] table[
            y=matrixbf16,
            x=gpu,
        ] {\teraflops};
        \addplot[fill=acces3] table[
            y=matrixtf32,
            x=gpu,
        ] {\teraflops};
        \legend{Float16, Bfloat16, TensorFloat32}
    \end{axis}
    \begin{axis}[
        at={(mainplot.north east)},
        anchor=north west,
        xshift=25pt,
        shared efficiency chart,
        width=0.55\textwidth,
        title=\textbf{Floating Point Performance},
        symbolic x coords={A5000, A6000, 2070, 2080Ti, 1080Ti, P5000, TitanXPascal, TitanX},
        xticklabels={5000\\Ada, A6000, 2070\\SUPER, 2080 Ti, 1080 Ti, P5000, Titan X\\(Pascal), Titan X},
        xticklabel style={font=\small, rotate=0, xshift=0, yshift=10, align=center},
    ]
        \addplot[fill=acces1] table[
            y=float16,
            x=gpu,
        ] {\teraflopsold};
        \addplot[fill=acces1!60] table[
            y=float32,
            x=gpu,
        ] {\teraflopsold};
        \legend{Float16, Float32}
    \end{axis}
 
    \draw (rel axis cs:0.36,-0.12) -- (rel axis cs:0.60,-0.12) node[midway, below, black] {\footnotesize \textbf{Turing}};
    \draw (rel axis cs:0.97,-0.12) -- (rel axis cs:1.19,-0.12) node[midway, below, black] {\footnotesize \textbf{Turing}};
    \draw (rel axis cs:1.21,-0.12) -- (rel axis cs:1.55,-0.12) node[midway, below, black] {\footnotesize \textbf{Pascal}};
    
\end{tikzpicture}
    \caption{\textbf{Theoretical GPU performance.} We report performance in floating point operations per second (FLOP/s), sourced from official NVIDIA documentation. Tensor Core performance corresponds to dense operations with FP32 accumulation. GPUs are ordered by release date from newer (left) to older (right). Hardware support for Bfloat16 and TensorFloat32 formats was introduced after the Turing generation. NVIDIA GPUs of the Pascal generation and older do not possess Tensor Cores. For reference, our baseline ViT-B/16 requires 17.56 GFLOP per image during a forward pass~\cite{Vit_b_16TorchvisionMain}.}
    \label{fig:gpu_flops}
    \Description{Performance in floating point operations per second of the GPUs in our inventory. Floating point performance has increased exponentially throughout GPU generations, but the largest improvements come from the introduction of Tensor Cores, which provide an increase of a factor of four with respect to pure floating point operations for our newest GPUs (5000 Ada and A6000).}
\end{figure}
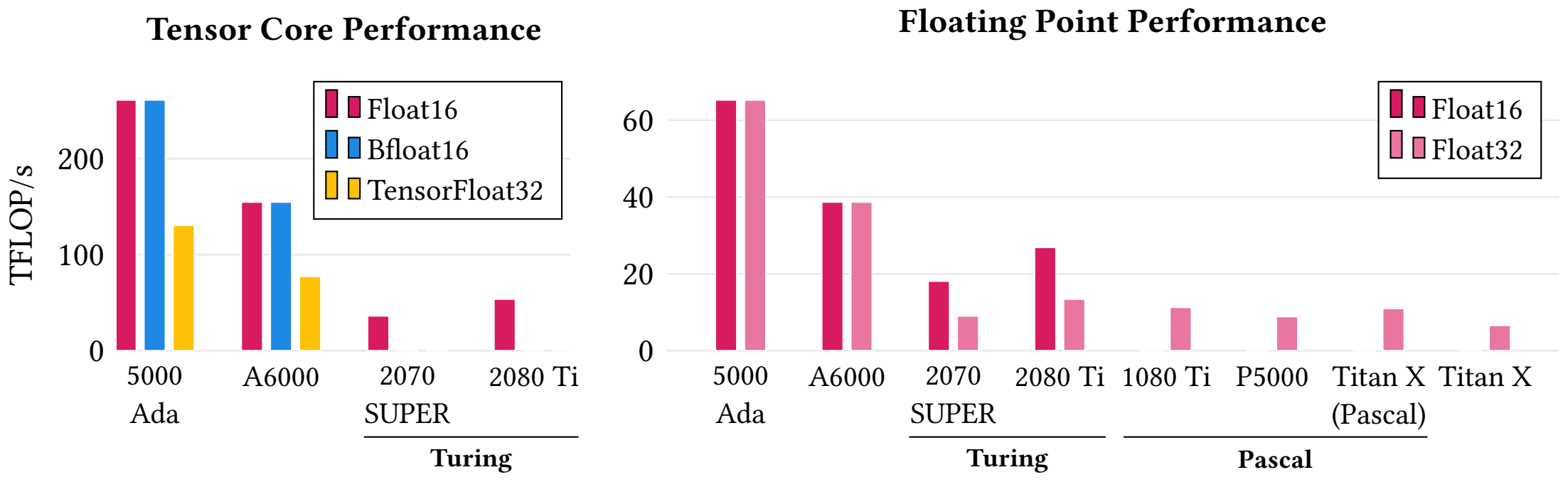

\begin{table}[ht]
    \caption{\textbf{Disk inventory.} Our baseline is the NVMe SSD. \emph{Max} read speed is the maximum rated speed of the disk. \emph{Sequential} and \emph{Random} read speeds are benchmarked locally with Fio~\cite{FioFlexible} for buffered 4 KB reads at queue depth 1, reflecting image sample retrieval during training.}
    \label{tab:disk_speed}
    \centering
    \resizebox{\columnwidth}{!}{
    \begin{tabular}{lllcccc}
    \toprule
    \multirow{2}[1]{*}{Name} & \multirow{2}[1]{*}{Disk Model} & \multirow{2}[1]{*}{Interface} & \multirow{2}[1]{*}{\makecell[c]{Capacity\\(TB)}} & \multicolumn{3}{c}{Read Speed (MB/s)}  \\ \cmidrule(lr){5-7}
    & & & & Max & Sequential & Random \\ \midrule
    \textbf{NVMe SSD} & SK Hynix PC801 & NVMe PCIe Gen 4 x4 & 2.0 & 7000 & 1011 & 49 \\
    SATA SSD & Samsung SSD 850 EVO & SATA 6 Gb/s & 0.5 & 540 & 509 & 26 \\
    HDD & Dell 6KR2M Seagate Exos 7E10 & SATA 6 Gb/s & 4.0 & 226 & 239 & 0.7 \\
    \bottomrule
\end{tabular}

    }
\end{table}

\begin{table}[ht]
    \centering
    \caption{\textbf{CPU inventory.} Our baseline is the Intel Xeon Silver 4208. TDP stands for Thermal Design Power, the theoretical average power consumption of the CPU when operating at base frequency with all cores active under a heavy load (according to \href{https://www.intel.com}{Intel}). All CPU models are manufactured using 14 nm lithography, except the Intel Core i7-5820K, which is based on a 22 nm process~\cite{IntelCoreI75820K}. \emph{Avg. Power} is average power consumption measured during ViT training for all CPUs in the server.}
    \begin{tabular}{lrcccccc}
    \toprule
    \multirow{2}[1]{*}{CPU Model} & \multirow{2}[1]{*}{Release Date} & \multicolumn{2}{c}{Cores} & \multicolumn{2}{c}{Freq. (GHz)} & \multirow{2}[1]{*}{\makecell[c]{TDP\\(W)}} & \multirow{2}[1]{*}{\makecell[c]{Avg. Power\\(W)}} \\ \cmidrule(lr){3-4} \cmidrule(lr){5-6}
    & & Physical & Logical & Base & Max & & \\ \midrule
    Intel Xeon Silver 4210R & Feb 2020 & 10 & 20 & 2.4 & 3.2 & 100 & 89 \\
    \textbf{Intel Xeon Silver 4208} & Apr 2019 & 8 & 16 & 2.1 & 3.2 & 85 & 83 \\
    Intel Core i9-9900X & Oct 2018 & 10 & 20 & 3.5 & 4.5 & 165 & 89 \\
    Intel Core i7-7820X & Jun 2017 & 8 & 16 & 3.6 & 4.5 & 140 & 77 \\
    Intel Core i7-6900K & May 2016 & 8 & 16 & 3.2 & 4.0 & 140 & 65 \\
    Intel Xeon E5-2609 v4 & Mar 2016 & 8 & 8 & 1.7 & 1.7 & 85 & 58 \\
    Intel Core i7-5820K & Sep 2014 & 6 & 12 & 3.3 & 3.6  & 140 & 92 \\
    \bottomrule
\end{tabular}

    \label{tab:cpu_inventory}
\end{table}

\begin{table}[ht]
    \centering
    \caption{\textbf{Server inventory.} Our baseline is S1. RAM speed is the effective running frequency determined by the CPU in megatransfers per second (MT/s), not the maximum rated speed of the modules. PCIe bandwidth is the effective host-to-device bandwidth benchmarked locally with NVIDIA NVBandwidth~\cite{NVIDIANvbandwidth2026}. CPU specifications can be found in \cref{tab:cpu_inventory}.}
    \begin{tabular}{lllccc}
    \toprule
    \multirow{2}{*}{Server} & \multirow{2}{*}{CPU} & \multirow{2}{*}{Motherboard} & \multirow{2}{*}{RAM} & RAM Speed & PCIe Bandwidth\\ 
    & & & & (MT/s) & (GB/s) \\ \midrule
    S0 & 2 $\times$ Intel Xeon Silver 4210R & Dell Precision 7920 & 8 $\times$ 16 GB & 2400 & 8.2 \\
    \textbf{S1} & 2 $\times$ Intel Xeon Silver 4208 & Dell Precision 7920 & 4 $\times$ 16 GB & 2400 & 8.0\\
    S2 & 1 $\times$ Intel Core i9-9900X & X299 AORUS MASTER & 6 $\times$ 16 GB & 2400 & 8.1 \\
    S3 & 1 $\times$ Intel Core i7-7820X & ASUS WS X299 SAGE & 8 $\times$ 16 GB & 2400 & 8.1 \\
    S4 & 1 $\times$ Intel Core i7-6900K & X99A GODLIKE GAMING & 8 $\times$ 16 GB & 2133 & 11.2 \\
    S5 & 2 $\times$ Intel Xeon E5-2609 v4 & Dell Precision 7810 & 4 $\times$ 16 GB & 1866 & 11.0 \\
    S6 & 1 $\times$ Intel Core i7-5820K & ASRock X99 WS-E & 8 $\times$ \hphantom{.} 8 GB & 2133 & 6.2 \\
    \bottomrule
\end{tabular}

    \label{tab:server_inventory}
\end{table}

\section{Hardware Specifications}\label{sec:B_hardware_specifications}

\Cref{tab:gpu_inventory,tab:cpu_inventory,tab:server_inventory,tab:disk_speed} describe the technical specifications of the GPUs (\cref{tab:gpu_inventory}), disks (\cref{tab:disk_speed}), CPUs (\cref{tab:cpu_inventory}), and servers (\cref{tab:server_inventory}) that we evaluate in this study. \Cref{fig:gpu_flops} illustrates theoretical performance differences between the GPU models in \cref{tab:gpu_inventory}, as well as architectural differences: our NVIDIA GPUs prior to the Turing generation are not equipped with Tensor Cores for faster matrix multiplication, and by extension also do not support TensorFloat32. Regarding mixed precision formats, our Pascal cards support Float16 arithmetic, but with drastically lower performance than Float32~\cite{PascalTuning}.

\begin{figure}[ht]
    \centering
    \input{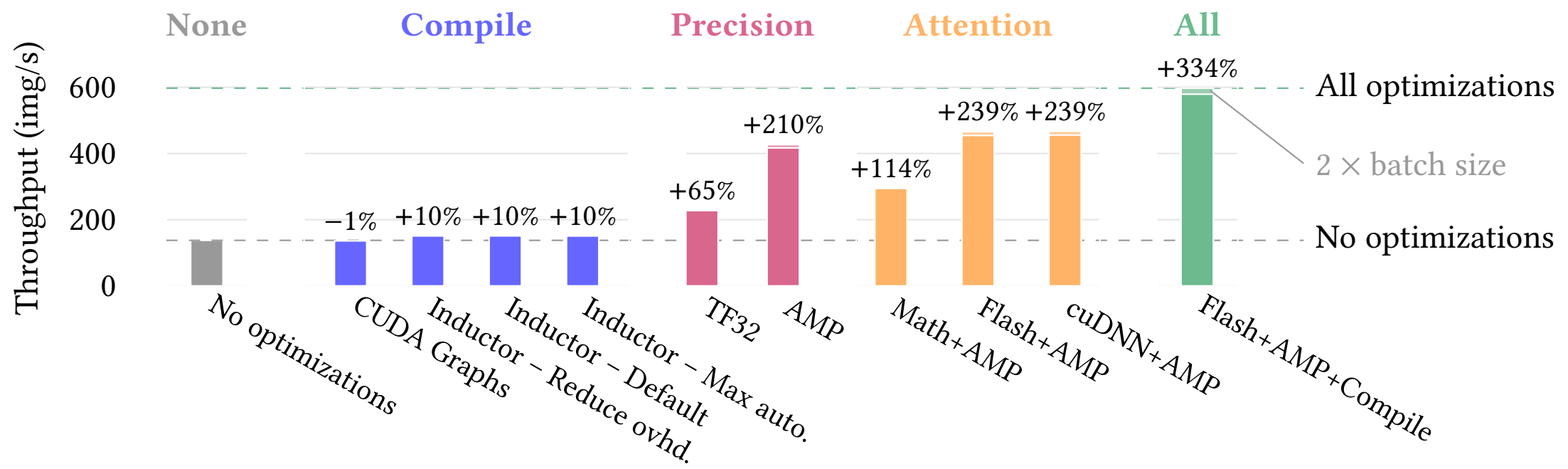}

\pgfplotsset{
    optimization panel/.append style={ymax=\optiYmaxThr, table/x=exp_name, table/y=img_per_s},
    pct labels/.append style={table/meta expr={\pctOf{img_per_s}{\noOptiThr}}},
}

\begin{tikzpicture}
    \begin{axis}[
        no opti panel,
        ylabel={Throughput (img/s)},
    ]
        \addplot[fill=gray!80] table {\baselinedata};
    \end{axis}

    \draw[dashed, gray!80] (rel axis cs:0.5,\noOptiThr/\optiYmaxThr) -- (rel axis cs:14.5,\noOptiThr/\optiYmaxThr) node[right, black] {No optimizations};
    \draw[dashed, ForestGreen!60] (rel axis cs:0.5,\allOptiThr/\optiYmaxThr) -- (rel axis cs:14.5,\allOptiThr/\optiYmaxThr) node[right, black] {All optimizations};

    \begin{axis}[
        compile panel,
    ]
        \addplot[fill=blue!60, pct labels] table {\compiledata};
    \end{axis}

    \begin{axis}[
        precision panel,
    ]
        \addplot[fill=purple!40, pct labels]
            table[y=img_per_s_large_bs, meta expr={\pctOf{img_per_s_large_bs}{\noOptiThr}}] {\precisiondata};
        \addplot[fill=purple!60] table {\precisiondata};
    \end{axis}

    \begin{axis}[
        attention panel,
    ]
        \addplot[fill=orange!40, pct labels]
            table[y=img_per_s_large_bs, meta expr={\pctOf{img_per_s_large_bs}{\noOptiThr}}] {\attentiondata};
        \addplot[fill=orange!60] table {\attentiondata};
    \end{axis}

    \begin{axis}[
        all opti panel,
    ]
        \addplot[fill=ForestGreen!40, pct labels]
            table[y=img_per_s_large_bs, meta expr={\pctOf{img_per_s_large_bs}{\noOptiThr}}] {\optidata};
        \addplot[fill=ForestGreen!60] table {\optidata};
    \end{axis}

    \draw[gray!80] (rel axis cs:13.35,590/\optiYmaxThr) -- (rel axis cs:14.5,370/\optiYmaxThr) node[right, gray!80] {2 $\times$ batch size};
\end{tikzpicture}
    \caption{\textbf{Impact of PyTorch optimizations on training throughput.} We measure throughput at batch size 256, and additionally at batch size 512 when made possible by the lower memory usage of Automatic Mixed Precision (AMP) (bar extensions).
    }
    \label{fig:optimization_throughput}
    \Description{Fully described in the text.}
\end{figure}

\begin{figure}
    \centering
    \input{tikz/common_styles/pytorch_cuda_versions_common}

\begin{tikzpicture}
    \begin{axis}[
        release year axis,
        ylabel={Throughput (img/s)},
    ]
    \end{axis}
    \begin{axis}[
        release plot axis,
        title={\textbf{PyTorch Version Throughput}},
        ymax=700,
    ]
        \pytorchOptiSmallBatch{img_per_s}
        \pytorchDefault{img_per_s}
        \pytorchOptiLargeBatch{img_per_s}
        \pytorchNoOpti{img_per_s}
        \legend{{All optim. on, BS 256}, Default optim. on, {All optim. on, BS 512}, No optimizations}
    \end{axis}
\end{tikzpicture}
    \caption{\textbf{Impact of PyTorch versions on training throughput.} We compare default settings to all applicable optimizations enabled or disabled. Larger green markers indicate a batch size of 512, which we use whenever possible instead of 256. Overlapping markers correspond to different CUDA builds of the same version.}
    \label{fig:pytorch_throughput}
    \Description{Fully described in the text.}
\end{figure}

\begin{figure}[ht]
    \centering
    \input{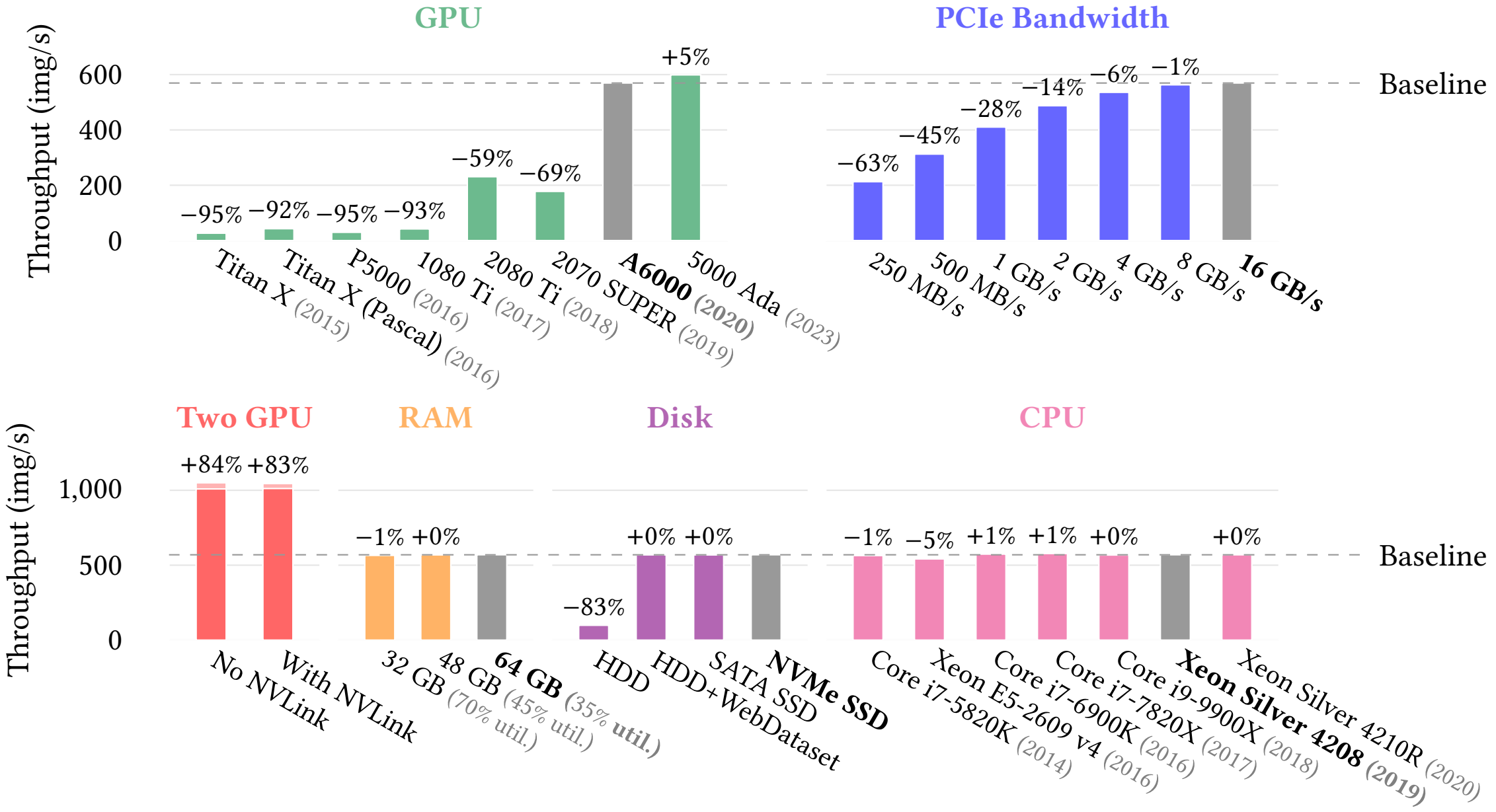}

\pgfplotsset{
    hardware panel/.append style={table/x=exp_name, table/y=img_per_s},
    pct labels/.append style={table/meta expr={\pctOf{img_per_s}{\baselineThr}}},
}

\begin{tikzpicture}
    \begin{axis}[
        gpu panel,
        ymax=\thrYmaxSmall,
        ylabel={Throughput (img/s)},
    ]
        \addplot[fill=ForestGreen!60, pct labels] table {\gpudata};
        \addplot[reference bar] coordinates {(a6000_512,\baselineThr)};
        \coordinate (baseline_start) at (rel axis cs: 0., \baselineThr/\thrYmaxSmall);
    \end{axis}

    \begin{axis}[
        two gpu panel,
        ymax=\thrYmaxLarge,
        ylabel={Throughput (img/s)},
    ]
        \addplot[fill=red!30, pct labels]
            table[y=img_per_s_large_bs, meta expr={\pctOf{img_per_s_large_bs}{\baselineThr}}] {\twogpudata};
        \addplot[fill=red!60] table {\twogpudata};
        \coordinate (baseline_start2) at (rel axis cs: 0., \baselineThr/\thrYmaxLarge);
    \end{axis}

    \begin{axis}[
        ram panel,
        ymax=\thrYmaxLarge,
    ]
        \addplot[fill=orange!60, pct labels] table {\ramdata};
        \addplot[reference bar] coordinates {(64gb,\baselineThr)};
    \end{axis}

    \begin{axis}[
        disk panel,
        ymax=\thrYmaxLarge,
    ]
        \addplot[fill=violet!60, pct labels] table {\diskdata};
        \addplot[reference bar] coordinates {(nvme_ssd,\baselineThr)};
    \end{axis}

    \begin{axis}[
        cpu panel,
        ymax=\thrYmaxLarge,
    ]
        \addplot[fill=magenta!60, pct labels] table {\cpudata};
        \addplot[reference bar] coordinates {(moravia,\baselineThr)};
    \end{axis}

    \begin{axis}[
        pcie panel,
        ymax=\thrYmaxSmall,
    ]
        \addplot[fill=blue!60, pct labels] table {\pciedata};
        \addplot[reference bar] coordinates {(x16gen3,\baselineThr)};
    \end{axis}

    \draw[dashed, gray!80] (baseline_start) -- ++(12, 0) node[right, black] {Baseline};
    \draw[dashed, gray!80] (baseline_start2) -- ++(12, 0) node[right, black] {Baseline};
\end{tikzpicture}
    \caption{\textbf{Impact of hardware on training throughput.} Our baseline is the \emph{power-capped RTX A6000} with all PyTorch \emph{optimizations enabled} at \emph{batch size 512}. In the \textcolor{ForestGreen!60}{\textbf{GPU}} plot, we show power-capped performance at best batch size with all supported optimizations enabled (peak efficiency in \cref{fig:gpu_power_capping}). \textcolor{blue!60}{\textbf{PCIe Bandwidth}} bar labels indicate the nominal unidirectional bandwidth as dictated by PCIe generation and number of lanes. Bar extensions in the \textcolor{red!60}{\textbf{Two GPU}} plot show throughput when doubling the batch size.}
    \label{fig:hardware_throughput}
    \Description{Fully described in the text.}
\end{figure}

\section{Throughput} \label{sec:C_throughput}

\Cref{fig:optimization_throughput,fig:pytorch_throughput,fig:hardware_throughput} are complementary to \cref{fig:optimization_efficiency,fig:pytorch_efficiency,fig:hardware_efficiency}, showing the throughput achieved by different levels of PyTorch optimizations (\cref{fig:optimization_throughput}), PyTorch versions (\cref{fig:pytorch_throughput}), and hardware (\cref{fig:hardware_throughput}).

For PyTorch optimizations and PyTorch versions (\cref{fig:optimization_throughput,fig:pytorch_throughput}), trends are similar to efficiency, although gains in throughput are larger than efficiency gains: at higher throughput, CPU power consumption increases to ensure data-loading speed keeps up with GPU computation. On the flip side, large throughput drops due to lower-performing hardware (low PCIe bandwidth or slow disks, as illustrated in \cref{fig:hardware_throughput}) have a less noticeable impact on efficiency (\cref{fig:hardware_efficiency}), as GPU utilization and power consumption decrease.

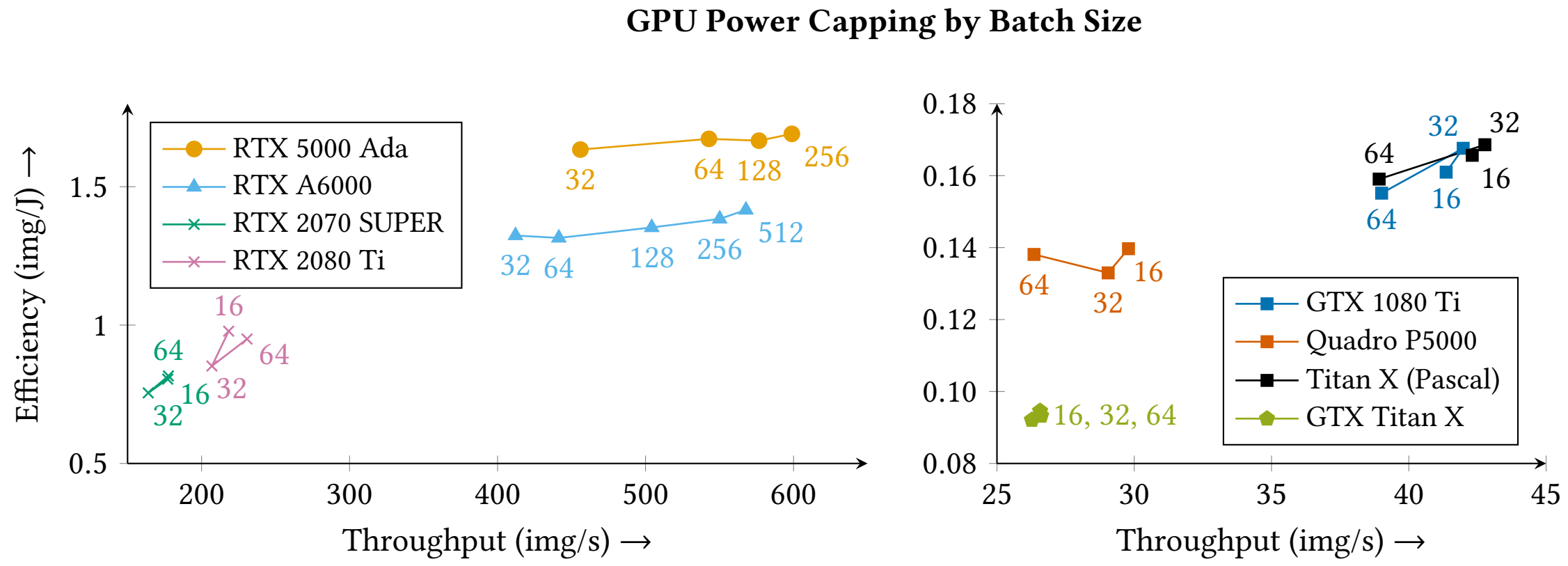
\begin{figure}
    \centering
    \begin{tikzpicture}

\pgfplotstableread{
exp_name img_per_j img_per_s
16   0.16097     41.36
32   0.16760     41.98
64   0.15513     39.01
}\teneighty

\pgfplotstableread{
exp_name img_per_j img_per_s
16   0.80474    177.14
32   0.75504    163.98
64   0.81625    177.59
}\seventy

\pgfplotstableread{
exp_name img_per_j img_per_s
16   0.97802    218.31
32   0.85212    207.00
64   0.94956    230.71
}\twentyeighty

\pgfplotstableread{
exp_name img_per_j img_per_s
32    1.63420    455.96
64    1.67280    542.83
128   1.66630    576.65
256   1.69100    598.86
}\afive

\pgfplotstableread{
exp_name img_per_j img_per_s
32    1.32370    412.13
64    1.31490    441.40
128   1.35270    504.18
256   1.38370    550.10
512   1.41600    567.83
}\asix

\pgfplotstableread{
exp_name img_per_j img_per_s
16   0.13969     29.79
32   0.13297     29.05
64   0.13810     26.35
}\pfive

\pgfplotstableread{
exp_name img_per_j img_per_s
16   0.09459     26.57
32   0.09326     26.61
64   0.09210     26.27
}\titanx

\pgfplotstableread{
exp_name img_per_j img_per_s
16   0.16565     42.31
32   0.16859     42.77
64   0.15904     38.92
}\titanxp

\begin{axis}[
    name=mainplot,
    width={0.57\textwidth},
    height=5cm,
    legend cell align=left,
    ymin=0.5,
    ymax=1.8,
    xmin=150,
    xmax=650,
    legend style={
        /tikz/column 2/.style={
            column sep=0.2cm,
        },
        at={(0.03,0.95)},
        anchor=north west,
        nodes={scale=0.9}
    },
    axis lines=left,
    xlabel={Throughput (img/s) $\rightarrow$},
    ylabel={Efficiency (img/J) $\rightarrow$},
]
\addplot[mark=*, mark size=2pt, line width=0.5pt, A5000color] table [
    x=img_per_s,
    y=img_per_j,
    meta=exp_name,
]{\afive};
\addplot[mark=triangle*, mark size=2pt, line width=0.5pt, A6000color] table [
    x=img_per_s,
    y=img_per_j,
    meta=exp_name,
]{\asix};
\addplot[mark=x, mark size=2pt, line width=0.5pt, 2070color] table [
    x=img_per_s,
    y=img_per_j,
    meta=exp_name,
]{\seventy};
\addplot[mark=x, mark size=2pt, line width=0.5pt, 2080color] table [
    x=img_per_s,
    y=img_per_j,
    meta=exp_name,
]{\twentyeighty};
\node[anchor=north, yshift=-2pt, A5000color] at (axis cs:455.96,1.6342) {32};
\node[anchor=north, yshift=-2pt, A5000color] at (axis cs:542.83,1.6728) {64};
\node[anchor=north, yshift=-2pt, A5000color] at (axis cs:576.65,1.6663) {128};
\node[anchor=north west, A5000color] at (axis cs:598.86,1.691) {256};

\node[anchor=north, yshift=-2pt, A6000color] at (axis cs:412.13,1.3237) {32};
\node[anchor=north, yshift=-2pt, A6000color] at (axis cs:441.4,1.3149) {64};
\node[anchor=north, yshift=-2pt, A6000color] at (axis cs:504.18,1.3527) {128};
\node[anchor=north, yshift=-2pt, A6000color] at (axis cs:550.1,1.3837) {256};
\node[anchor=north west, A6000color] at (axis cs:567.83,1.416) {512};

\node[anchor=north west, yshift=2pt, 2070color] at (axis cs:177.14,0.80474) {16};
\node[anchor=north west, xshift=-2pt, 2070color] at (axis cs:163.98,0.75504) {32};
\node[anchor=south, yshift=1pt, 2070color] at (axis cs:177.59,0.81625) {64};

\node[anchor=south, yshift=1pt, 2080color] at (axis cs:218.31,0.97802) {16};
\node[anchor=north west, xshift=-2pt, 2080color] at (axis cs:207,0.85212) {32};
\node[anchor=north west, yshift=2pt, 2080color] at (axis cs:230.71,0.94956) {64};

\legend{RTX 5000 Ada, RTX A6000, RTX 2070 SUPER, RTX 2080 Ti}
\end{axis}

\begin{axis}[
    name=secondplot,
    at={(mainplot.east)},
    anchor=west,
    xshift=35pt,
    width={0.45\textwidth},
    height=5cm,
    legend cell align=left,
    legend style={
        /tikz/column 2/.style={
            column sep=0.2cm,
        },
        at = {(0.95, 0.05)},
        anchor=south east,
        nodes={scale=0.9}
    },
    ymin=0.08,
    ymax=0.18,
    xmin=25,
    xmax=45,
    axis lines=left,
    xlabel={Throughput (img/s) $\rightarrow$},
    yticklabel style={
        /pgf/number format/.cd,
        fixed,
        precision=2,
        fixed zerofill
    },
]
\addplot[mark=square*, mark size=1.5pt, line width=0.5pt, 1080color] table [
    x=img_per_s,
    y=img_per_j,
    meta=exp_name,
]{\teneighty};

\addplot[mark=square*, mark size=1.5pt, line width=0.5pt, P5000color] table [
    x=img_per_s,
    y=img_per_j,
    meta=exp_name,
]{\pfive};

\addplot[mark=square*, mark size=1.5pt, line width=0.5pt, TitanXPcolor] table [
    x=img_per_s,
    y=img_per_j,
    meta=exp_name,
]{\titanxp};

\addplot[mark=pentagon*, mark size=2pt, line width=0.5pt, TitanXcolor] table [
    x=img_per_s,
    y=img_per_j,
    meta=exp_name,
]{\titanx};
\node[anchor=north, yshift=-1pt, 1080color] at (axis cs:41.36,0.16097) {16};
\node[anchor=south east, xshift=2pt, 1080color] at (axis cs:41.98,0.1676) {32};
\node[anchor=north, yshift=-1pt, 1080color] at (axis cs:39.01,0.15513) {64};

\node[anchor=north west, xshift=-2pt, P5000color] at (axis cs:29.79,0.13969) {16};
\node[anchor=north, yshift=-2pt, P5000color] at (axis cs:29.05,0.13297) {32};
\node[anchor=north, yshift=-2pt, P5000color] at (axis cs:26.35,0.1381) {64};

\node[anchor=north west, xshift=-1pt, TitanXPcolor] at (axis cs:42.31,0.16565) {16};
\node[anchor=south west, xshift=-2pt, TitanXPcolor] at (axis cs:42.77,0.16859) {32};
\node[anchor=south, yshift=1pt, TitanXPcolor] at (axis cs:38.92,0.15904) {64};

\node[anchor=west, TitanXcolor] at (axis cs:26.6,0.093) {16, 32, 64};
\legend{GTX 1080 Ti, Quadro P5000, Titan X (Pascal), GTX Titan X}

\end{axis}
\node[anchor=south] at ($(mainplot.north)!0.5!(secondplot.north) + (0, 0.5cm)$) {\textbf{GPU Power Capping by Batch Size}};
\end{tikzpicture}
    \caption{\textbf{GPU power capping by batch size.} Throughput and efficiency of \emph{power-capped} GPUs for a range of batch sizes until saturating GPU memory, with all supported PyTorch \emph{optimizations enabled}. Marker shapes indicate different NVIDIA generations: Ada Lovelace, Ampere, Turing, Pascal, and Maxwell (see \cref{fig:gpu_power_capping}).}
    \Description{Two line graphs showing throughput in images per second on the X axis against efficiency in images per joule on the Y axis for eight NVIDIA GPUs, with point labels indicating batch size. The left plot contains data for the RTX 5000 Ada, RTX A6000, RTX 2070 SUPER, and RTX 2080 Ti. The right plot contains data for the GTX 1080 Ti, Quadro P5000, Titan X (Pascal), and GTX Titan X. The newer cards on the left plot are typically faster and more efficient at larger batch sizes. The older cards on the right plot are more efficient at batch sizes that do not saturate GPU memory.}
    \label{fig:gpu_efficiency_bsize}
\end{figure}

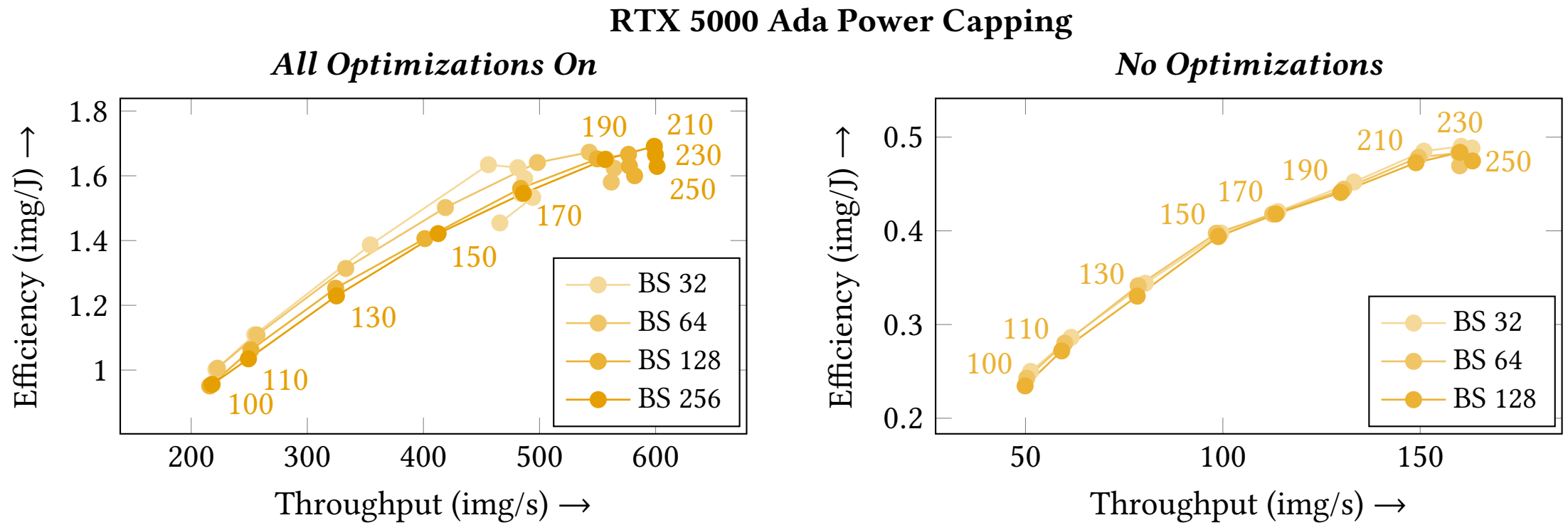
\begin{figure}[ht]
    \centering    
        \begin{tikzpicture}
    \begin{axis}[
        name=mainplot,
        title={\textbf{\emph{All Optimizations On}}},
        title style={yshift=-5pt},
        width={0.49\textwidth},
        height=4.7cm,
        xlabel={Throughput (img/s) $\rightarrow$},
        ylabel={Efficiency (img/J) $\rightarrow$},
        legend columns=1,
        enlarge x limits = 0.2,
        enlarge y limits = 0.2,
        legend cell align=left,
        legend style={
            /tikz/column 2/.style={
                column sep=0.2cm,
            },
            cells={align=left},
            at={(0.99,0.02)},
            anchor=south east,
            nodes={scale=0.9}
        },
        point meta=explicit,
        every node near coord/.style={anchor=south east, rotate=0},
    ]
    \addplot[mark=*, line width=0.5pt, A5000color!40] table [
        x=img_per_s,
        y=img_per_j,
        meta=power_limit_w,
        col sep=comma,
        discard if two={batch_size}{32}{compile}{True},
    ]{data/A5000.csv};
    \addplot[mark=*, line width=0.5pt, A5000color!60] table [
        x=img_per_s,
        y=img_per_j,
        meta=power_limit_w,
        col sep=comma,
        discard if two={batch_size}{64}{compile}{True},
    ]{data/A5000.csv};
    \addplot[mark=*, line width=0.5pt, A5000color!80] table [
        x=img_per_s,
        y=img_per_j,
        meta=power_limit_w,
        col sep=comma,
        discard if two={batch_size}{128}{compile}{True},
    ]{data/A5000.csv};
    \addplot[mark=*, line width=0.5pt, A5000color] table [
        x=img_per_s,
        y=img_per_j,
        meta=power_limit_w,
        col sep=comma,
        discard if two={batch_size}{256}{compile}{True},
    ]{data/A5000.csv};
    \node[anchor=north west, text=A5000color] at (axis cs:220.23,0.96531) {100};
    \node[anchor=north west, text=A5000color] at (axis cs:250,1.0406) {110};
    \node[anchor=north west, text=A5000color] at (axis cs:326,1.2335) {130};
    \node[anchor=north west, text=A5000color] at (axis cs:412.58,1.422) {150};
    \node[anchor=north west, text=A5000color] at (axis cs:485.86,1.5492) {170};
    \node[anchor=south, yshift=3pt, text=A5000color] at (axis cs:555.63,1.6474) {190};
    \node[anchor=south west, text=A5000color] at (axis cs:598.81,1.6897) {210};
    \node[anchor=west, xshift=2pt, text=A5000color] at (axis cs:599.58,1.6654) {230};
    \node[anchor=north west, text=A5000color] at (axis cs:601.14,1.6293) {250};
    \legend{BS 32, BS 64, BS 128, BS 256}
    \end{axis}

    \begin{axis}[
        name=secondplot,
        at={(mainplot.north east)},
        anchor=north west,
        xshift=50pt,
        enlarge x limits = 0.2,
        enlarge y limits = 0.2,
        title={\textbf{\emph{No Optimizations}}},
        title style={yshift=-5pt},
        width={0.49\textwidth},
        height=4.7cm,
        xlabel={Throughput (img/s) $\rightarrow$},
        ylabel={Efficiency (img/J) $\rightarrow$},
        legend cell align=left,
        legend columns=1,
        legend style={
            /tikz/column 2/.style={
                column sep=0.2cm,
            },
            cells={align=left},
            at={(0.99,0.02)},
            anchor=south east,
            nodes={scale=0.9}
        },
        point meta=explicit,
        every node near coord/.style={anchor=south east, rotate=0},
    ]

    \addplot[mark=*, line width=0.5pt, A5000color!40] table [
        x=img_per_s,
        y=img_per_j,
        meta=power_limit_w,
        col sep=comma,
        discard if two={batch_size}{32}{compile}{False},
    ]{data/A5000.csv};
    \addplot[mark=*, line width=0.5pt, A5000color!60] table [
        x=img_per_s,
        y=img_per_j,
        meta=power_limit_w,
        col sep=comma,
        discard if two={batch_size}{64}{compile}{False},
    ]{data/A5000.csv};
    \addplot[mark=*, line width=0.5pt, A5000color!80] table [
        x=img_per_s,
        y=img_per_j,
        meta=power_limit_w,
        col sep=comma,
        discard if two={batch_size}{128}{compile}{False},
    ]{data/A5000.csv};
    \node[anchor=south east, text=A5000color!80] at (axis cs:49.897,0.23448) {100};
    \node[anchor=south east, text=A5000color!80] at (axis cs:59.161,0.27184) {110};
    \node[anchor=south east, text=A5000color!80] at (axis cs:78.301,0.33029) {130};
    \node[anchor=south east, text=A5000color!80] at (axis cs:98.849,0.39384) {150};
    \node[anchor=south east, text=A5000color!80] at (axis cs:113.46,0.41801) {170};
    \node[anchor=south east, text=A5000color!80] at (axis cs:129.87,0.44093) {190};
    \node[anchor=south east, text=A5000color!80] at (axis cs:148.9,0.47268) {210};
    \node[anchor=south, yshift=2pt, text=A5000color!80] at (axis cs:160.07,0.4838) {230};
    \node[anchor=west, text=A5000color!80] at (axis cs:163.25,0.47459) {250};

    \legend{BS 32, BS 64, BS 128}
    \end{axis}
    \node at ($(mainplot.north)!0.5!(secondplot.north) + (0, 0.7cm)$) {\textbf{RTX 5000 Ada Power Capping}};
    \end{tikzpicture}%
    \caption{\textbf{RTX 5000 Ada power capping.} Curve annotations indicate power limits in watts. \emph{BS} stands for \emph{batch size}.}
    \label{fig:A5000_power_capping}
    \Description{Two line graphs showing throughput in images per second on the X axis against efficiency in images per joule on the Y axis, for the RTX 5000 Ada GPU. The line graph on the left plots four lines for batch sizes 32, 64, 128, and 256 for code with all PyTorch optimizations enabled. The line graph on the right plots three lines for batch sizes 32, 64, and 128 for code with no optimizations enabled. Values along each line indicate power limits in watts, ranging from 250 on the top right to 100 on the bottom left. Throughput ranges from 215 to 600 with optimizations, and from 50 to 163 without optimizations. Efficiency ranges from 0.95 to 1.7 with optimizations, and from 0.2 to 0.5 without optimizations. The RTX 5000 Ada is most efficient at 210 watts for large batch sizes, and at even lower power as batch size decreases. Without optimizations, it is most efficient at 230 watts. The batch size curves align without optimizations, but they do not when optimizations are enabled.}
\end{figure}

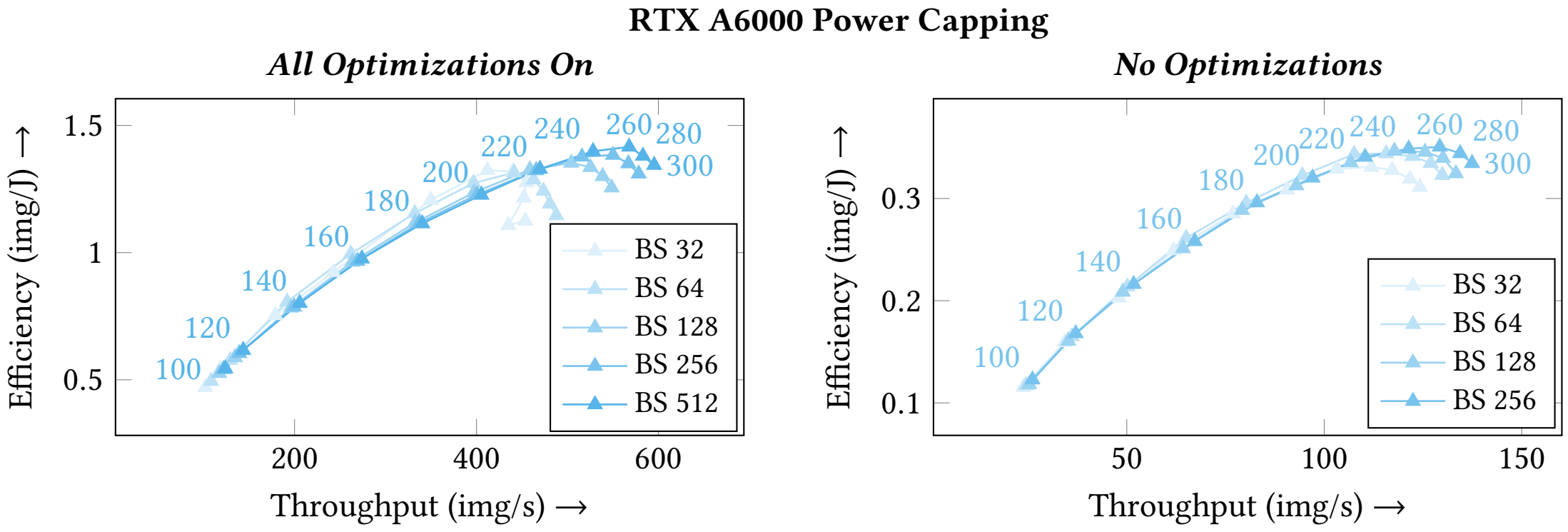
\begin{figure}[ht]
    \centering    
        \begin{tikzpicture}
    \begin{axis}[
        name=mainplot,
        title={\textbf{\emph{All Optimizations On}}},
        title style={yshift=-5pt},
        width={0.49\textwidth},
        height=4.7cm,
        enlarge x limits = 0.2,
        enlarge y limits = 0.2,
        xlabel={Throughput (img/s) $\rightarrow$},
        ylabel={Efficiency (img/J) $\rightarrow$},
        legend cell align=left,
        legend columns=1,
        legend style={
            /tikz/column 2/.style={
                column sep=0.2cm,
            },
            cells={align=left},
            at={(0.99,0.01)},
            anchor=south east,
            nodes={scale=0.9}
        },
        point meta=explicit,
        every node near coord/.style={anchor=south east, rotate=0},
    ]
    \addplot[mark=triangle*, line width=0.5pt, A6000color!20] table [
        x=img_per_s,
        y=img_per_j,
        meta=power_limit_w,
        col sep=comma,
        discard if two={batch_size}{32}{compile}{True},
    ]{data/A6000.csv};
    \addplot[mark=triangle*, line width=0.5pt, A6000color!40] table [
        x=img_per_s,
        y=img_per_j,
        meta=power_limit_w,
        col sep=comma,
        discard if two={batch_size}{64}{compile}{True},
    ]{data/A6000.csv};
    \addplot[mark=triangle*, line width=0.5pt, A6000color!60] table [
        x=img_per_s,
        y=img_per_j,
        meta=power_limit_w,
        col sep=comma,
        discard if two={batch_size}{128}{compile}{True},
    ]{data/A6000.csv};
    \addplot[mark=triangle*, line width=0.5pt, A6000color!80] table [
        x=img_per_s,
        y=img_per_j,
        meta=power_limit_w,
        col sep=comma,
        discard if two={batch_size}{256}{compile}{True},
    ]{data/A6000.csv};
    \addplot[mark=triangle*, line width=0.5pt, A6000color] table [
        x=img_per_s,
        y=img_per_j,
        meta=power_limit_w,
        col sep=comma,
        discard if two={batch_size}{512}{compile}{True},
    ]{data/A6000.csv};
    \node[anchor=east, xshift=-3pt, text=A6000color] at (axis cs:123.98,0.54415) {100};
    \node[anchor=south east, text=A6000color] at (axis cs:143.56,0.61672) {120};
    \node[anchor=south east, text=A6000color] at (axis cs:205.65,0.80137) {140};
    \node[anchor=south east, text=A6000color] at (axis cs:274.07,0.97596) {160};
    \node[anchor=south east, text=A6000color] at (axis cs:340.42,1.1145) {180};
    \node[anchor=south east, text=A6000color] at (axis cs:405.39,1.227) {200};
    \node[anchor=south east, text=A6000color] at (axis cs:469.85,1.329) {220};
    \node[anchor=south east, text=A6000color] at (axis cs:528.27,1.3991) {240};
    \node[anchor=south, text=A6000color] at (axis cs:567.83,1.416) {260};
    \node[anchor=south west, text=A6000color] at (axis cs:582.96,1.38) {280};
    \node[anchor=west, text=A6000color] at (axis cs:595.29,1.3443) {300};
    \legend{BS 32, BS 64, BS 128, BS 256, BS 512}
    \end{axis}

    \begin{axis}[
        at={(mainplot.north east)}, anchor=north west, xshift=50pt,
        title={\textbf{\emph{No Optimizations}}},
        title style={yshift=-5pt},
        width={0.49\textwidth},
        height=4.7cm,
        enlarge x limits = 0.2,
        enlarge y limits = 0.2,
        name=secondplot,
        xlabel={Throughput (img/s) $\rightarrow$},
        ylabel={Efficiency (img/J) $\rightarrow$},
        legend cell align=left,
        legend columns=1,
        legend style={
            /tikz/column 2/.style={
                column sep=0.2cm,
            },
            cells={align=left},
            at={(0.99,0.02)},
            anchor=south east,
            nodes={scale=0.9}
        },
        point meta=explicit,
        every node near coord/.style={anchor=south east, rotate=0},
    ]
    \addplot[mark=triangle*, line width=0.5pt, A6000color!20] table [
        x=img_per_s,
        y=img_per_j,
        meta=power_limit_w,
        col sep=comma,
        discard if two={batch_size}{32}{compile}{False},
    ]{data/A6000.csv};
    \addplot[mark=triangle*, line width=0.5pt, A6000color!40] table [
        x=img_per_s,
        y=img_per_j,
        meta=power_limit_w,
        col sep=comma,
        discard if two={batch_size}{64}{compile}{False},
    ]{data/A6000.csv};
    \addplot[mark=triangle*, line width=0.5pt, A6000color!60] table [
        x=img_per_s,
        y=img_per_j,
        meta=power_limit_w,
        col sep=comma,
        discard if two={batch_size}{128}{compile}{False},
    ]{data/A6000.csv};
    \addplot[mark=triangle*, line width=0.5pt, A6000color!80] table [
        x=img_per_s,
        y=img_per_j,
        meta=power_limit_w,
        col sep=comma,
        discard if two={batch_size}{256}{compile}{False},
    ]{data/A6000.csv};
    \node[anchor=south east, text=A6000color!80] at (axis cs:26.13,0.12277) {100};
    \node[anchor=south east, text=A6000color!80] at (axis cs:37.077,0.16798) {120};
    \node[anchor=south east, text=A6000color!80] at (axis cs:51.745,0.21619) {140};
    \node[anchor=south east, text=A6000color!80] at (axis cs:67.22,0.25811) {160};
    \node[anchor=south east, text=A6000color!80] at (axis cs:82.97,0.29605) {180};
    \node[anchor=south east, text=A6000color!80] at (axis cs:97.028,0.32023) {200};
    \node[anchor=south east, xshift=-2pt, yshift=-1pt, text=A6000color!80] at (axis cs:110.35,0.34064) {220};
    \node[anchor=south east, text=A6000color!80] at (axis cs:121.37,0.3487) {240};
    \node[anchor=south, text=A6000color!80] at (axis cs:129.25,0.35055) {260};
    \node[anchor=south west, text=A6000color!80] at (axis cs:134.42,0.3442) {280};
    \node[anchor=west, text=A6000color!80] at (axis cs:137.44,0.33448) {300};

    \legend{BS 32, BS 64, BS 128, BS 256}
    \end{axis}
    \node at ($(mainplot.north)!0.5!(secondplot.north) + (0, 0.7cm)$) {\textbf{RTX A6000 Power Capping}};
    \end{tikzpicture}%
    \caption{\textbf{RTX A6000 power capping.} Curve annotations indicate power limits in watts. \emph{BS} stands for \emph{batch size}.}
    \label{fig:A6000_power_capping}
    \Description{Two line graphs showing throughput in images per second on the X axis against efficiency in images per joule on the Y axis, for the RTX A6000 GPU. The line graph on the left plots five lines for batch sizes 32, 64, 128, 256, and 512 for code with all PyTorch optimizations enabled. The line graph on the right plots four lines for batch sizes 32, 64, 128, and 256 for code with no optimizations enabled. Values along each line indicate power limits in watts, ranging from 300 on the top right to 100 on the bottom left. Throughput ranges from 100 to 600 with optimizations, and from 24 to 137 without optimizations. Efficiency ranges from 0.47 to 1.4 with optimizations, and from 0.1 to 0.35 without optimizations. The RTX A6000 is most efficient at 260 watts for large batch sizes. Trends are similar with all PyTorch optimizations enabled and with no optimizations enabled.}
\end{figure}

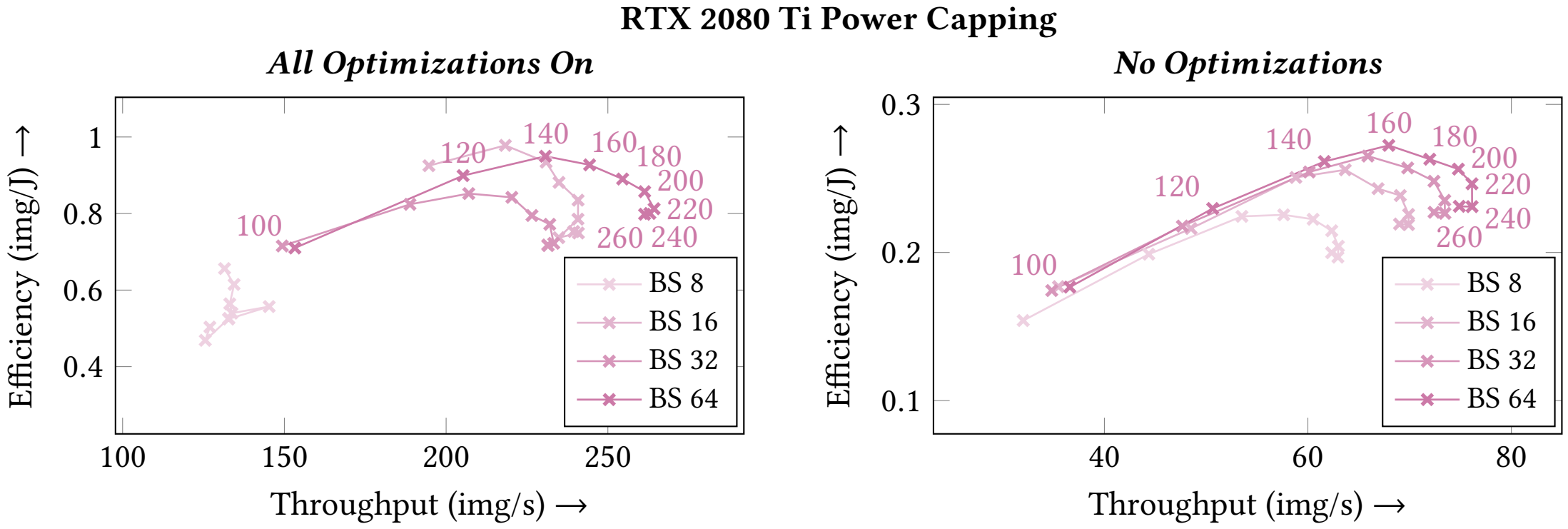
\begin{figure}[ht]
    \centering    
        \begin{tikzpicture}
        \begin{axis}[
        name=mainplot,
        title={\textbf{\emph{All Optimizations On}}},
        width={0.49\textwidth},
        title style={yshift=-5pt},
        height=4.7cm,
        xlabel={Throughput (img/s) $\rightarrow$},
        ylabel={Efficiency (img/J) $\rightarrow$},
        legend cell align=left,
        legend columns=1,
        ymin=0.35,
        enlarge x limits=0.2,
        enlarge y limits=0.2,
        mark options={line width=1pt},
        legend style={
            /tikz/column 2/.style={
                column sep=0.2cm,
            },
            cells={align=left},
            at={(0.99,0.02)},
            anchor=south east,
            nodes={scale=0.9}
        },
        point meta=explicit,
        every node near coord/.style={anchor=south, rotate=0},
    ]
    \addplot[mark=x, line width=0.5pt, 2080color!34] table [
        x=img_per_s,
        y=img_per_j,
        meta=power_limit_w,
        col sep=comma,
        discard if two={batch_size}{8}{compile}{True},
    ]{data/2080Ti.csv};
    \addplot[mark=x, line width=0.5pt, 2080color!56] table [
        x=img_per_s,
        y=img_per_j,
        meta=power_limit_w,
        col sep=comma,
        discard if two={batch_size}{16}{compile}{True},
    ]{data/2080Ti.csv};
    \addplot[mark=x, line width=0.5pt, 2080color!78] table [
        x=img_per_s,
        y=img_per_j,
        meta=power_limit_w,
        col sep=comma,
        discard if two={batch_size}{32}{compile}{True},
    ]{data/2080Ti.csv};
    \addplot[mark=x, line width=0.5pt, 2080color] table [
        x=img_per_s,
        y=img_per_j,
        meta=power_limit_w,
        col sep=comma,
        discard if two={batch_size}{64}{compile}{True},
    ]{data/2080Ti.csv};
    \node[anchor=south east, text=2080color] at (axis cs:153.24,0.71001) {100};
    \node[anchor=south, text=2080color] at (axis cs:205.25,0.89902) {120};
    \node[anchor=south, text=2080color] at (axis cs:230.71,0.94955) {140};
    \node[anchor=south west, xshift=-3pt, text=2080color] at (axis cs:244.35,0.92712) {160};
    \node[anchor=south west, text=2080color] at (axis cs:254.56,0.89011) {180};
    \node[anchor=west, yshift=3pt, text=2080color] at (axis cs:261.37,0.85714) {200};
    \node[anchor=west, text=2080color] at (axis cs:264.28,0.81254) {220};
    \node[anchor=north west, xshift=-3pt, text=2080color] at (axis cs:262.98,0.80018) {240};
    \node[anchor=north east, xshift=3pt, text=2080color] at (axis cs:261.35,0.79806) {260};
    \legend{BS 8, BS 16, BS 32, BS 64}
    \end{axis}

    \begin{axis}[
        at={(mainplot.north east)}, anchor=north west, xshift=50pt,
        title={\textbf{\emph{No Optimizations}}},
        name=secondplot,
        width={0.49\textwidth},
        height=4.7cm,
        title style={yshift=-5pt},
        mark options={line width=1pt},
        xlabel={Throughput (img/s) $\rightarrow$},
        ylabel={Efficiency (img/J) $\rightarrow$},
        legend cell align=left,
        legend columns=1,
        ymin=0.11,
        enlarge x limits=0.2,
        enlarge y limits=0.2,
        point meta=explicit,
        legend style={
            /tikz/column 2/.style={
                column sep=0.2cm,
            },
            cells={align=left},
            at={(0.99,0.02)},
            anchor=south east,
            nodes={scale=0.9}
        },
    ]
    \addplot[mark=x, line width=0.5pt, 2080color!34] table [
        x=img_per_s,
        y=img_per_j,
        meta=power_limit_w,
        col sep=comma,
        discard if two={batch_size}{8}{compile}{False},
    ]{data/2080Ti.csv};
        \addplot[mark=x, line width=0.5pt, 2080color!56] table [
            x=img_per_s,
            y=img_per_j,
            meta=power_limit_w,
            col sep=comma,
            discard if two={batch_size}{16}{compile}{False},
        ]{data/2080Ti.csv};
    \addplot[mark=x, line width=0.5pt, 2080color!78] table [
        x=img_per_s,
        y=img_per_j,
        meta=power_limit_w,
        col sep=comma,
        discard if two={batch_size}{32}{compile}{False},
    ]{data/2080Ti.csv};
    \addplot[mark=x, line width=0.5pt, 2080color] table [
        x=img_per_s,
        y=img_per_j,
        meta=power_limit_w,
        col sep=comma,
        discard if two={batch_size}{64}{compile}{False},
    ]{data/2080Ti.csv};
    \node[anchor=south east, text=2080color] at (axis cs:36.604,0.17663) {100};
    \node[anchor=south east, text=2080color] at (axis cs:50.644,0.22969) {120};
    \node[anchor=south east, text=2080color] at (axis cs:61.624,0.26141) {140};
    \node[anchor=south, text=2080color] at (axis cs:67.953,0.27233) {160};
    \node[anchor=south west, xshift=-3pt, text=2080color] at (axis cs:71.982,0.26297) {180};
    \node[anchor=west, yshift=3pt, text=2080color] at (axis cs:74.793,0.25629) {200};
    \node[anchor=west, text=2080color] at (axis cs:76.131,0.24637) {220};
    \node[anchor=west, yshift=-3pt, text=2080color] at (axis cs:76.143,0.23094) {240};
    \node[anchor=north, yshift=-2pt, text=2080color] at (axis cs:74.887,0.23116) {260};
    \legend{BS 8, BS 16, BS 32, BS 64}
    \end{axis}
    \node at ($(mainplot.north)!0.5!(secondplot.north) + (0, 0.7cm)$) {\textbf{RTX 2080 Ti Power Capping}};
    \end{tikzpicture}%
    \caption{\textbf{RTX 2080 Ti power capping.} Curve annotations indicate power limits in watts. \emph{BS} stands for \emph{batch size}.}
    \label{fig:2080_power_capping}
    \Description{Two line graphs showing throughput in images per second on the X axis against efficiency in images per joule on the Y axis, for the RTX 2080 Ti GPU. The line graph on the left plots four lines for batch sizes 8, 16, 32, and 64 for code with all PyTorch optimizations enabled. The line graph on the right plots four lines for batch sizes 8, 16, 32, and 64 for code with no optimizations enabled. Values along each line indicate power limits in watts, ranging from 260 on the top right to 100 on the bottom left. Throughput ranges from 125 to 265 with optimizations, and from 32 to 76 without optimizations. Efficiency ranges from 0.47 to 0.98 with optimizations, and from 0.15 to 0.27 without optimizations. The 2080 Ti is most efficient at 140 watts for large batch sizes, with efficiency differences among batch sizes being more noticeable than for the RTX 5000 Ada and A6000. Without optimizations enabled, the card is most efficient at 160 watts.}
\end{figure}

\begin{figure}[ht]
    \centering    
        \begin{tikzpicture}
    \begin{axis}[
        yshift=-5pt,
        name=mainplot,
        title={\textbf{\emph{All Optimizations On}}},
        title style={yshift=-5pt},
        width={0.49\textwidth},
        height=4.8cm,
        enlarge x limits=0.2,
        enlarge y limits=0.2,
        xlabel={Throughput (img/s) $\rightarrow$},
        ylabel={Efficiency (img/J) $\rightarrow$},
        legend cell align=left,
        legend columns=1,
        legend style={
            /tikz/column 2/.style={
                column sep=0.2cm,
            },
            at={(0.01,0.99)},
            anchor=north west,
            nodes={scale=0.9}
        },
        point meta=explicit,
        mark options={line width=1pt},
        every node near coord/.style={anchor=south west, rotate=0},
    ]
    \addplot[mark=x, line width=0.5pt, 2070color!40] table [
        x=img_per_s,
        y=img_per_j,
        meta=power_limit_w,
        col sep=comma,
        discard if two={batch_size}{8}{compile}{True},
    ]{data/2070_super.csv};
    \addplot[mark=x, line width=0.5pt, 2070color!60] table [
        x=img_per_s,
        y=img_per_j,
        meta=power_limit_w,
        col sep=comma,
        discard if two={batch_size}{16}{compile}{True},
    ]{data/2070_super.csv};
    \addplot[mark=x, line width=0.5pt, 2070color!80] table [
        x=img_per_s,
        y=img_per_j,
        meta=power_limit_w,
        col sep=comma,
        discard if two={batch_size}{32}{compile}{True},
    ]{data/2070_super.csv};
    \addplot[mark=x, line width=0.5pt, 2070color] table [
        x=img_per_s,
        y=img_per_j,
        meta=power_limit_w,
        col sep=comma,
        discard if two={batch_size}{64}{compile}{True},
    ]{data/2070_super.csv};
    \node[anchor=south, yshift=1pt, text=2070color] at (axis cs:177.59,0.81625) {125};
    \node[anchor=south west, text=2070color] at (axis cs:180.87,0.80045) {135};
    \node[anchor=south west, text=2070color] at (axis cs:189.51,0.77055) {155};
    \node[anchor=south west, text=2070color] at (axis cs:195.39,0.74423) {175};
    \node[anchor=south west, text=2070color] at (axis cs:198.32,0.70109) {195};
    \node[anchor=south west, text=2070color] at (axis cs:199.84,0.65725) {215};
    \legend{BS 8, BS 16, BS 32, BS 64}
    \end{axis}

    \begin{axis}[
        at={(mainplot.north east)}, anchor=north west, xshift=50pt,
        title={\textbf{\emph{No Optimizations}}},
        title style={yshift=-5pt},
        name=secondplot,
        width={0.49\textwidth},
        height=4.8cm,
        enlarge x limits=0.2,
        ymin = 0.15,
        ymax=0.24,
        xlabel={Throughput (img/s) $\rightarrow$},
        ylabel={Efficiency (img/J) $\rightarrow$},
        legend cell align=left,
        legend columns=1,
        mark options={line width=1pt},
        legend style={
            /tikz/column 2/.style={
                column sep=0.2cm,
            },
            cells={align=left},
            at={(0.99,0.02)},
            anchor=south east,
            nodes={scale=0.9}
        },
        point meta=explicit,
        every node near coord/.style={anchor=south west, rotate=0},
    ]

    \addplot[mark=x, line width=0.5pt, 2070color!40] table [
        x=img_per_s,
        y=img_per_j,
        meta=power_limit_w,
        col sep=comma,
        discard if two={batch_size}{8}{compile}{False},
    ]{data/2070_super.csv};
    \addplot[mark=x, line width=0.5pt, 2070color!60] table [
        x=img_per_s,
        y=img_per_j,
        meta=power_limit_w,
        col sep=comma,
        discard if two={batch_size}{16}{compile}{False},
    ]{data/2070_super.csv};
    \addplot[mark=x, line width=0.5pt, 2070color!80] table [
        x=img_per_s,
        y=img_per_j,
        meta=power_limit_w,
        col sep=comma,
        discard if two={batch_size}{32}{compile}{False},
    ]{data/2070_super.csv};

    \node[anchor=south, yshift=1pt, text=2070color!80] at (axis cs:47.909,0.22454) {125};
    \node[anchor=south west, text=2070color!80] at (axis cs:49.123,0.2217) {135};
    \node[anchor=south west, text=2070color!80] at (axis cs:52.426,0.22139) {155};
    \node[anchor=south west, text=2070color!80] at (axis cs:54.397,0.21531) {175};
    \node[anchor=south west, text=2070color!80] at (axis cs:56.024,0.20598) {195};
    \node[anchor=south west, text=2070color!80] at (axis cs:57.005,0.19601) {215};
    \legend{BS 8, BS 16, BS 32}
    \end{axis}
    \node at ($(mainplot.north)!0.5!(secondplot.north) + (0, 0.7cm)$) {\textbf{RTX 2070 SUPER Power Capping}};
    \end{tikzpicture}%
    \caption{\textbf{RTX 2070 SUPER power capping.} Curve annotations indicate power limits in watts. \emph{BS} stands for \emph{batch size}.}
    \label{fig:2070super_power_capping}
    \Description{Two line graphs showing throughput in images per second on the X axis against efficiency in images per joule on the Y axis, for the RTX 2070 SUPER GPU. The line graph on the left plots four lines for batch sizes 8, 16, 32, and 64 for code with all PyTorch optimizations enabled. The line graph on the right plots three lines for batch sizes 8, 16, and 32 for code with no optimizations enabled. Values along each line indicate power limits in watts, ranging from 215 on the bottom right to 125 on the top left. Throughput ranges from 120 to 200 with optimizations, and from 42 to 57 without optimizations. Efficiency ranges from 0.57 to 0.82 with optimizations, and from 0.17 to 0.23 without optimizations. The 2070 SUPER is most efficient at minimum power (125 watts) for large batch sizes, with efficiency differences among batch sizes being more noticeable than for the RTX 5000 Ada and A6000.}
\end{figure}
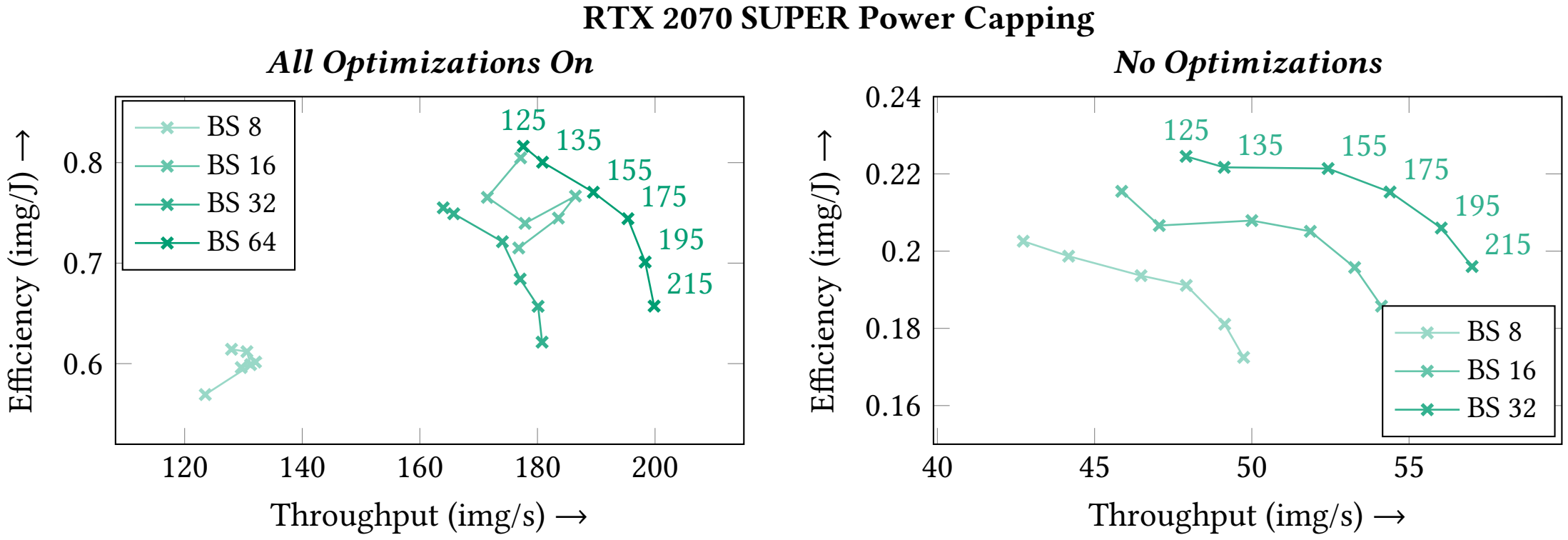

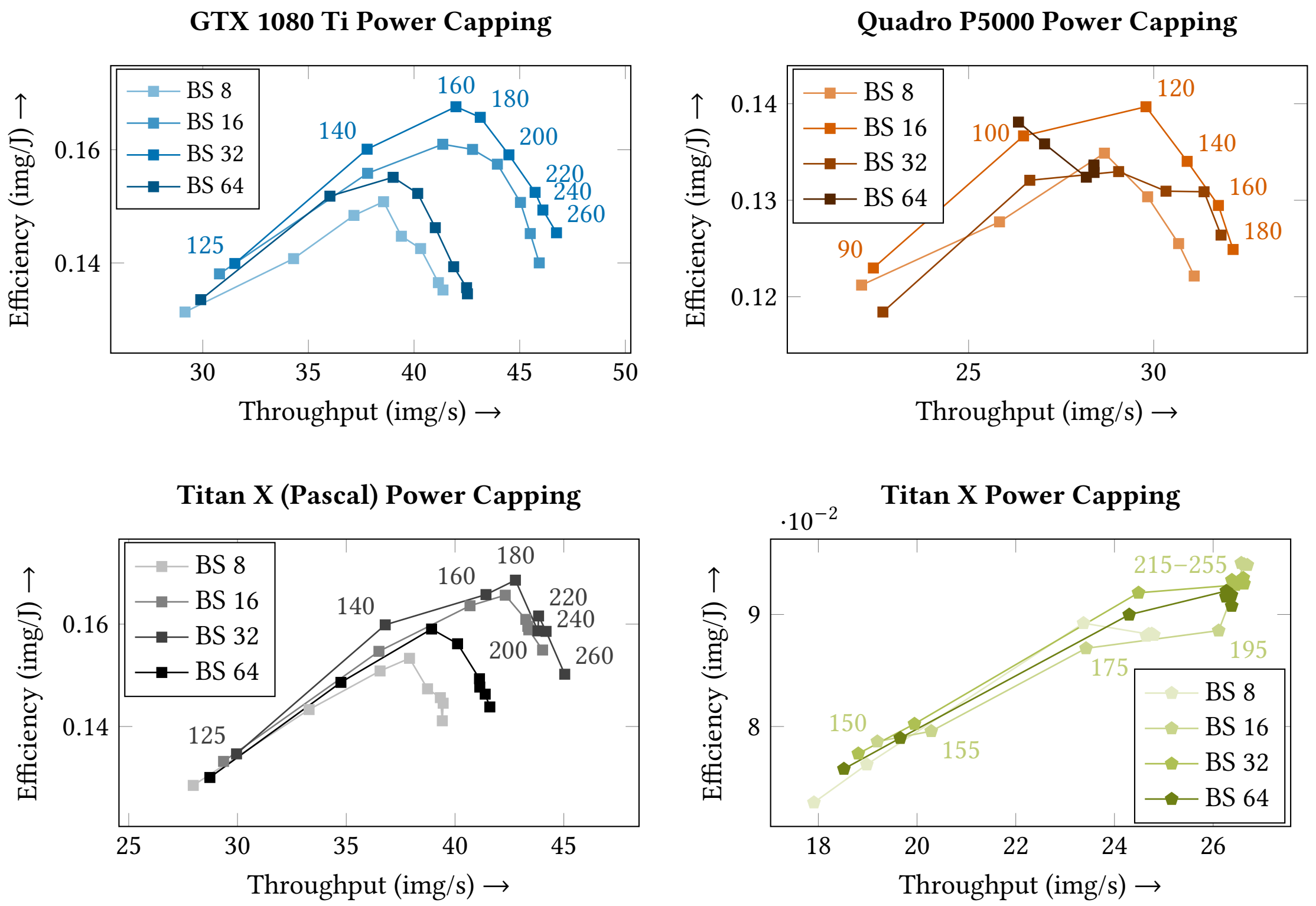
\begin{figure}[ht]
    \centering
    \begin{subfigure}[b]{0.49\textwidth}
        \centering
            \begin{tikzpicture}
    \begin{axis}[
        name=mainplot,
        title={\textbf{GTX 1080 Ti Power Capping}},
        width={\textwidth},
        height=4.8cm,
        xlabel={Throughput (img/s) $\rightarrow$},
        ylabel={Efficiency (img/J) $\rightarrow$},
        legend cell align=left,
        enlarge x limits = 0.2,
        enlarge y limits = 0.2,
        legend columns=1,
        legend style={
            /tikz/column 2/.style={
                column sep=0.2cm,
            },
            cells={align=left},
            at={(0.01,0.99)},
            anchor=north west,
            nodes={scale=0.9},
        },
        point meta=explicit,
        every node near coord/.style={anchor=south west, rotate=0},
    ]
    
    \addplot[mark=square*, mark size=1.5pt, line width=0.5pt, 1080color!50] table [
        x=img_per_s,
        y=img_per_j,
        meta=power_limit_w,
        col sep=comma,
        discard if not={batch_size}{8},
    ]{data/1080Ti.csv};
    \addplot[mark=square*, mark size=1.5pt, line width=0.5pt, 1080color!75] table [
        x=img_per_s,
        y=img_per_j,
        meta=power_limit_w,
        col sep=comma,
        discard if not={batch_size}{16},
    ]{data/1080Ti.csv};
    \addplot[mark=square*, mark size=1.5pt, line width=0.5pt, 1080color] table [
        x=img_per_s,
        y=img_per_j,
        meta=power_limit_w,
        col sep=comma,
        discard if not={batch_size}{32},
    ]{data/1080Ti.csv};
    \node[anchor=south east, text=1080color] at (axis cs:31.518,0.1399) {125};
    \node[anchor=south east, text=1080color] at (axis cs:37.779,0.1601) {140};
    \node[anchor=south, yshift=1pt, text=1080color] at (axis cs:41.983,0.1676) {160};
    \node[anchor=south west, text=1080color] at (axis cs:43.134,0.16574) {180};
    \node[anchor=south west, text=1080color] at (axis cs:44.49,0.1591) {200};
    \node[anchor=south west, text=1080color] at (axis cs:45.729,0.15247) {220};
    \node[anchor=south west, text=1080color] at (axis cs:46.095,0.14934) {240};
    \node[anchor=south west, text=1080color] at (axis cs:46.742,0.14534) {260};
    \addplot[mark=square*, mark size=1.5pt, line width=0.5pt, 1080color!75!black] table [
        x=img_per_s,
        y=img_per_j,
        meta=power_limit_w,
        col sep=comma,
        discard if not={batch_size}{64},
    ]{data/1080Ti.csv};
    \legend{BS 8, BS 16, BS 32, BS 64}
    \end{axis}
    \end{tikzpicture}%
        \end{subfigure}
        \begin{subfigure}[b]{0.49\textwidth}  
            \centering 
            \begin{tikzpicture}
\begin{axis}[
    title={\textbf{Quadro P5000 Power Capping}},
    width={\textwidth},
    height=4.8cm,
    xlabel={Throughput (img/s) $\rightarrow$},
    ylabel={Efficiency (img/J) $\rightarrow$},
    legend cell align=left,
    enlarge x limits = 0.2,
    enlarge y limits = 0.2,
    legend style={
        at={(0.01,0.99)},
        anchor=north west,
    },
    point meta=explicit,
    every node near coord/.style={anchor=south west, rotate=0},
]
\addplot[mark=square*, mark size=1.5pt, line width=0.5pt, P5000color!70] table [
    x=img_per_s,
    y=img_per_j,
    meta=power_limit_w,
    col sep=comma,
    discard if not={batch_size}{8},
]{data/P5000.csv};
\addplot[mark=square*, mark size=1.5pt, line width=0.5pt, P5000color] table [
    x=img_per_s,
    y=img_per_j,
    meta=power_limit_w,
    col sep=comma,
    discard if not={batch_size}{16},
]{data/P5000.csv};
\node[anchor=south east, text=P5000color] at (axis cs:22.428,0.12298) {90};
\node[anchor=east, xshift=-1pt, yshift=1pt, text=P5000color] at (axis cs:26.484,0.13667) {100};
\node[anchor=south west, text=P5000color] at (axis cs:29.791,0.13969) {120};
\node[anchor=south west, text=P5000color] at (axis cs:30.904,0.13403) {140};
\node[anchor=south west, text=P5000color] at (axis cs:31.748,0.12946) {160};
\node[anchor=south west, text=P5000color] at (axis cs:32.144,0.1249) {180};
\addplot[mark=square*, mark size=1.5pt, line width=0.5pt, P5000color!70!black] table [
    x=img_per_s,
    y=img_per_j,
    meta=power_limit_w,
    col sep=comma,
    discard if not={batch_size}{32},
]{data/P5000.csv};
\addplot[mark=square*, mark size=1.5pt, line width=0.5pt, P5000color!40!black] table [
    x=img_per_s,
    y=img_per_j,
    meta=power_limit_w,
    col sep=comma,
    discard if not={batch_size}{64},
]{data/P5000.csv};
\legend{BS 8, BS 16, BS 32, BS 64}
\end{axis}
\end{tikzpicture}%
        \end{subfigure}
        \vskip\baselineskip
        \begin{subfigure}[b]{0.49\textwidth}   
            \centering 
            \begin{tikzpicture}
\begin{axis}[
    title={\textbf{Titan X (Pascal) Power Capping}},
    width={\textwidth},
    height=4.8cm,
    xlabel={Throughput (img/s) $\rightarrow$},
    ylabel={Efficiency (img/J) $\rightarrow$},
    enlarge x limits = 0.2,
    enlarge y limits = 0.2,
    legend cell align=left,
    legend style={
    cells={align=left},
    at={(0.01,0.99)},
    anchor=north west,
    },
    point meta=explicit,
    every node near coord/.style={anchor=south west, rotate=0},
]
\addplot[mark=square*, mark size=1.5pt, line width=0.5pt, TitanXPcolor!25] table [
    x=img_per_s,
    y=img_per_j,
    meta=power_limit_w,
    col sep=comma,
    discard if not={batch_size}{8},
]{data/TitanXPascal.csv};
\addplot[mark=square*, mark size=1.5pt, line width=0.5pt, TitanXPcolor!50] table [
    x=img_per_s,
    y=img_per_j,
    meta=power_limit_w,
    col sep=comma,
    discard if not={batch_size}{16},
]{data/TitanXPascal.csv};
\addplot[mark=square*, mark size=1.5pt, line width=0.5pt, TitanXPcolor!75] table [
    x=img_per_s,
    y=img_per_j,
    meta=power_limit_w,
    col sep=comma,
    discard if not={batch_size}{32},
]{data/TitanXPascal.csv};
\addplot[mark=square*, mark size=1.5pt, line width=0.5pt, TitanXPcolor] table [
    x=img_per_s,
    y=img_per_j,
    meta=power_limit_w,
    col sep=comma,
    discard if not={batch_size}{64},
]{data/TitanXPascal.csv};
\node[anchor=south east, TitanXPcolor!75] at (axis cs:29.958,0.13465) {125};
\node[anchor=south east, TitanXPcolor!75] at (axis cs:36.792,0.15987) {140};
\node[anchor=south east, TitanXPcolor!75] at (axis cs:41.424,0.16577) {160};
\node[anchor=south, TitanXPcolor!75, yshift=2pt] at (axis cs:42.775,0.16859) {180};
\node[anchor=north east, TitanXPcolor!75] at (axis cs:43.812,0.15868) {200};
\node[anchor=south west, TitanXPcolor!75] at (axis cs:43.839,0.16158) {220};
\node[anchor=south west, yshift=-2pt, TitanXPcolor!75] at (axis cs:44.181,0.15857) {240};
\node[anchor=south west, TitanXPcolor!75] at (axis cs:45.042,0.15023) {260};
\legend{BS 8, BS 16, BS 32, BS 64}
\end{axis}
\end{tikzpicture}%
        \end{subfigure}
        \begin{subfigure}[b]{0.49\textwidth}   
            \centering 
            \begin{tikzpicture}
\begin{axis}[
    title={\textbf{Titan X Power Capping}},
    width={\textwidth},
    height=4.8cm,
    xlabel={Throughput (img/s) $\rightarrow$},
    ylabel={Efficiency (img/J) $\rightarrow$},
    legend cell align=left,
    legend style={
        at={(0.99,0.01)},
        anchor=south east,
    },
    point meta=explicit,
    every node near coord/.style={anchor=south east, rotate=0},
]
\addplot[mark=pentagon*, mark size=2pt, line width=0.5pt, TitanXcolor!25] table [
    x=img_per_s,
    y=img_per_j,
    meta=power_limit_w,
    col sep=comma,
    discard if not={batch_size}{8},
]{data/TitanX.csv};
\addplot[mark=pentagon*, mark size=2pt, line width=0.5pt, TitanXcolor!50] table [
    x=img_per_s,
    y=img_per_j,
    meta=power_limit_w,
    col sep=comma,
    discard if not={batch_size}{16},
]{data/TitanX.csv};
\addplot[mark=pentagon*, mark size=2pt, line width=0.5pt, TitanXcolor!75] table [
    x=img_per_s,
    y=img_per_j,
    meta=power_limit_w,
    col sep=comma,
    discard if not={batch_size}{32},
]{data/TitanX.csv};
\addplot[mark=pentagon*, mark size=2pt, line width=0.5pt, TitanXcolor!75!black] table [
    x=img_per_s,
    y=img_per_j,
    meta=power_limit_w,
    col sep=comma,
    discard if not={batch_size}{64},
]{data/TitanX.csv};
\node[anchor=south east, TitanXcolor!60] at (axis cs:19.196,0.078658) {150};
\node[anchor=north west, TitanXcolor!60] at (axis cs:20.288,0.079584) {155};
\node[anchor=north west, xshift=-2pt, TitanXcolor!60] at (axis cs:23.424,0.08698) {175};
\node[anchor=north west, TitanXcolor!60] at (axis cs:26.111,0.088562) {195};
\node[anchor=east, xshift=-2pt, TitanXcolor!60] at (axis cs:26.626,0.094483) {215--255};
\legend{BS 8, BS 16, BS 32, BS 64}
\end{axis}
\end{tikzpicture}%
        \end{subfigure}
        
    \caption{\textbf{Pascal and Maxwell GPU power capping.} Efficiency and throughput of the GTX 1080 Ti, Quadro P5000, Titan X (Pascal), and GTX Titan X power-capped at different batch sizes with \emph{no PyTorch optimizations} enabled. Curve annotations indicate power limits in watts. \emph{BS} stands for \emph{batch size}.}
    \label{fig:old_gpu_power_capping}
    \Description{Four line graphs showing throughput in images per second on the X axis against efficiency in images per joule on the Y axis for Pascal and Maxwell GPUs. Each graph shows batch sizes 8, 16, 32, and 64; labels along the curves indicate power limits in watts. The Titan X of the Maxwell generation is most efficient close to maximum power and efficiency shows no significant variation with batch size. The remaining three GPUs, of the Pascal generation (GTX 1080 Ti, Quadro P5000, and Titan X (Pascal)), are clearly more efficient for a specific batch size, always smaller than the maximum size that fits in GPU memory. The Pascal GPUs are more efficient at lower power limits, ranging between 120 and 180 watts, as opposed to over 200 watts for the Maxwell card.}
\end{figure}

\section{All GPU Power Capping Results} \label{sec:D_gpu_power_capping_all}

\Cref{fig:gpu_efficiency_bsize} shows the efficiency and throughput per batch size for all the GPUs in \cref{tab:gpu_inventory}, when power-capped to maximize efficiency. \Cref{fig:A5000_power_capping,fig:A6000_power_capping,fig:2070super_power_capping,fig:2080_power_capping,fig:old_gpu_power_capping} show the efficiency and throughput at every power limit for each batch size on each GPU.

For recent GPUs, the peak-efficiency batch size is typically the largest one that fits in memory, but this is not true for older GPUs. In the main text, we define the best batch size as the one that achieves highest throughput and efficiency, which usually aligns with the batch size that achieves peak efficiency. One notable exception is the RTX 2080 Ti, for which the \emph{best} batch size is 64 (\cref{fig:2080_power_capping}), even if batch size 16 technically achieves peak efficiency for this card.
\section{Environmental Assessment}\label{sec:E_environmental_assessment}

This appendix provides details on embodied impact modeling (\cref{sec:E1_embodied_impacts}) and complete results for our environmental assessment of upgrade and workload evolution scenarios presented in \cref{sec:hardware_replacement} (\cref{sec:E2_all_env_assessment_results}).

\subsection{Embodied Impacts}\label{sec:E1_embodied_impacts}

We source global warming potential and water use impacts for the production of graphics cards and servers from the Resilio database~\cite{ResilioDatabaseIT}. To estimate the abiotic depletion potential of GPUs, we use the calculator provided by the ADEME agency~\cite{lees-perasso2026lcagpu, HubbloGraphicsCard}. Due to a lack of detailed data, we assume a weight of 150 g for the casing and the default $53\%$ radiator weight for all cards. We fill in the remaining GPU specifications with data from TechPowerUp~\cite{TechPowerUp2026}. 

More precisely, we use as reference the Dell PowerEdge T640 model from ResilioDB~\cite{ResilioDatabaseIT} and modify it to fit the configuration of our default S1 server, outlined in \cref{tab:server_inventory}. We add a power supply of 1{,}400 W, two 2 TB NVMe M.2 SSDs (176-layer TLC), and one 4 TB HDD, reflecting the full configuration of our server. We do not directly model the Dell Precision 7920 as it is not currently available in the database. To model the server in Datavizta~\cite{Datavizta}, we configure a rack server with the following components: two 8-core Cascade Lake CPUs with a TDP of 85 W, four 16 GB SK Hynix RAM modules, two 2 TB Micron SSDs\footnote{The vendor of the SSDs in our server is SK Hynix, but Datavizta does not provide this option.}, one HDD, and one power supply. Datavizta only considers servers of type rack and blade, and not tower servers like ours. We choose the rack format arbitrarily, as this choice varies embodied abiotic depletion potential impacts by less than $1\%$. 

\begin{table}[ht]
    \centering
    \caption{\textbf{GPU specifications for abiotic depletion potential estimation.} We omit two values required by the calculator~\cite{HubbloGraphicsCard}: the number of VRAM dies and their surface, since finding reliable sources for these specifications proved challenging and they do not influence abiotic depletion impacts. We assume a weight of 150 g for the casing and the default $53\%$ radiator weight for all cards due to a lack of data.}
    \label{tab:gpu_calculator_values}
    \begin{tabular}{llrrr}
    \toprule
    Specification & Unit & RTX 2080 Ti & RTX A6000 & RTX 5000 Ada \\ \midrule
    Total weight & g & 1270.00 & 1179.00 & 1079.70 \\
    Heatsink weight & g & 673.10 & 624.87& 572.24 \\
    Casing weight & g & 150.00 & 150.00 & 150.00 \\
    Graphics card surface & cm$^2$ & 296.37 & 299.04 & 299.04 \\
    GPU chip surface & mm$^2$ & 754 & 628 & 609 \\
    VRAM & GB & 11 & 48 & 32 \\ \midrule
    \multicolumn{2}{l}{Sources} & \cite{NVIDIAGeForceRTX2026} & \cite{NVIDIARTXA60002026,mktNVIDIARTXA6000} & \cite{NVIDIARTX50002026,mktNVIDIARTX5000} \\
    \bottomrule
\end{tabular}

\end{table}

\subsection{All Environmental Assessment Results}\label{sec:E2_all_env_assessment_results}

\Cref{tab:constant_workload,tab:workload_exponential,tab:adjusted_workload} gather the full results of our environmental assessment for the hardware upgrade scenarios introduced in \cref{sec:hardware_replacement}, for the \emph{constant}, \emph{exponential}, and \emph{adjusted} workload evolution scenarios, respectively. Abiotic depletion potential is the same in all cases, so we only include it in \cref{tab:germany_impacts}.

\begin{figure}[htb]
    \centering
    \input{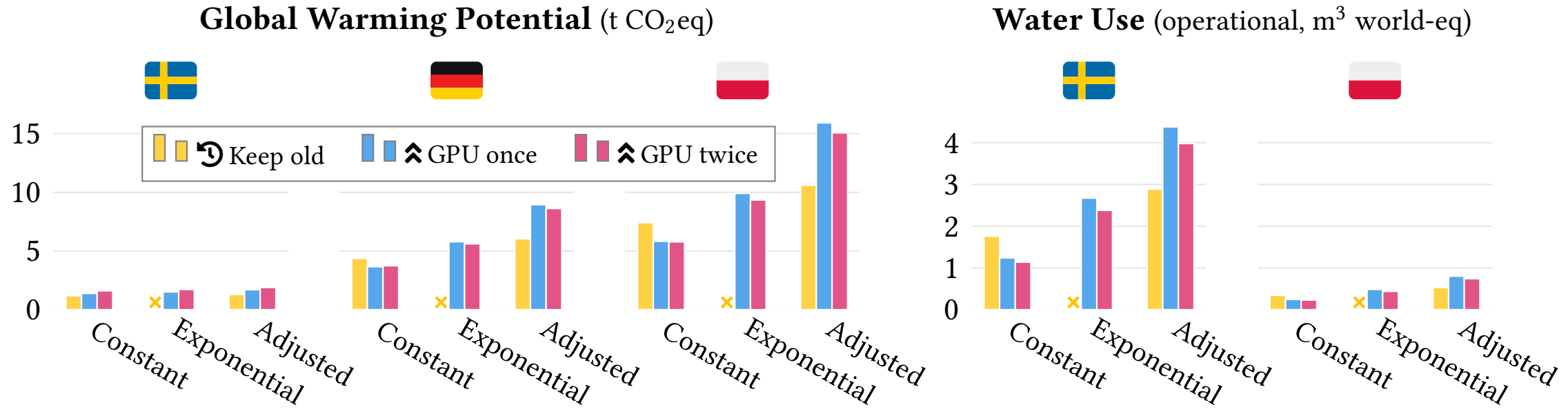}
\pgfplotstableread{
scenario poland germany sweden
constant              7.42141   4.36897   1.18937
exponential                 0         0         0
adjusted              10.6051   6.04111   1.28905
}\keepoldcarbon

\pgfplotstableread{
scenario poland germany sweden
constant             0.345541  0.224275   1.76189
exponential                 0         0         0
adjusted             0.527076  0.339794    2.8895
}\keepoldwater

\pgfplotstableread{
scenario poland germany sweden
constant              5.82886   3.65836   1.38651
exponential           9.91032   5.78815   1.51485
adjusted              15.9202   8.94699   1.69791
}\upgradeoncecarbon

\pgfplotstableread{
scenario poland germany sweden
constant             0.245955  0.159919   1.24156
exponential          0.484436  0.308003   2.67406
adjusted             0.802956  0.517783   4.38088
}\upgradeoncewater

\pgfplotstableread{
scenario poland germany sweden
constant              5.77923   3.74289   1.60439
exponential           9.34867   5.60991   1.71639
adjusted               15.072   8.62168    1.8907
}\upgradetwicecarbon

\pgfplotstableread{
scenario poland germany sweden
constant             0.230504  0.150228   1.13861
exponential          0.439372  0.280317   2.38089
adjusted             0.742026  0.479292   3.98205
}\upgradetwicewater

\begin{tikzpicture}
    \begin{axis}[gwp sweden panel, ymax=16]
        \addplot[fill=acces3!75] table[
            y expr={\thisrow{sweden}},
            x=scenario,
        ] {\keepoldcarbon};
        \addplot[fill=acces2!75] table[
            y expr={\thisrow{sweden}},
            x=scenario,
        ] {\upgradeoncecarbon};
        \addplot[fill=acces1!75] table[
            y expr={\thisrow{sweden}},
            x=scenario,
        ] {\upgradetwicecarbon};
        \addplot[infeasible scenario cross] coordinates {(exponential,0)};
    \end{axis}

    \begin{axis}[gwp germany panel, ymax=16]
        \addplot[fill=acces3!75] table[
            y expr={\thisrow{germany}},
            x=scenario,
        ] {\keepoldcarbon};
        \addplot[fill=acces2!75] table[
            y expr={\thisrow{germany}},
            x=scenario,
        ] {\upgradeoncecarbon};
        \addplot[fill=acces1!75] table[
            y expr={\thisrow{germany}},
            x=scenario,
        ] {\upgradetwicecarbon};
        \addplot[infeasible scenario cross] coordinates {(exponential,0)};
    \end{axis}

    \begin{axis}[gwp poland panel, ymax=16]
        \addplot[fill=acces3!75] table[
            y expr={\thisrow{poland}},
            x=scenario,
        ] {\keepoldcarbon};
        \addplot[fill=acces2!75] table[
            y expr={\thisrow{poland}},
            x=scenario,
        ] {\upgradeoncecarbon};
        \addplot[fill=acces1!75] table[
            y expr={\thisrow{poland}},
            x=scenario,
        ] {\upgradetwicecarbon};
        \addplot[infeasible scenario cross] coordinates {(exponential,0)};
        \carbonLegend
    \end{axis}
    \gwpHeading

    \begin{axis}[water sweden panel, ymax=4.5]
        \addplot[fill=acces3!75] table[
            y expr={\thisrow{sweden}},
            x=scenario,
        ] {\keepoldwater};
        \addplot[fill=acces2!75] table[
            y expr={\thisrow{sweden}},
            x=scenario,
        ] {\upgradeoncewater};
        \addplot[fill=acces1!75] table[
            y expr={\thisrow{sweden}},
            x=scenario,
        ] {\upgradetwicewater};
        \addplot[infeasible scenario cross] coordinates {(exponential,0)};
    \end{axis}

    \begin{axis}[water poland panel, ymax=4.5]
        \addplot[fill=acces3!75] table[
            y expr={\thisrow{poland}},
            x=scenario,
        ] {\keepoldwater};
        \addplot[fill=acces2!75] table[
            y expr={\thisrow{poland}},
            x=scenario,
        ] {\upgradeoncewater};
        \addplot[fill=acces1!75] table[
            y expr={\thisrow{poland}},
            x=scenario,
        ] {\upgradetwicewater};
        \addplot[infeasible scenario cross] coordinates {(exponential,0)};
    \end{axis}
    \waterHeading
    \trimCarbonPlotBoundingBox
\end{tikzpicture}
    \caption{\textbf{Global warming potential and operational water use across GPU upgrade scenarios, no GPU power capping.} Analogous to \cref{fig:carbon_all_scenarios}, but without capping GPU power consumption.}
    \label{fig:carbon_all_scenarios_max_power}
    \Description{Three bar plots on the left side under the title "Global Warming Potential (tonnes of carbon dioxide equivalent)", and two bar plots on the right side under the title "Water Use (operational, cubic metres world equivalent)". The three plots on the left show accumulated greenhouse gas emissions in Sweden, Germany and Poland for three workload evolution scenarios (constant, exponential, and adjusted) and three GPU upgrade scenarios (keep old, upgrade once, and upgrade twice). The Y axis ranges from 0 to 16. Emissions in Sweden stay under 2 in all scenarios. Emissions in Germany range between 3.7 and 8.9 tonnes of carbon dioxide equivalent. Emissions in Poland range between 5.8 and 15.9 tonnes of carbon dioxide equivalent. The two plots on the right show accumulated water use in Sweden and Poland for the same scenarios. The Y axis ranges from 0 to 4.5. Water use in Sweden ranges between 1 and 4.5. Water use in Poland stays under 0.81 in all scenarios. Figure analyzed in the text.}
\end{figure}

\Cref{fig:carbon_all_scenarios_max_power,tab:constant_workload_max_power,tab:workload_exponential_max_power,tab:adjusted_workload_max_power,fig:carbon_all_scenarios_max_power} simulate the same three workload evolution scenarios when not applying power capping, i.e., when the maximum allowed operating power of GPUs matches their thermal design power (TDP). Omitting power capping increases overall operational impacts by $6\%$ to $19\%$ and makes GPU upgrades more advantageous, but the main conclusions outlined in \cref{sec:hardware_replacement} still hold.

\begin{table}[ht]
    \centering
    \caption{\textbf{Constant workload between 2021 and 2026.} Environmental impacts of a constant daily Vision Transformer training workload sustained over six years on a local server. We estimate operational global warming potential and water use impacts using the average daily carbon intensity and water scarcity footprint in Poland, Germany, and south-central Sweden. \emph{Low}, \emph{Medium}, and \emph{High} utilization correspond to using the starting GPU $30\%$, $60\%$, and $90\%$ of the time, respectively. Upon upgrading hardware or software, daily utilization decreases since the workload remains constant.}
    \label{tab:constant_workload}
    \resizebox{\columnwidth}{!}{
    \begin{tabular}{clr@{ }l!{\hspace{6pt}}r@{ }l!{\hspace{6pt}}r@{ }l!{\hspace{6pt}}r@{ }l!{\hspace{6pt}}r@{ }l!{\hspace{6pt}}r@{ }l!{\hspace{4pt}}}
\multicolumn{14}{c}{\textbf{Global Warming Potential} (kg CO$_2$eq)} \\ \toprule
\multirow{3}[2]{*}{Country} &
\multirow{3}[2]{*}{\makecell[c]{Initial\\Util.}} & 
\multicolumn{12}{c}{Upgrade Scenario} \\ \cmidrule(lr){3-14}
 &  & \multicolumn{2}{c}{\makebox[0pt][c]{\hspace*{-3pt}\makecell[c]{\faHistory\hspace{2pt}Keep\\old GPU}\hspace*{3pt}}} & \multicolumn{2}{c}{\makebox[0pt][c]{\hspace*{-3pt}\makecell[c]{\faHistory\hspace{2pt}Keep old\\PyTorch}\hspace*{3pt}}} & \multicolumn{2}{c}{\makebox[0pt][c]{\hspace*{-3pt}\makecell[c]{\faPowerOff\hspace{2pt}Keep\\server on}\hspace*{3pt}}} & \multicolumn{2}{c}{\makebox[0pt][c]{\hspace*{-3pt}\makecell[c]{\faAngleDoubleUp\hspace{2pt}Upgrade\\GPU once}\hspace*{3pt}}} & \multicolumn{2}{c}{\makebox[0pt][c]{\hspace*{-3pt}\makecell[c]{\faAngleDoubleUp\hspace{2pt}Upgrade\\GPU twice}\hspace*{3pt}}} & \multicolumn{2}{c}{\makebox[0pt][c]{\hspace*{-2pt}\makecell[c]{\faAngleDoubleUp\hspace{2pt}Upgrade\\Server+GPU}\hspace*{2pt}}} \\
\midrule
\multirow{3}{*}{\twemoji[height=0.45cm]{flag: Poland}} & Low & 2,828 & \textcolor{gray}{\small{(ref)}} & 3,690 & \textcolor{Red}{\small{$+30\%$}} & 5,223 & \textcolor{Red}{\small{$+85\%$}} & 2,599 & \textcolor{ForestGreen}{\small{$-8\%$}} & 2,749 & \textcolor{ForestGreen}{\small{$-3\%$}} & 5,109 & \textcolor{Red}{\small{$+81\%$}} \\
 & Med. & 4,654 & \textcolor{gray}{\small{(ref)}} & 6,378 & \textcolor{Red}{\small{$+37\%$}} & 6,440 & \textcolor{Red}{\small{$+38\%$}} & 3,942 & \textcolor{ForestGreen}{\small{$-15\%$}} & 4,016 & \textcolor{ForestGreen}{\small{$-14\%$}} & 6,376 & \textcolor{Red}{\small{$+37\%$}} \\
 & High & 6,480 & \textcolor{gray}{\small{(ref)}} & 9,066 & \textcolor{Red}{\small{$+40\%$}} & 7,658 & \textcolor{Red}{\small{$+18\%$}} & 5,286 & \textcolor{ForestGreen}{\small{$-18\%$}} & 5,284 & \textcolor{ForestGreen}{\small{$-18\%$}} & 7,644 & \textcolor{Red}{\small{$+18\%$}} \\
\cmidrule(lr){1-14}
\multirow{3}{*}{\twemoji[height=0.45cm]{flag: Germany}} & Low & 1,956 & \textcolor{gray}{\small{(ref)}} & 2,409 & \textcolor{Red}{\small{$+23\%$}} & 3,213 & \textcolor{Red}{\small{$+64\%$}} & 1,958 & \textcolor{gray}{\small{$+0\%$}} & 2,146 & \textcolor{Red}{\small{$+10\%$}} & 4,506 & \textcolor{Red}{\small{$+130\%$}} \\
 & Med. & 2,914 & \textcolor{gray}{\small{(ref)}} & 3,820 & \textcolor{Red}{\small{$+31\%$}} & 3,852 & \textcolor{Red}{\small{$+32\%$}} & 2,666 & \textcolor{ForestGreen}{\small{$-9\%$}} & 2,814 & \textcolor{ForestGreen}{\small{$-3\%$}} & 5,174 & \textcolor{Red}{\small{$+78\%$}} \\
 & High & 3,872 & \textcolor{gray}{\small{(ref)}} & 5,232 & \textcolor{Red}{\small{$+35\%$}} & 4,490 & \textcolor{Red}{\small{$+16\%$}} & 3,373 & \textcolor{ForestGreen}{\small{$-13\%$}} & 3,482 & \textcolor{ForestGreen}{\small{$-10\%$}} & 5,842 & \textcolor{Red}{\small{$+51\%$}} \\
\cmidrule(lr){1-14}
\multirow{3}{*}{\twemoji[height=0.45cm]{flag: Sweden}} & Low & 1,049 & \textcolor{gray}{\small{(ref)}} & 1,076 & \textcolor{Red}{\small{$+3\%$}} & 1,123 & \textcolor{Red}{\small{$+7\%$}} & 1,288 & \textcolor{Red}{\small{$+23\%$}} & 1,513 & \textcolor{Red}{\small{$+44\%$}} & 3,873 & \textcolor{Red}{\small{$+269\%$}} \\
 & Med. & 1,105 & \textcolor{gray}{\small{(ref)}} & 1,159 & \textcolor{Red}{\small{$+5\%$}} & 1,160 & \textcolor{Red}{\small{$+5\%$}} & 1,329 & \textcolor{Red}{\small{$+20\%$}} & 1,551 & \textcolor{Red}{\small{$+40\%$}} & 3,911 & \textcolor{Red}{\small{$+254\%$}} \\
 & High & 1,161 & \textcolor{gray}{\small{(ref)}} & 1,241 & \textcolor{Red}{\small{$+7\%$}} & 1,197 & \textcolor{Red}{\small{$+3\%$}} & 1,370 & \textcolor{Red}{\small{$+18\%$}} & 1,589 & \textcolor{Red}{\small{$+37\%$}} & 3,949 & \textcolor{Red}{\small{$+240\%$}} \\
\midrule
\multicolumn{2}{l}{Of which embodied:} & 994 & \textcolor{gray}{\small{(ref)}} & 994 & \textcolor{gray}{\small{$+0\%$}} & 994 & \textcolor{gray}{\small{$+0\%$}} & 1,247 & \textcolor{Red}{\small{$+26\%$}} & 1,474 & \textcolor{Red}{\small{$+48\%$}} & 3,834 & \textcolor{Red}{\small{$+286\%$}}\\
\bottomrule
\end{tabular}

    }%
    \vspace{1em}
    \resizebox{\columnwidth}{!}{
    \begin{tabular}{clr@{ }l!{\hspace{6pt}}r@{ }l!{\hspace{6pt}}r@{ }l!{\hspace{6pt}}r@{ }l!{\hspace{6pt}}r@{ }l!{\hspace{6pt}}r@{ }l!{\hspace{4pt}}}
\multicolumn{14}{c}{\textbf{Water Use} (dm$^3$ world-eq)} \\ \toprule
\multirow{3}[2]{*}{Country} &
\multirow{3}[2]{*}{\makecell[c]{Initial\\Util.}} & 
\multicolumn{12}{c}{Upgrade Scenario} \\ \cmidrule(lr){3-14}
 &  & \multicolumn{2}{c}{\makebox[0pt][c]{\hspace*{-3pt}\makecell[c]{\faHistory\hspace{2pt}Keep\\old GPU}\hspace*{3pt}}} & \multicolumn{2}{c}{\makebox[0pt][c]{\hspace*{-3pt}\makecell[c]{\faHistory\hspace{2pt}Keep old\\PyTorch}\hspace*{3pt}}} & \multicolumn{2}{c}{\makebox[0pt][c]{\hspace*{-3pt}\makecell[c]{\faPowerOff\hspace{2pt}Keep\\server on}\hspace*{3pt}}} & \multicolumn{2}{c}{\makebox[0pt][c]{\hspace*{-3pt}\makecell[c]{\faAngleDoubleUp\hspace{2pt}Upgrade\\GPU once}\hspace*{3pt}}} & \multicolumn{2}{c}{\makebox[0pt][c]{\hspace*{-3pt}\makecell[c]{\faAngleDoubleUp\hspace{2pt}Upgrade\\GPU twice}\hspace*{3pt}}} & \multicolumn{2}{c}{\makebox[0pt][c]{\hspace*{-2pt}\makecell[c]{\faAngleDoubleUp\hspace{2pt}Upgrade\\Server+GPU}\hspace*{2pt}}} \\
\midrule
\multirow{3}{*}{\twemoji[height=0.45cm]{flag: Poland}} & Low & 98 & \textcolor{gray}{\small{(ref)}} & 147 & \textcolor{Red}{\small{$+50\%$}} & 230 & \textcolor{Red}{\small{$+134\%$}} & 73 & \textcolor{ForestGreen}{\small{$-26\%$}} & 68 & \textcolor{ForestGreen}{\small{$-30\%$}} & 68 & \textcolor{ForestGreen}{\small{$-30\%$}} \\
 & Med. & 196 & \textcolor{gray}{\small{(ref)}} & 294 & \textcolor{Red}{\small{$+50\%$}} & 295 & \textcolor{Red}{\small{$+50\%$}} & 145 & \textcolor{ForestGreen}{\small{$-26\%$}} & 136 & \textcolor{ForestGreen}{\small{$-30\%$}} & 136 & \textcolor{ForestGreen}{\small{$-30\%$}} \\
 & High & 294 & \textcolor{gray}{\small{(ref)}} & 441 & \textcolor{Red}{\small{$+50\%$}} & 360 & \textcolor{Red}{\small{$+23\%$}} & 217 & \textcolor{ForestGreen}{\small{$-26\%$}} & 204 & \textcolor{ForestGreen}{\small{$-31\%$}} & 204 & \textcolor{ForestGreen}{\small{$-31\%$}} \\
\cmidrule(lr){1-14}
\multirow{3}{*}{\twemoji[height=0.45cm]{flag: Germany}} & Low & 64 & \textcolor{gray}{\small{(ref)}} & 95 & \textcolor{Red}{\small{$+49\%$}} & 149 & \textcolor{Red}{\small{$+133\%$}} & 47 & \textcolor{ForestGreen}{\small{$-26\%$}} & 45 & \textcolor{ForestGreen}{\small{$-30\%$}} & 45 & \textcolor{ForestGreen}{\small{$-30\%$}} \\
 & Med. & 127 & \textcolor{gray}{\small{(ref)}} & 190 & \textcolor{Red}{\small{$+49\%$}} & 191 & \textcolor{Red}{\small{$+50\%$}} & 94 & \textcolor{ForestGreen}{\small{$-26\%$}} & 89 & \textcolor{ForestGreen}{\small{$-30\%$}} & 89 & \textcolor{ForestGreen}{\small{$-30\%$}} \\
 & High & 191 & \textcolor{gray}{\small{(ref)}} & 285 & \textcolor{Red}{\small{$+49\%$}} & 233 & \textcolor{Red}{\small{$+22\%$}} & 141 & \textcolor{ForestGreen}{\small{$-26\%$}} & 133 & \textcolor{ForestGreen}{\small{$-30\%$}} & 133 & \textcolor{ForestGreen}{\small{$-30\%$}} \\
\cmidrule(lr){1-14}
\multirow{3}{*}{\twemoji[height=0.45cm]{flag: Sweden}} & Low & 497 & \textcolor{gray}{\small{(ref)}} & 801 & \textcolor{Red}{\small{$+61\%$}} & 1,225 & \textcolor{Red}{\small{$+147\%$}} & 368 & \textcolor{ForestGreen}{\small{$-26\%$}} & 339 & \textcolor{ForestGreen}{\small{$-32\%$}} & 339 & \textcolor{ForestGreen}{\small{$-32\%$}} \\
 & Med. & 991 & \textcolor{gray}{\small{(ref)}} & 1,600 & \textcolor{Red}{\small{$+61\%$}} & 1,554 & \textcolor{Red}{\small{$+57\%$}} & 733 & \textcolor{ForestGreen}{\small{$-26\%$}} & 677 & \textcolor{ForestGreen}{\small{$-32\%$}} & 677 & \textcolor{ForestGreen}{\small{$-32\%$}} \\
 & High & 1,485 & \textcolor{gray}{\small{(ref)}} & 2,398 & \textcolor{Red}{\small{$+61\%$}} & 1,884 & \textcolor{Red}{\small{$+27\%$}} & 1,099 & \textcolor{ForestGreen}{\small{$-26\%$}} & 1,014 & \textcolor{ForestGreen}{\small{$-32\%$}} & 1,014 & \textcolor{ForestGreen}{\small{$-32\%$}} \\
\midrule
\multicolumn{2}{l}{+ Embodied:} & 3.3e5 & \textcolor{gray}{\small{(ref)}} & 3.3e5 & \textcolor{gray}{\small{$+0\%$}} & 3.3e5 & \textcolor{gray}{\small{$+0\%$}} & 4.0e5 & \textcolor{Red}{\small{$+21\%$}} & 4.5e5 & \textcolor{Red}{\small{$+38\%$}} & 1.2e6 & \textcolor{Red}{\small{$+279\%$}}\\
\bottomrule
\end{tabular}

    }%
\end{table}

\begin{table}[ht]
    \centering
    \caption{\textbf{Constant workload between 2021 and 2026, no GPU power capping.} Analogous to \cref{tab:constant_workload}, but without capping GPU power consumption.}
    \label{tab:constant_workload_max_power}
    \resizebox{\columnwidth}{!}{
    \begin{tabular}{clr@{ }l!{\hspace{6pt}}r@{ }l!{\hspace{6pt}}r@{ }l!{\hspace{6pt}}r@{ }l!{\hspace{6pt}}r@{ }l!{\hspace{6pt}}r@{ }l!{\hspace{4pt}}}
\multicolumn{14}{c}{\textbf{Global Warming Potential} (kg CO$_2$eq)} \\ \toprule
\multirow{3}[2]{*}{Country} &
\multirow{3}[2]{*}{\makecell[c]{Initial\\Util.}} & 
\multicolumn{12}{c}{Upgrade Scenario} \\ \cmidrule(lr){3-14}
 &  & \multicolumn{2}{c}{\makebox[0pt][c]{\hspace*{-3pt}\makecell[c]{\faHistory\hspace{2pt}Keep\\old GPU}\hspace*{3pt}}} & \multicolumn{2}{c}{\makebox[0pt][c]{\hspace*{-3pt}\makecell[c]{\faHistory\hspace{2pt}Keep old\\PyTorch}\hspace*{3pt}}} & \multicolumn{2}{c}{\makebox[0pt][c]{\hspace*{-3pt}\makecell[c]{\faPowerOff\hspace{2pt}Keep\\server on}\hspace*{3pt}}} & \multicolumn{2}{c}{\makebox[0pt][c]{\hspace*{-3pt}\makecell[c]{\faAngleDoubleUp\hspace{2pt}Upgrade\\GPU once}\hspace*{3pt}}} & \multicolumn{2}{c}{\makebox[0pt][c]{\hspace*{-3pt}\makecell[c]{\faAngleDoubleUp\hspace{2pt}Upgrade\\GPU twice}\hspace*{3pt}}} & \multicolumn{2}{c}{\makebox[0pt][c]{\hspace*{-2pt}\makecell[c]{\faAngleDoubleUp\hspace{2pt}Upgrade\\Server+GPU}\hspace*{2pt}}} \\
\midrule
\multirow{3}{*}{\twemoji[height=0.45cm]{flag: Poland}} & Low & 3,141 & \textcolor{gray}{\small{(ref)}} & 3,988 & \textcolor{Red}{\small{$+27\%$}} & 5,538 & \textcolor{Red}{\small{$+76\%$}} & 2,779 & \textcolor{ForestGreen}{\small{$-12\%$}} & 2,914 & \textcolor{ForestGreen}{\small{$-7\%$}} & 5,274 & \textcolor{Red}{\small{$+68\%$}} \\
 & Med. & 5,281 & \textcolor{gray}{\small{(ref)}} & 6,975 & \textcolor{Red}{\small{$+32\%$}} & 7,071 & \textcolor{Red}{\small{$+34\%$}} & 4,304 & \textcolor{ForestGreen}{\small{$-19\%$}} & 4,347 & \textcolor{ForestGreen}{\small{$-18\%$}} & 6,707 & \textcolor{Red}{\small{$+27\%$}} \\
 & High & 7,421 & \textcolor{gray}{\small{(ref)}} & 9,962 & \textcolor{Red}{\small{$+34\%$}} & 8,604 & \textcolor{Red}{\small{$+16\%$}} & 5,829 & \textcolor{ForestGreen}{\small{$-21\%$}} & 5,779 & \textcolor{ForestGreen}{\small{$-22\%$}} & 8,139 & \textcolor{Red}{\small{$+10\%$}} \\
\cmidrule(lr){1-14}
\multirow{3}{*}{\twemoji[height=0.45cm]{flag: Germany}} & Low & 2,121 & \textcolor{gray}{\small{(ref)}} & 2,566 & \textcolor{Red}{\small{$+21\%$}} & 3,380 & \textcolor{Red}{\small{$+59\%$}} & 2,054 & \textcolor{ForestGreen}{\small{$-3\%$}} & 2,233 & \textcolor{Red}{\small{$+5\%$}} & 4,593 & \textcolor{Red}{\small{$+117\%$}} \\
 & Med. & 3,245 & \textcolor{gray}{\small{(ref)}} & 4,134 & \textcolor{Red}{\small{$+27\%$}} & 4,185 & \textcolor{Red}{\small{$+29\%$}} & 2,856 & \textcolor{ForestGreen}{\small{$-12\%$}} & 2,988 & \textcolor{ForestGreen}{\small{$-8\%$}} & 5,348 & \textcolor{Red}{\small{$+65\%$}} \\
 & High & 4,369 & \textcolor{gray}{\small{(ref)}} & 5,702 & \textcolor{Red}{\small{$+31\%$}} & 4,990 & \textcolor{Red}{\small{$+14\%$}} & 3,658 & \textcolor{ForestGreen}{\small{$-16\%$}} & 3,743 & \textcolor{ForestGreen}{\small{$-14\%$}} & 6,103 & \textcolor{Red}{\small{$+40\%$}} \\
\cmidrule(lr){1-14}
\multirow{3}{*}{\twemoji[height=0.45cm]{flag: Sweden}} & Low & 1,059 & \textcolor{gray}{\small{(ref)}} & 1,086 & \textcolor{Red}{\small{$+3\%$}} & 1,133 & \textcolor{Red}{\small{$+7\%$}} & 1,294 & \textcolor{Red}{\small{$+22\%$}} & 1,518 & \textcolor{Red}{\small{$+43\%$}} & 3,878 & \textcolor{Red}{\small{$+266\%$}} \\
 & Med. & 1,124 & \textcolor{gray}{\small{(ref)}} & 1,177 & \textcolor{Red}{\small{$+5\%$}} & 1,180 & \textcolor{Red}{\small{$+5\%$}} & 1,340 & \textcolor{Red}{\small{$+19\%$}} & 1,561 & \textcolor{Red}{\small{$+39\%$}} & 3,921 & \textcolor{Red}{\small{$+249\%$}} \\
 & High & 1,189 & \textcolor{gray}{\small{(ref)}} & 1,269 & \textcolor{Red}{\small{$+7\%$}} & 1,226 & \textcolor{Red}{\small{$+3\%$}} & 1,387 & \textcolor{Red}{\small{$+17\%$}} & 1,604 & \textcolor{Red}{\small{$+35\%$}} & 3,964 & \textcolor{Red}{\small{$+233\%$}} \\
\midrule
\multicolumn{2}{l}{Of which embodied:} & 994 & \textcolor{gray}{\small{(ref)}} & 994 & \textcolor{gray}{\small{$+0\%$}} & 994 & \textcolor{gray}{\small{$+0\%$}} & 1,247 & \textcolor{Red}{\small{$+26\%$}} & 1,474 & \textcolor{Red}{\small{$+48\%$}} & 3,834 & \textcolor{Red}{\small{$+286\%$}}\\
\bottomrule
\end{tabular}

    }%
    \vspace{1em}
    \resizebox{\columnwidth}{!}{
    \begin{tabular}{clr@{ }l!{\hspace{6pt}}r@{ }l!{\hspace{6pt}}r@{ }l!{\hspace{6pt}}r@{ }l!{\hspace{6pt}}r@{ }l!{\hspace{6pt}}r@{ }l!{\hspace{4pt}}}
\multicolumn{14}{c}{\textbf{Water Use} (dm$^3$ world-eq)} \\ \toprule
\multirow{3}[2]{*}{Country} &
\multirow{3}[2]{*}{\makecell[c]{Initial\\Util.}} & 
\multicolumn{12}{c}{Upgrade Scenario} \\ \cmidrule(lr){3-14}
 &  & \multicolumn{2}{c}{\makebox[0pt][c]{\hspace*{-3pt}\makecell[c]{\faHistory\hspace{2pt}Keep\\old GPU}\hspace*{3pt}}} & \multicolumn{2}{c}{\makebox[0pt][c]{\hspace*{-3pt}\makecell[c]{\faHistory\hspace{2pt}Keep old\\PyTorch}\hspace*{3pt}}} & \multicolumn{2}{c}{\makebox[0pt][c]{\hspace*{-3pt}\makecell[c]{\faPowerOff\hspace{2pt}Keep\\server on}\hspace*{3pt}}} & \multicolumn{2}{c}{\makebox[0pt][c]{\hspace*{-3pt}\makecell[c]{\faAngleDoubleUp\hspace{2pt}Upgrade\\GPU once}\hspace*{3pt}}} & \multicolumn{2}{c}{\makebox[0pt][c]{\hspace*{-3pt}\makecell[c]{\faAngleDoubleUp\hspace{2pt}Upgrade\\GPU twice}\hspace*{3pt}}} & \multicolumn{2}{c}{\makebox[0pt][c]{\hspace*{-2pt}\makecell[c]{\faAngleDoubleUp\hspace{2pt}Upgrade\\Server+GPU}\hspace*{2pt}}} \\
\midrule
\multirow{3}{*}{\twemoji[height=0.45cm]{flag: Poland}} & Low & 115 & \textcolor{gray}{\small{(ref)}} & 164 & \textcolor{Red}{\small{$+42\%$}} & 247 & \textcolor{Red}{\small{$+114\%$}} & 82 & \textcolor{ForestGreen}{\small{$-29\%$}} & 77 & \textcolor{ForestGreen}{\small{$-33\%$}} & 77 & \textcolor{ForestGreen}{\small{$-33\%$}} \\
 & Med. & 230 & \textcolor{gray}{\small{(ref)}} & 327 & \textcolor{Red}{\small{$+42\%$}} & 330 & \textcolor{Red}{\small{$+43\%$}} & 164 & \textcolor{ForestGreen}{\small{$-29\%$}} & 154 & \textcolor{ForestGreen}{\small{$-33\%$}} & 154 & \textcolor{ForestGreen}{\small{$-33\%$}} \\
 & High & 346 & \textcolor{gray}{\small{(ref)}} & 490 & \textcolor{Red}{\small{$+42\%$}} & 412 & \textcolor{Red}{\small{$+19\%$}} & 246 & \textcolor{ForestGreen}{\small{$-29\%$}} & 231 & \textcolor{ForestGreen}{\small{$-33\%$}} & 231 & \textcolor{ForestGreen}{\small{$-33\%$}} \\
\cmidrule(lr){1-14}
\multirow{3}{*}{\twemoji[height=0.45cm]{flag: Germany}} & Low & 75 & \textcolor{gray}{\small{(ref)}} & 106 & \textcolor{Red}{\small{$+41\%$}} & 160 & \textcolor{Red}{\small{$+113\%$}} & 53 & \textcolor{ForestGreen}{\small{$-29\%$}} & 50 & \textcolor{ForestGreen}{\small{$-33\%$}} & 50 & \textcolor{ForestGreen}{\small{$-33\%$}} \\
 & Med. & 150 & \textcolor{gray}{\small{(ref)}} & 211 & \textcolor{Red}{\small{$+41\%$}} & 213 & \textcolor{Red}{\small{$+43\%$}} & 107 & \textcolor{ForestGreen}{\small{$-29\%$}} & 100 & \textcolor{ForestGreen}{\small{$-33\%$}} & 100 & \textcolor{ForestGreen}{\small{$-33\%$}} \\
 & High & 224 & \textcolor{gray}{\small{(ref)}} & 316 & \textcolor{Red}{\small{$+41\%$}} & 267 & \textcolor{Red}{\small{$+19\%$}} & 160 & \textcolor{ForestGreen}{\small{$-29\%$}} & 150 & \textcolor{ForestGreen}{\small{$-33\%$}} & 150 & \textcolor{ForestGreen}{\small{$-33\%$}} \\
\cmidrule(lr){1-14}
\multirow{3}{*}{\twemoji[height=0.45cm]{flag: Sweden}} & Low & 589 & \textcolor{gray}{\small{(ref)}} & 890 & \textcolor{Red}{\small{$+51\%$}} & 1,318 & \textcolor{Red}{\small{$+124\%$}} & 415 & \textcolor{ForestGreen}{\small{$-29\%$}} & 381 & \textcolor{ForestGreen}{\small{$-35\%$}} & 381 & \textcolor{ForestGreen}{\small{$-35\%$}} \\
 & Med. & 1,175 & \textcolor{gray}{\small{(ref)}} & 1,777 & \textcolor{Red}{\small{$+51\%$}} & 1,740 & \textcolor{Red}{\small{$+48\%$}} & 828 & \textcolor{ForestGreen}{\small{$-30\%$}} & 760 & \textcolor{ForestGreen}{\small{$-35\%$}} & 760 & \textcolor{ForestGreen}{\small{$-35\%$}} \\
 & High & 1,762 & \textcolor{gray}{\small{(ref)}} & 2,664 & \textcolor{Red}{\small{$+51\%$}} & 2,163 & \textcolor{Red}{\small{$+23\%$}} & 1,242 & \textcolor{ForestGreen}{\small{$-30\%$}} & 1,139 & \textcolor{ForestGreen}{\small{$-35\%$}} & 1,139 & \textcolor{ForestGreen}{\small{$-35\%$}} \\
\midrule
\multicolumn{2}{l}{+ Embodied:} & 3.3e5 & \textcolor{gray}{\small{(ref)}} & 3.3e5 & \textcolor{gray}{\small{$+0\%$}} & 3.3e5 & \textcolor{gray}{\small{$+0\%$}} & 4.0e5 & \textcolor{Red}{\small{$+21\%$}} & 4.5e5 & \textcolor{Red}{\small{$+38\%$}} & 1.2e6 & \textcolor{Red}{\small{$+279\%$}}\\
\bottomrule
\end{tabular}

    }%
\end{table}

\begin{table}[ht]
    \centering
    \caption{\textbf{Exponential workload between 2021 and 2026.} Environmental impacts of an exponentially growing daily Vision Transformer training workload over six years on a local server. We estimate operational global warming potential and water use impacts using the average daily carbon intensity and water scarcity footprint in Poland, Germany, and south-central Sweden. \emph{Low}, \emph{Medium}, and \emph{High} utilization correspond to using the starting GPU $30\%$, $60\%$, and $90\%$ of the time, respectively. The daily workload increases every day, doubling every three years.}
    \label{tab:workload_exponential}
    \resizebox{0.9\columnwidth}{!}{
    \begin{tabular}{clr@{ }l!{\hspace{6pt}}r@{ }l!{\hspace{6pt}}r@{ }l!{\hspace{6pt}}r@{ }l!{\hspace{6pt}}r@{ }l!{\hspace{4pt}}}
\multicolumn{12}{c}{\textbf{Global Warming Potential} (kg CO$_2$eq)} \\ \toprule
\multirow{3}[2]{*}{Country} &
\multirow{3}[2]{*}{\makecell[c]{Initial\\Util.}} & 
\multicolumn{10}{c}{Upgrade Scenario} \\ \cmidrule(lr){3-12}
 &  & \multicolumn{2}{c}{\makebox[0pt][c]{\hspace*{-3pt}\makecell[c]{\faHistory\hspace{2pt}Keep\\old GPU}\hspace*{3pt}}} & \multicolumn{2}{c}{\makebox[0pt][c]{\hspace*{-3pt}\makecell[c]{\faPowerOff\hspace{2pt}Keep\\server on}\hspace*{3pt}}} & \multicolumn{2}{c}{\makebox[0pt][c]{\hspace*{-3pt}\makecell[c]{\faAngleDoubleUp\hspace{2pt}Upgrade\\GPU once}\hspace*{3pt}}} & \multicolumn{2}{c}{\makebox[0pt][c]{\hspace*{-3pt}\makecell[c]{\faAngleDoubleUp\hspace{2pt}Upgrade\\GPU twice}\hspace*{3pt}}} & \multicolumn{2}{c}{\makebox[0pt][c]{\hspace*{-2pt}\makecell[c]{\faAngleDoubleUp\hspace{2pt}Upgrade\\Server+GPU}\hspace*{2pt}}} \\
\midrule
\multirow{3}{*}{\twemoji[height=0.45cm]{flag: Poland}} & Low & 4,535 & \textcolor{gray}{\small{(ref)}} & 6,366 & \textcolor{Red}{\small{$+40\%$}} & 3,807 & \textcolor{ForestGreen}{\small{$-16\%$}} & 3,816 & \textcolor{ForestGreen}{\small{$-16\%$}} & 6,176 & \textcolor{Red}{\small{$+36\%$}} \\
 & Med. & \multicolumn{2}{c}{-} & \multicolumn{2}{c}{-} & 6,359 & \textcolor{gray}{\small{(ref)}} & 6,150 & \textcolor{ForestGreen}{\small{$-3\%$}} & 8,510 & \textcolor{Red}{\small{$+34\%$}} \\
 & High & \multicolumn{2}{c}{-} & \multicolumn{2}{c}{-} & 8,911 & \textcolor{gray}{\small{(ref)}} & 8,485 & \textcolor{ForestGreen}{\small{$-5\%$}} & 10,845 & \textcolor{Red}{\small{$+22\%$}} \\
\cmidrule(lr){1-12}
\multirow{3}{*}{\twemoji[height=0.45cm]{flag: Germany}} & Low & 2,844 & \textcolor{gray}{\small{(ref)}} & 3,808 & \textcolor{Red}{\small{$+34\%$}} & 2,589 & \textcolor{ForestGreen}{\small{$-9\%$}} & 2,704 & \textcolor{ForestGreen}{\small{$-5\%$}} & 5,064 & \textcolor{Red}{\small{$+78\%$}} \\
 & Med. & \multicolumn{2}{c}{-} & \multicolumn{2}{c}{-} & 3,927 & \textcolor{gray}{\small{(ref)}} & 3,930 & \textcolor{gray}{\small{$+0\%$}} & 6,290 & \textcolor{Red}{\small{$+60\%$}} \\
 & High & \multicolumn{2}{c}{-} & \multicolumn{2}{c}{-} & 5,264 & \textcolor{gray}{\small{(ref)}} & 5,156 & \textcolor{ForestGreen}{\small{$-2\%$}} & 7,516 & \textcolor{Red}{\small{$+43\%$}} \\
\cmidrule(lr){1-12}
\multirow{3}{*}{\twemoji[height=0.45cm]{flag: Sweden}} & Low & 1,103 & \textcolor{gray}{\small{(ref)}} & 1,159 & \textcolor{Red}{\small{$+5\%$}} & 1,326 & \textcolor{Red}{\small{$+20\%$}} & 1,546 & \textcolor{Red}{\small{$+40\%$}} & 3,906 & \textcolor{Red}{\small{$+254\%$}} \\
 & Med. & \multicolumn{2}{c}{-} & \multicolumn{2}{c}{-} & 1,405 & \textcolor{gray}{\small{(ref)}} & 1,618 & \textcolor{Red}{\small{$+15\%$}} & 3,978 & \textcolor{Red}{\small{$+183\%$}} \\
 & High & \multicolumn{2}{c}{-} & \multicolumn{2}{c}{-} & 1,484 & \textcolor{gray}{\small{(ref)}} & 1,690 & \textcolor{Red}{\small{$+14\%$}} & 4,050 & \textcolor{Red}{\small{$+173\%$}} \\
\midrule
\multicolumn{2}{l}{Of which embodied:} & 994 & \textcolor{gray}{\small{(ref)}} & 994 & \textcolor{gray}{\small{$+0\%$}} & 1,247 & \textcolor{Red}{\small{$+26\%$}} & 1,474 & \textcolor{Red}{\small{$+48\%$}} & 3,834 & \textcolor{Red}{\small{$+286\%$}}\\
\bottomrule
\end{tabular}
    }%
    \vspace{1em}
    \resizebox{0.9\columnwidth}{!}{
    \begin{tabular}{clr@{ }l!{\hspace{6pt}}r@{ }l!{\hspace{6pt}}r@{ }l!{\hspace{6pt}}r@{ }l!{\hspace{6pt}}r@{ }l!{\hspace{4pt}}}
\multicolumn{12}{c}{\textbf{Water Use} (dm$^3$ world-eq)} \\ \toprule
\multirow{3}[2]{*}{Country} &
\multirow{3}[2]{*}{\makecell[c]{Initial\\Util.}} & 
\multicolumn{10}{c}{Upgrade Scenario} \\ \cmidrule(lr){3-12}
 &  & \multicolumn{2}{c}{\makebox[0pt][c]{\hspace*{-3pt}\makecell[c]{\faHistory\hspace{2pt}Keep\\old GPU}\hspace*{3pt}}} & \multicolumn{2}{c}{\makebox[0pt][c]{\hspace*{-3pt}\makecell[c]{\faPowerOff\hspace{2pt}Keep\\server on}\hspace*{3pt}}} & \multicolumn{2}{c}{\makebox[0pt][c]{\hspace*{-3pt}\makecell[c]{\faAngleDoubleUp\hspace{2pt}Upgrade\\GPU once}\hspace*{3pt}}} & \multicolumn{2}{c}{\makebox[0pt][c]{\hspace*{-3pt}\makecell[c]{\faAngleDoubleUp\hspace{2pt}Upgrade\\GPU twice}\hspace*{3pt}}} & \multicolumn{2}{c}{\makebox[0pt][c]{\hspace*{-2pt}\makecell[c]{\faAngleDoubleUp\hspace{2pt}Upgrade\\Server+GPU}\hspace*{2pt}}} \\
\midrule
\multirow{3}{*}{\twemoji[height=0.45cm]{flag: Poland}} & Low & 198 & \textcolor{gray}{\small{(ref)}} & 297 & \textcolor{Red}{\small{$+50\%$}} & 143 & \textcolor{ForestGreen}{\small{$-28\%$}} & 131 & \textcolor{ForestGreen}{\small{$-34\%$}} & 131 & \textcolor{ForestGreen}{\small{$-34\%$}} \\
 & Med. & \multicolumn{2}{c}{-} & \multicolumn{2}{c}{-} & 286 & \textcolor{gray}{\small{(ref)}} & 261 & \textcolor{ForestGreen}{\small{$-9\%$}} & 261 & \textcolor{ForestGreen}{\small{$-9\%$}} \\
 & High & \multicolumn{2}{c}{-} & \multicolumn{2}{c}{-} & 429 & \textcolor{gray}{\small{(ref)}} & 392 & \textcolor{ForestGreen}{\small{$-9\%$}} & 392 & \textcolor{ForestGreen}{\small{$-9\%$}} \\
\cmidrule(lr){1-12}
\multirow{3}{*}{\twemoji[height=0.45cm]{flag: Germany}} & Low & 126 & \textcolor{gray}{\small{(ref)}} & 190 & \textcolor{Red}{\small{$+51\%$}} & 91 & \textcolor{ForestGreen}{\small{$-27\%$}} & 83 & \textcolor{ForestGreen}{\small{$-34\%$}} & 83 & \textcolor{ForestGreen}{\small{$-34\%$}} \\
 & Med. & \multicolumn{2}{c}{-} & \multicolumn{2}{c}{-} & 182 & \textcolor{gray}{\small{(ref)}} & 167 & \textcolor{ForestGreen}{\small{$-8\%$}} & 167 & \textcolor{ForestGreen}{\small{$-8\%$}} \\
 & High & \multicolumn{2}{c}{-} & \multicolumn{2}{c}{-} & 273 & \textcolor{gray}{\small{(ref)}} & 250 & \textcolor{ForestGreen}{\small{$-8\%$}} & 250 & \textcolor{ForestGreen}{\small{$-8\%$}} \\
\cmidrule(lr){1-12}
\multirow{3}{*}{\twemoji[height=0.45cm]{flag: Sweden}} & Low & 1,098 & \textcolor{gray}{\small{(ref)}} & 1,627 & \textcolor{Red}{\small{$+48\%$}} & 792 & \textcolor{ForestGreen}{\small{$-28\%$}} & 711 & \textcolor{ForestGreen}{\small{$-35\%$}} & 711 & \textcolor{ForestGreen}{\small{$-35\%$}} \\
 & Med. & \multicolumn{2}{c}{-} & \multicolumn{2}{c}{-} & 1,582 & \textcolor{gray}{\small{(ref)}} & 1,420 & \textcolor{ForestGreen}{\small{$-10\%$}} & 1,420 & \textcolor{ForestGreen}{\small{$-10\%$}} \\
 & High & \multicolumn{2}{c}{-} & \multicolumn{2}{c}{-} & 2,372 & \textcolor{gray}{\small{(ref)}} & 2,129 & \textcolor{ForestGreen}{\small{$-10\%$}} & 2,129 & \textcolor{ForestGreen}{\small{$-10\%$}} \\
\midrule
\multicolumn{2}{l}{+ Embodied:} & 3.3e5 & \textcolor{gray}{\small{(ref)}} & 3.3e5 & \textcolor{gray}{\small{$+0\%$}} & 4.0e5 & \textcolor{Red}{\small{$+21\%$}} & 4.5e5 & \textcolor{Red}{\small{$+38\%$}} & 1.2e6 & \textcolor{Red}{\small{$+279\%$}}\\
\bottomrule
\end{tabular}
    }
\end{table}

\begin{table}[ht]
    \centering
    \caption{\textbf{Exponential workload between 2021 and 2026, no GPU power capping.} Analogous to \cref{tab:workload_exponential}, but without capping GPU power consumption.}
    \label{tab:workload_exponential_max_power}
    \resizebox{0.9\columnwidth}{!}{
        \begin{tabular}{clr@{ }l!{\hspace{6pt}}r@{ }l!{\hspace{6pt}}r@{ }l!{\hspace{6pt}}r@{ }l!{\hspace{6pt}}r@{ }l!{\hspace{4pt}}}
\multicolumn{12}{c}{\textbf{Global Warming Potential} (kg CO$_2$eq)} \\ \toprule
\multirow{3}[2]{*}{Country} &
\multirow{3}[2]{*}{\makecell[c]{Initial\\Util.}} & 
\multicolumn{10}{c}{Upgrade Scenario} \\ \cmidrule(lr){3-12}
 &  & \multicolumn{2}{c}{\makebox[0pt][c]{\hspace*{-3pt}\makecell[c]{\faHistory\hspace{2pt}Keep\\old GPU}\hspace*{3pt}}} & \multicolumn{2}{c}{\makebox[0pt][c]{\hspace*{-3pt}\makecell[c]{\faPowerOff\hspace{2pt}Keep\\server on}\hspace*{3pt}}} & \multicolumn{2}{c}{\makebox[0pt][c]{\hspace*{-3pt}\makecell[c]{\faAngleDoubleUp\hspace{2pt}Upgrade\\GPU once}\hspace*{3pt}}} & \multicolumn{2}{c}{\makebox[0pt][c]{\hspace*{-3pt}\makecell[c]{\faAngleDoubleUp\hspace{2pt}Upgrade\\GPU twice}\hspace*{3pt}}} & \multicolumn{2}{c}{\makebox[0pt][c]{\hspace*{-2pt}\makecell[c]{\faAngleDoubleUp\hspace{2pt}Upgrade\\Server+GPU}\hspace*{2pt}}} \\
\midrule
\multirow{3}{*}{\twemoji[height=0.45cm]{flag: Poland}} & Low & 5,171 & \textcolor{gray}{\small{(ref)}} & 7,006 & \textcolor{Red}{\small{$+35\%$}} & 4,140 & \textcolor{ForestGreen}{\small{$-20\%$}} & 4,104 & \textcolor{ForestGreen}{\small{$-21\%$}} & 6,464 & \textcolor{Red}{\small{$+25\%$}} \\
 & Med. & \multicolumn{2}{c}{-} & \multicolumn{2}{c}{-} & 7,025 & \textcolor{gray}{\small{(ref)}} & 6,726 & \textcolor{ForestGreen}{\small{$-4\%$}} & 9,086 & \textcolor{Red}{\small{$+29\%$}} \\
 & High & \multicolumn{2}{c}{-} & \multicolumn{2}{c}{-} & 9,910 & \textcolor{gray}{\small{(ref)}} & 9,349 & \textcolor{ForestGreen}{\small{$-6\%$}} & 11,709 & \textcolor{Red}{\small{$+18\%$}} \\
\cmidrule(lr){1-12}
\multirow{3}{*}{\twemoji[height=0.45cm]{flag: Germany}} & Low & 3,178 & \textcolor{gray}{\small{(ref)}} & 4,144 & \textcolor{Red}{\small{$+30\%$}} & 2,764 & \textcolor{ForestGreen}{\small{$-13\%$}} & 2,855 & \textcolor{ForestGreen}{\small{$-10\%$}} & 5,215 & \textcolor{Red}{\small{$+64\%$}} \\
 & Med. & \multicolumn{2}{c}{-} & \multicolumn{2}{c}{-} & 4,276 & \textcolor{gray}{\small{(ref)}} & 4,233 & \textcolor{ForestGreen}{\small{$-1\%$}} & 6,593 & \textcolor{Red}{\small{$+54\%$}} \\
 & High & \multicolumn{2}{c}{-} & \multicolumn{2}{c}{-} & 5,788 & \textcolor{gray}{\small{(ref)}} & 5,610 & \textcolor{ForestGreen}{\small{$-3\%$}} & 7,970 & \textcolor{Red}{\small{$+38\%$}} \\
\cmidrule(lr){1-12}
\multirow{3}{*}{\twemoji[height=0.45cm]{flag: Sweden}} & Low & 1,123 & \textcolor{gray}{\small{(ref)}} & 1,179 & \textcolor{Red}{\small{$+5\%$}} & 1,337 & \textcolor{Red}{\small{$+19\%$}} & 1,555 & \textcolor{Red}{\small{$+38\%$}} & 3,915 & \textcolor{Red}{\small{$+249\%$}} \\
 & Med. & \multicolumn{2}{c}{-} & \multicolumn{2}{c}{-} & 1,426 & \textcolor{gray}{\small{(ref)}} & 1,636 & \textcolor{Red}{\small{$+15\%$}} & 3,996 & \textcolor{Red}{\small{$+180\%$}} \\
 & High & \multicolumn{2}{c}{-} & \multicolumn{2}{c}{-} & 1,515 & \textcolor{gray}{\small{(ref)}} & 1,716 & \textcolor{Red}{\small{$+13\%$}} & 4,076 & \textcolor{Red}{\small{$+169\%$}} \\
\midrule
\multicolumn{2}{l}{Of which embodied:} & 994 & \textcolor{gray}{\small{(ref)}} & 994 & \textcolor{gray}{\small{$+0\%$}} & 1,247 & \textcolor{Red}{\small{$+26\%$}} & 1,474 & \textcolor{Red}{\small{$+48\%$}} & 3,834 & \textcolor{Red}{\small{$+286\%$}}\\
\bottomrule
\end{tabular}
        }%
    \vspace{1em}
    \resizebox{0.9\columnwidth}{!}{
    \begin{tabular}{clr@{ }l!{\hspace{6pt}}r@{ }l!{\hspace{6pt}}r@{ }l!{\hspace{6pt}}r@{ }l!{\hspace{6pt}}r@{ }l!{\hspace{4pt}}}
\multicolumn{12}{c}{\textbf{Water Use} (dm$^3$ world-eq)} \\ \toprule
\multirow{3}[2]{*}{Country} &
\multirow{3}[2]{*}{\makecell[c]{Initial\\Util.}} & 
\multicolumn{10}{c}{Upgrade Scenario} \\ \cmidrule(lr){3-12}
 &  & \multicolumn{2}{c}{\makebox[0pt][c]{\hspace*{-3pt}\makecell[c]{\faHistory\hspace{2pt}Keep\\old GPU}\hspace*{3pt}}} & \multicolumn{2}{c}{\makebox[0pt][c]{\hspace*{-3pt}\makecell[c]{\faPowerOff\hspace{2pt}Keep\\server on}\hspace*{3pt}}} & \multicolumn{2}{c}{\makebox[0pt][c]{\hspace*{-3pt}\makecell[c]{\faAngleDoubleUp\hspace{2pt}Upgrade\\GPU once}\hspace*{3pt}}} & \multicolumn{2}{c}{\makebox[0pt][c]{\hspace*{-3pt}\makecell[c]{\faAngleDoubleUp\hspace{2pt}Upgrade\\GPU twice}\hspace*{3pt}}} & \multicolumn{2}{c}{\makebox[0pt][c]{\hspace*{-2pt}\makecell[c]{\faAngleDoubleUp\hspace{2pt}Upgrade\\Server+GPU}\hspace*{2pt}}} \\
\midrule
\multirow{3}{*}{\twemoji[height=0.45cm]{flag: Poland}} & Low & 234 & \textcolor{gray}{\small{(ref)}} & 333 & \textcolor{Red}{\small{$+42\%$}} & 162 & \textcolor{ForestGreen}{\small{$-31\%$}} & 147 & \textcolor{ForestGreen}{\small{$-37\%$}} & 147 & \textcolor{ForestGreen}{\small{$-37\%$}} \\
 & Med. & \multicolumn{2}{c}{-} & \multicolumn{2}{c}{-} & 323 & \textcolor{gray}{\small{(ref)}} & 293 & \textcolor{ForestGreen}{\small{$-9\%$}} & 293 & \textcolor{ForestGreen}{\small{$-9\%$}} \\
 & High & \multicolumn{2}{c}{-} & \multicolumn{2}{c}{-} & 484 & \textcolor{gray}{\small{(ref)}} & 439 & \textcolor{ForestGreen}{\small{$-9\%$}} & 439 & \textcolor{ForestGreen}{\small{$-9\%$}} \\
\cmidrule(lr){1-12}
\multirow{3}{*}{\twemoji[height=0.45cm]{flag: Germany}} & Low & 148 & \textcolor{gray}{\small{(ref)}} & 213 & \textcolor{Red}{\small{$+44\%$}} & 103 & \textcolor{ForestGreen}{\small{$-31\%$}} & 94 & \textcolor{ForestGreen}{\small{$-37\%$}} & 94 & \textcolor{ForestGreen}{\small{$-37\%$}} \\
 & Med. & \multicolumn{2}{c}{-} & \multicolumn{2}{c}{-} & 205 & \textcolor{gray}{\small{(ref)}} & 187 & \textcolor{ForestGreen}{\small{$-9\%$}} & 187 & \textcolor{ForestGreen}{\small{$-9\%$}} \\
 & High & \multicolumn{2}{c}{-} & \multicolumn{2}{c}{-} & 308 & \textcolor{gray}{\small{(ref)}} & 280 & \textcolor{ForestGreen}{\small{$-9\%$}} & 280 & \textcolor{ForestGreen}{\small{$-9\%$}} \\
\cmidrule(lr){1-12}
\multirow{3}{*}{\twemoji[height=0.45cm]{flag: Sweden}} & Low & 1,305 & \textcolor{gray}{\small{(ref)}} & 1,836 & \textcolor{Red}{\small{$+41\%$}} & 893 & \textcolor{ForestGreen}{\small{$-32\%$}} & 795 & \textcolor{ForestGreen}{\small{$-39\%$}} & 795 & \textcolor{ForestGreen}{\small{$-39\%$}} \\
 & Med. & \multicolumn{2}{c}{-} & \multicolumn{2}{c}{-} & 1,783 & \textcolor{gray}{\small{(ref)}} & 1,588 & \textcolor{ForestGreen}{\small{$-11\%$}} & 1,588 & \textcolor{ForestGreen}{\small{$-11\%$}} \\
 & High & \multicolumn{2}{c}{-} & \multicolumn{2}{c}{-} & 2,674 & \textcolor{gray}{\small{(ref)}} & 2,381 & \textcolor{ForestGreen}{\small{$-11\%$}} & 2,381 & \textcolor{ForestGreen}{\small{$-11\%$}} \\
\midrule
\multicolumn{2}{l}{+ Embodied:} & 3.3e5 & \textcolor{gray}{\small{(ref)}} & 3.3e5 & \textcolor{gray}{\small{$+0\%$}} & 4.0e5 & \textcolor{Red}{\small{$+21\%$}} & 4.5e5 & \textcolor{Red}{\small{$+38\%$}} & 1.2e6 & \textcolor{Red}{\small{$+279\%$}}\\
\bottomrule
\end{tabular}
    }
\end{table}

\begin{table}[ht]
    \centering
    \caption{\textbf{Adjusted workload between 2021 and 2026.} Environmental impacts of a daily Vision Transformer training workload over six years on a local server under a rebound effect. We estimate operational global warming potential and water use impacts using the average daily carbon intensity and water scarcity footprint in Poland, Germany, and south-central Sweden. \emph{Low}, \emph{Medium}, and \emph{High} utilization correspond to using the starting GPU $30\%$, $60\%$, and $90\%$ of the time, respectively. Daily utilization remains constant upon upgrading hardware or software, leading to an increase in the completed workload and in environmental impacts, especially when newer hardware operates more efficiently at higher power draw levels.}
    \label{tab:adjusted_workload}
    \resizebox{\columnwidth}{!}{
    \begin{tabular}{clr@{ }l!{\hspace{6pt}}r@{ }l!{\hspace{6pt}}r@{ }l!{\hspace{6pt}}r@{ }l!{\hspace{6pt}}r@{ }l!{\hspace{6pt}}r@{ }l!{\hspace{4pt}}}
\multicolumn{14}{c}{\textbf{Global Warming Potential} (kg CO$_2$eq)} \\ \toprule
\multirow{3}[2]{*}{Country} &
\multirow{3}[2]{*}{\makecell[c]{Initial\\Util.}} & 
\multicolumn{12}{c}{Upgrade Scenario} \\ \cmidrule(lr){3-14}
 &  & \multicolumn{2}{c}{\makebox[0pt][c]{\hspace*{-3pt}\makecell[c]{\faHistory\hspace{2pt}Keep\\old GPU}\hspace*{3pt}}} & \multicolumn{2}{c}{\makebox[0pt][c]{\hspace*{-3pt}\makecell[c]{\faHistory\hspace{2pt}Keep old\\PyTorch}\hspace*{3pt}}} & \multicolumn{2}{c}{\makebox[0pt][c]{\hspace*{-3pt}\makecell[c]{\faPowerOff\hspace{2pt}Keep\\server on}\hspace*{3pt}}} & \multicolumn{2}{c}{\makebox[0pt][c]{\hspace*{-3pt}\makecell[c]{\faAngleDoubleUp\hspace{2pt}Upgrade\\GPU once}\hspace*{3pt}}} & \multicolumn{2}{c}{\makebox[0pt][c]{\hspace*{-3pt}\makecell[c]{\faAngleDoubleUp\hspace{2pt}Upgrade\\GPU twice}\hspace*{3pt}}} & \multicolumn{2}{c}{\makebox[0pt][c]{\hspace*{-2pt}\makecell[c]{\faAngleDoubleUp\hspace{2pt}Upgrade\\Server+GPU}\hspace*{2pt}}} \\
\midrule
\multirow{3}{*}{\twemoji[height=0.45cm]{flag: Poland}} & Low & 3,710 & \textcolor{gray}{\small{(ref)}} & 3,690 & \textcolor{ForestGreen}{\small{$-1\%$}} & 5,813 & \textcolor{Red}{\small{$+57\%$}} & 5,731 & \textcolor{Red}{\small{$+54\%$}} & 5,712 & \textcolor{Red}{\small{$+54\%$}} & 8,072 & \textcolor{Red}{\small{$+118\%$}} \\
 & Med. & 6,418 & \textcolor{gray}{\small{(ref)}} & 6,378 & \textcolor{ForestGreen}{\small{$-1\%$}} & 7,620 & \textcolor{Red}{\small{$+19\%$}} & 10,206 & \textcolor{Red}{\small{$+59\%$}} & 9,942 & \textcolor{Red}{\small{$+55\%$}} & 12,302 & \textcolor{Red}{\small{$+92\%$}} \\
 & High & 9,126 & \textcolor{gray}{\small{(ref)}} & 9,066 & \textcolor{ForestGreen}{\small{$-1\%$}} & 9,427 & \textcolor{Red}{\small{$+3\%$}} & 14,682 & \textcolor{Red}{\small{$+61\%$}} & 14,172 & \textcolor{Red}{\small{$+55\%$}} & 16,532 & \textcolor{Red}{\small{$+81\%$}} \\
\cmidrule(lr){1-14}
\multirow{3}{*}{\twemoji[height=0.45cm]{flag: Germany}} & Low & 2,419 & \textcolor{gray}{\small{(ref)}} & 2,409 & \textcolor{gray}{\small{$-0\%$}} & 3,523 & \textcolor{Red}{\small{$+46\%$}} & 3,601 & \textcolor{Red}{\small{$+49\%$}} & 3,702 & \textcolor{Red}{\small{$+53\%$}} & 6,062 & \textcolor{Red}{\small{$+151\%$}} \\
 & Med. & 3,840 & \textcolor{gray}{\small{(ref)}} & 3,820 & \textcolor{ForestGreen}{\small{$-1\%$}} & 4,471 & \textcolor{Red}{\small{$+16\%$}} & 5,951 & \textcolor{Red}{\small{$+55\%$}} & 5,925 & \textcolor{Red}{\small{$+54\%$}} & 8,285 & \textcolor{Red}{\small{$+116\%$}} \\
 & High & 5,261 & \textcolor{gray}{\small{(ref)}} & 5,232 & \textcolor{ForestGreen}{\small{$-1\%$}} & 5,419 & \textcolor{Red}{\small{$+3\%$}} & 8,301 & \textcolor{Red}{\small{$+58\%$}} & 8,149 & \textcolor{Red}{\small{$+55\%$}} & 10,509 & \textcolor{Red}{\small{$+100\%$}} \\
\cmidrule(lr){1-14}
\multirow{3}{*}{\twemoji[height=0.45cm]{flag: Sweden}} & Low & 1,077 & \textcolor{gray}{\small{(ref)}} & 1,076 & \textcolor{gray}{\small{$-0\%$}} & 1,142 & \textcolor{Red}{\small{$+6\%$}} & 1,385 & \textcolor{Red}{\small{$+29\%$}} & 1,604 & \textcolor{Red}{\small{$+49\%$}} & 3,964 & \textcolor{Red}{\small{$+268\%$}} \\
 & Med. & 1,160 & \textcolor{gray}{\small{(ref)}} & 1,159 & \textcolor{gray}{\small{$-0\%$}} & 1,197 & \textcolor{Red}{\small{$+3\%$}} & 1,522 & \textcolor{Red}{\small{$+31\%$}} & 1,734 & \textcolor{Red}{\small{$+49\%$}} & 4,094 & \textcolor{Red}{\small{$+253\%$}} \\
 & High & 1,243 & \textcolor{gray}{\small{(ref)}} & 1,241 & \textcolor{gray}{\small{$-0\%$}} & 1,253 & \textcolor{Red}{\small{$+1\%$}} & 1,660 & \textcolor{Red}{\small{$+34\%$}} & 1,863 & \textcolor{Red}{\small{$+50\%$}} & 4,223 & \textcolor{Red}{\small{$+240\%$}} \\
\midrule
\multicolumn{2}{l}{Of which embodied:} & 994 & \textcolor{gray}{\small{(ref)}} & 994 & \textcolor{gray}{\small{$+0\%$}} & 994 & \textcolor{gray}{\small{$+0\%$}} & 1,247 & \textcolor{Red}{\small{$+26\%$}} & 1,474 & \textcolor{Red}{\small{$+48\%$}} & 3,834 & \textcolor{Red}{\small{$+286\%$}}\\
\bottomrule
\end{tabular}

    }%
    \vspace{1em}
    \resizebox{\columnwidth}{!}{
    \begin{tabular}{clr@{ }l!{\hspace{6pt}}r@{ }l!{\hspace{6pt}}r@{ }l!{\hspace{6pt}}r@{ }l!{\hspace{6pt}}r@{ }l!{\hspace{6pt}}r@{ }l!{\hspace{4pt}}}
\multicolumn{14}{c}{\textbf{Water Use} (dm$^3$ world-eq)} \\ \toprule
\multirow{3}[2]{*}{Country} &
\multirow{3}[2]{*}{\makecell[c]{Initial\\Util.}} & 
\multicolumn{12}{c}{Upgrade Scenario} \\ \cmidrule(lr){3-14}
 &  & \multicolumn{2}{c}{\makebox[0pt][c]{\hspace*{-3pt}\makecell[c]{\faHistory\hspace{2pt}Keep\\old GPU}\hspace*{3pt}}} & \multicolumn{2}{c}{\makebox[0pt][c]{\hspace*{-3pt}\makecell[c]{\faHistory\hspace{2pt}Keep old\\PyTorch}\hspace*{3pt}}} & \multicolumn{2}{c}{\makebox[0pt][c]{\hspace*{-3pt}\makecell[c]{\faPowerOff\hspace{2pt}Keep\\server on}\hspace*{3pt}}} & \multicolumn{2}{c}{\makebox[0pt][c]{\hspace*{-3pt}\makecell[c]{\faAngleDoubleUp\hspace{2pt}Upgrade\\GPU once}\hspace*{3pt}}} & \multicolumn{2}{c}{\makebox[0pt][c]{\hspace*{-3pt}\makecell[c]{\faAngleDoubleUp\hspace{2pt}Upgrade\\GPU twice}\hspace*{3pt}}} & \multicolumn{2}{c}{\makebox[0pt][c]{\hspace*{-2pt}\makecell[c]{\faAngleDoubleUp\hspace{2pt}Upgrade\\Server+GPU}\hspace*{2pt}}} \\
\midrule
\multirow{3}{*}{\twemoji[height=0.45cm]{flag: Poland}} & Low & 149 & \textcolor{gray}{\small{(ref)}} & 147 & \textcolor{ForestGreen}{\small{$-1\%$}} & 264 & \textcolor{Red}{\small{$+77\%$}} & 245 & \textcolor{Red}{\small{$+65\%$}} & 232 & \textcolor{Red}{\small{$+56\%$}} & 232 & \textcolor{Red}{\small{$+56\%$}} \\
 & Med. & 297 & \textcolor{gray}{\small{(ref)}} & 294 & \textcolor{ForestGreen}{\small{$-1\%$}} & 362 & \textcolor{Red}{\small{$+22\%$}} & 490 & \textcolor{Red}{\small{$+65\%$}} & 463 & \textcolor{Red}{\small{$+56\%$}} & 463 & \textcolor{Red}{\small{$+56\%$}} \\
 & High & 445 & \textcolor{gray}{\small{(ref)}} & 441 & \textcolor{ForestGreen}{\small{$-1\%$}} & 461 & \textcolor{Red}{\small{$+4\%$}} & 735 & \textcolor{Red}{\small{$+65\%$}} & 694 & \textcolor{Red}{\small{$+56\%$}} & 694 & \textcolor{Red}{\small{$+56\%$}} \\
\cmidrule(lr){1-14}
\multirow{3}{*}{\twemoji[height=0.45cm]{flag: Germany}} & Low & 96 & \textcolor{gray}{\small{(ref)}} & 95 & \textcolor{ForestGreen}{\small{$-1\%$}} & 170 & \textcolor{Red}{\small{$+77\%$}} & 158 & \textcolor{Red}{\small{$+65\%$}} & 150 & \textcolor{Red}{\small{$+56\%$}} & 150 & \textcolor{Red}{\small{$+56\%$}} \\
 & Med. & 191 & \textcolor{gray}{\small{(ref)}} & 190 & \textcolor{ForestGreen}{\small{$-1\%$}} & 234 & \textcolor{Red}{\small{$+22\%$}} & 316 & \textcolor{Red}{\small{$+65\%$}} & 299 & \textcolor{Red}{\small{$+56\%$}} & 299 & \textcolor{Red}{\small{$+56\%$}} \\
 & High & 287 & \textcolor{gray}{\small{(ref)}} & 285 & \textcolor{ForestGreen}{\small{$-1\%$}} & 298 & \textcolor{Red}{\small{$+4\%$}} & 474 & \textcolor{Red}{\small{$+65\%$}} & 448 & \textcolor{Red}{\small{$+56\%$}} & 448 & \textcolor{Red}{\small{$+56\%$}} \\
\cmidrule(lr){1-14}
\multirow{3}{*}{\twemoji[height=0.45cm]{flag: Sweden}} & Low & 808 & \textcolor{gray}{\small{(ref)}} & 801 & \textcolor{ForestGreen}{\small{$-1\%$}} & 1,433 & \textcolor{Red}{\small{$+77\%$}} & 1,339 & \textcolor{Red}{\small{$+66\%$}} & 1,248 & \textcolor{Red}{\small{$+54\%$}} & 1,248 & \textcolor{Red}{\small{$+54\%$}} \\
 & Med. & 1,614 & \textcolor{gray}{\small{(ref)}} & 1,600 & \textcolor{ForestGreen}{\small{$-1\%$}} & 1,971 & \textcolor{Red}{\small{$+22\%$}} & 2,677 & \textcolor{Red}{\small{$+66\%$}} & 2,494 & \textcolor{Red}{\small{$+54\%$}} & 2,494 & \textcolor{Red}{\small{$+54\%$}} \\
 & High & 2,421 & \textcolor{gray}{\small{(ref)}} & 2,398 & \textcolor{ForestGreen}{\small{$-1\%$}} & 2,510 & \textcolor{Red}{\small{$+4\%$}} & 4,014 & \textcolor{Red}{\small{$+66\%$}} & 3,740 & \textcolor{Red}{\small{$+55\%$}} & 3,740 & \textcolor{Red}{\small{$+55\%$}} \\
\midrule
\multicolumn{2}{l}{+ Embodied:} & 3.3e5 & \textcolor{gray}{\small{(ref)}} & 3.3e5 & \textcolor{gray}{\small{$+0\%$}} & 3.3e5 & \textcolor{gray}{\small{$+0\%$}} & 4.0e5 & \textcolor{Red}{\small{$+21\%$}} & 4.5e5 & \textcolor{Red}{\small{$+38\%$}} & 1.2e6 & \textcolor{Red}{\small{$+279\%$}}\\
\bottomrule
\end{tabular}

    }
\end{table}

\begin{table}[ht]
    \centering
    \caption{\textbf{Adjusted workload between 2021 and 2026, no GPU power capping.} Analogous to \cref{tab:adjusted_workload}, but without capping GPU power consumption.}
    \label{tab:adjusted_workload_max_power}
    \resizebox{\columnwidth}{!}{
        \begin{tabular}{clr@{ }l!{\hspace{6pt}}r@{ }l!{\hspace{6pt}}r@{ }l!{\hspace{6pt}}r@{ }l!{\hspace{6pt}}r@{ }l!{\hspace{6pt}}r@{ }l!{\hspace{4pt}}}
\multicolumn{14}{c}{\textbf{Global Warming Potential} (kg CO$_2$eq)} \\ \toprule
\multirow{3}[2]{*}{Country} &
\multirow{3}[2]{*}{\makecell[c]{Initial\\Util.}} & 
\multicolumn{12}{c}{Upgrade Scenario} \\ \cmidrule(lr){3-14}
 &  & \multicolumn{2}{c}{\makebox[0pt][c]{\hspace*{-3pt}\makecell[c]{\faHistory\hspace{2pt}Keep\\old GPU}\hspace*{3pt}}} & \multicolumn{2}{c}{\makebox[0pt][c]{\hspace*{-3pt}\makecell[c]{\faHistory\hspace{2pt}Keep old\\PyTorch}\hspace*{3pt}}} & \multicolumn{2}{c}{\makebox[0pt][c]{\hspace*{-3pt}\makecell[c]{\faPowerOff\hspace{2pt}Keep\\server on}\hspace*{3pt}}} & \multicolumn{2}{c}{\makebox[0pt][c]{\hspace*{-3pt}\makecell[c]{\faAngleDoubleUp\hspace{2pt}Upgrade\\GPU once}\hspace*{3pt}}} & \multicolumn{2}{c}{\makebox[0pt][c]{\hspace*{-3pt}\makecell[c]{\faAngleDoubleUp\hspace{2pt}Upgrade\\GPU twice}\hspace*{3pt}}} & \multicolumn{2}{c}{\makebox[0pt][c]{\hspace*{-2pt}\makecell[c]{\faAngleDoubleUp\hspace{2pt}Upgrade\\Server+GPU}\hspace*{2pt}}} \\
\midrule
\multirow{3}{*}{\twemoji[height=0.45cm]{flag: Poland}} & Low & 4,202 & \textcolor{gray}{\small{(ref)}} & 3,988 & \textcolor{ForestGreen}{\small{$-5\%$}} & 6,305 & \textcolor{Red}{\small{$+50\%$}} & 6,143 & \textcolor{Red}{\small{$+46\%$}} & 6,012 & \textcolor{Red}{\small{$+43\%$}} & 8,372 & \textcolor{Red}{\small{$+99\%$}} \\
 & Med. & 7,404 & \textcolor{gray}{\small{(ref)}} & 6,975 & \textcolor{ForestGreen}{\small{$-6\%$}} & 8,605 & \textcolor{Red}{\small{$+16\%$}} & 11,032 & \textcolor{Red}{\small{$+49\%$}} & 10,542 & \textcolor{Red}{\small{$+42\%$}} & 12,902 & \textcolor{Red}{\small{$+74\%$}} \\
 & High & 10,605 & \textcolor{gray}{\small{(ref)}} & 9,962 & \textcolor{ForestGreen}{\small{$-6\%$}} & 10,905 & \textcolor{Red}{\small{$+3\%$}} & 15,920 & \textcolor{Red}{\small{$+50\%$}} & 15,072 & \textcolor{Red}{\small{$+42\%$}} & 17,432 & \textcolor{Red}{\small{$+64\%$}} \\
\cmidrule(lr){1-14}
\multirow{3}{*}{\twemoji[height=0.45cm]{flag: Germany}} & Low & 2,679 & \textcolor{gray}{\small{(ref)}} & 2,566 & \textcolor{ForestGreen}{\small{$-4\%$}} & 3,783 & \textcolor{Red}{\small{$+41\%$}} & 3,816 & \textcolor{Red}{\small{$+42\%$}} & 3,859 & \textcolor{Red}{\small{$+44\%$}} & 6,219 & \textcolor{Red}{\small{$+132\%$}} \\
 & Med. & 4,360 & \textcolor{gray}{\small{(ref)}} & 4,134 & \textcolor{ForestGreen}{\small{$-5\%$}} & 4,991 & \textcolor{Red}{\small{$+14\%$}} & 6,382 & \textcolor{Red}{\small{$+46\%$}} & 6,240 & \textcolor{Red}{\small{$+43\%$}} & 8,600 & \textcolor{Red}{\small{$+97\%$}} \\
 & High & 6,041 & \textcolor{gray}{\small{(ref)}} & 5,702 & \textcolor{ForestGreen}{\small{$-6\%$}} & 6,199 & \textcolor{Red}{\small{$+3\%$}} & 8,947 & \textcolor{Red}{\small{$+48\%$}} & 8,622 & \textcolor{Red}{\small{$+43\%$}} & 10,982 & \textcolor{Red}{\small{$+82\%$}} \\
\cmidrule(lr){1-14}
\multirow{3}{*}{\twemoji[height=0.45cm]{flag: Sweden}} & Low & 1,092 & \textcolor{gray}{\small{(ref)}} & 1,086 & \textcolor{ForestGreen}{\small{$-1\%$}} & 1,157 & \textcolor{Red}{\small{$+6\%$}} & 1,398 & \textcolor{Red}{\small{$+28\%$}} & 1,613 & \textcolor{Red}{\small{$+48\%$}} & 3,973 & \textcolor{Red}{\small{$+264\%$}} \\
 & Med. & 1,191 & \textcolor{gray}{\small{(ref)}} & 1,177 & \textcolor{ForestGreen}{\small{$-1\%$}} & 1,228 & \textcolor{Red}{\small{$+3\%$}} & 1,548 & \textcolor{Red}{\small{$+30\%$}} & 1,752 & \textcolor{Red}{\small{$+47\%$}} & 4,112 & \textcolor{Red}{\small{$+245\%$}} \\
 & High & 1,289 & \textcolor{gray}{\small{(ref)}} & 1,269 & \textcolor{ForestGreen}{\small{$-2\%$}} & 1,298 & \textcolor{Red}{\small{$+1\%$}} & 1,698 & \textcolor{Red}{\small{$+32\%$}} & 1,891 & \textcolor{Red}{\small{$+47\%$}} & 4,251 & \textcolor{Red}{\small{$+230\%$}} \\
\midrule
\multicolumn{2}{l}{Of which embodied:} & 994 & \textcolor{gray}{\small{(ref)}} & 994 & \textcolor{gray}{\small{$+0\%$}} & 994 & \textcolor{gray}{\small{$+0\%$}} & 1,247 & \textcolor{Red}{\small{$+26\%$}} & 1,474 & \textcolor{Red}{\small{$+48\%$}} & 3,834 & \textcolor{Red}{\small{$+286\%$}}\\
\bottomrule
\end{tabular}

    }%
    \vspace{1em}
    \resizebox{\columnwidth}{!}{
        \begin{tabular}{clr@{ }l!{\hspace{6pt}}r@{ }l!{\hspace{6pt}}r@{ }l!{\hspace{6pt}}r@{ }l!{\hspace{6pt}}r@{ }l!{\hspace{6pt}}r@{ }l!{\hspace{4pt}}}
\multicolumn{14}{c}{\textbf{Water Use} (dm$^3$ world-eq)} \\ \toprule
\multirow{3}[2]{*}{Country} &
\multirow{3}[2]{*}{\makecell[c]{Initial\\Util.}} & 
\multicolumn{12}{c}{Upgrade Scenario} \\ \cmidrule(lr){3-14}
 &  & \multicolumn{2}{c}{\makebox[0pt][c]{\hspace*{-3pt}\makecell[c]{\faHistory\hspace{2pt}Keep\\old GPU}\hspace*{3pt}}} & \multicolumn{2}{c}{\makebox[0pt][c]{\hspace*{-3pt}\makecell[c]{\faHistory\hspace{2pt}Keep old\\PyTorch}\hspace*{3pt}}} & \multicolumn{2}{c}{\makebox[0pt][c]{\hspace*{-3pt}\makecell[c]{\faPowerOff\hspace{2pt}Keep\\server on}\hspace*{3pt}}} & \multicolumn{2}{c}{\makebox[0pt][c]{\hspace*{-3pt}\makecell[c]{\faAngleDoubleUp\hspace{2pt}Upgrade\\GPU once}\hspace*{3pt}}} & \multicolumn{2}{c}{\makebox[0pt][c]{\hspace*{-3pt}\makecell[c]{\faAngleDoubleUp\hspace{2pt}Upgrade\\GPU twice}\hspace*{3pt}}} & \multicolumn{2}{c}{\makebox[0pt][c]{\hspace*{-2pt}\makecell[c]{\faAngleDoubleUp\hspace{2pt}Upgrade\\Server+GPU}\hspace*{2pt}}} \\
\midrule
\multirow{3}{*}{\twemoji[height=0.45cm]{flag: Poland}} & Low & 176 & \textcolor{gray}{\small{(ref)}} & 164 & \textcolor{ForestGreen}{\small{$-7\%$}} & 291 & \textcolor{Red}{\small{$+65\%$}} & 268 & \textcolor{Red}{\small{$+52\%$}} & 248 & \textcolor{Red}{\small{$+41\%$}} & 248 & \textcolor{Red}{\small{$+41\%$}} \\
 & Med. & 352 & \textcolor{gray}{\small{(ref)}} & 327 & \textcolor{ForestGreen}{\small{$-7\%$}} & 417 & \textcolor{Red}{\small{$+19\%$}} & 535 & \textcolor{Red}{\small{$+52\%$}} & 495 & \textcolor{Red}{\small{$+41\%$}} & 495 & \textcolor{Red}{\small{$+41\%$}} \\
 & High & 527 & \textcolor{gray}{\small{(ref)}} & 490 & \textcolor{ForestGreen}{\small{$-7\%$}} & 544 & \textcolor{Red}{\small{$+3\%$}} & 803 & \textcolor{Red}{\small{$+52\%$}} & 742 & \textcolor{Red}{\small{$+41\%$}} & 742 & \textcolor{Red}{\small{$+41\%$}} \\
\cmidrule(lr){1-14}
\multirow{3}{*}{\twemoji[height=0.45cm]{flag: Germany}} & Low & 113 & \textcolor{gray}{\small{(ref)}} & 106 & \textcolor{ForestGreen}{\small{$-7\%$}} & 188 & \textcolor{Red}{\small{$+65\%$}} & 173 & \textcolor{Red}{\small{$+52\%$}} & 160 & \textcolor{Red}{\small{$+41\%$}} & 160 & \textcolor{Red}{\small{$+41\%$}} \\
 & Med. & 227 & \textcolor{gray}{\small{(ref)}} & 211 & \textcolor{ForestGreen}{\small{$-7\%$}} & 269 & \textcolor{Red}{\small{$+19\%$}} & 345 & \textcolor{Red}{\small{$+52\%$}} & 320 & \textcolor{Red}{\small{$+41\%$}} & 320 & \textcolor{Red}{\small{$+41\%$}} \\
 & High & 340 & \textcolor{gray}{\small{(ref)}} & 316 & \textcolor{ForestGreen}{\small{$-7\%$}} & 350 & \textcolor{Red}{\small{$+3\%$}} & 518 & \textcolor{Red}{\small{$+52\%$}} & 479 & \textcolor{Red}{\small{$+41\%$}} & 479 & \textcolor{Red}{\small{$+41\%$}} \\
\cmidrule(lr){1-14}
\multirow{3}{*}{\twemoji[height=0.45cm]{flag: Sweden}} & Low & 965 & \textcolor{gray}{\small{(ref)}} & 890 & \textcolor{ForestGreen}{\small{$-8\%$}} & 1,589 & \textcolor{Red}{\small{$+65\%$}} & 1,462 & \textcolor{Red}{\small{$+52\%$}} & 1,329 & \textcolor{Red}{\small{$+38\%$}} & 1,329 & \textcolor{Red}{\small{$+38\%$}} \\
 & Med. & 1,927 & \textcolor{gray}{\small{(ref)}} & 1,777 & \textcolor{ForestGreen}{\small{$-8\%$}} & 2,284 & \textcolor{Red}{\small{$+19\%$}} & 2,921 & \textcolor{Red}{\small{$+52\%$}} & 2,655 & \textcolor{Red}{\small{$+38\%$}} & 2,655 & \textcolor{Red}{\small{$+38\%$}} \\
 & High & 2,890 & \textcolor{gray}{\small{(ref)}} & 2,664 & \textcolor{ForestGreen}{\small{$-8\%$}} & 2,979 & \textcolor{Red}{\small{$+3\%$}} & 4,381 & \textcolor{Red}{\small{$+52\%$}} & 3,982 & \textcolor{Red}{\small{$+38\%$}} & 3,982 & \textcolor{Red}{\small{$+38\%$}} \\
\midrule
\multicolumn{2}{l}{+ Embodied:} & 3.3e5 & \textcolor{gray}{\small{(ref)}} & 3.3e5 & \textcolor{gray}{\small{$+0\%$}} & 3.3e5 & \textcolor{gray}{\small{$+0\%$}} & 4.0e5 & \textcolor{Red}{\small{$+21\%$}} & 4.5e5 & \textcolor{Red}{\small{$+38\%$}} & 1.2e6 & \textcolor{Red}{\small{$+279\%$}}\\
\bottomrule
\end{tabular}

    }
\end{table}

\end{document}